\documentclass[a4paper,fleqn,usenatbib]{mnras}

\usepackage{mathptmx}
\usepackage[T1]{fontenc}
\usepackage{ae,aecompl}

\usepackage{graphicx}	% Including figure files
\usepackage{amsmath}	% Advanced maths commands
\usepackage{amssymb}	% Extra maths symbols

\newcommand{\cmnt}[1]{}
\newcommand{\review}[1]{#1}

\newcommand{\software}[1]{\texttt{\sc #1}}

\newcommand{\porb}{$P_\mathrm{orb}$}
\newcommand{\teff}{$T_\mathrm{eff}$}

\newcommand{\tlv}{$^1$}
\newcommand{\warwick}{$^2$}
\newcommand{\weissman}{$^3$}
\newcommand{\harvard}{$^4$}
\newcommand{\mpia}{$^5$}

\newcommand{\samplesize}{15\,779}
\newcommand{\inputsize}{4\,301\,148}
\newcommand{\longinputsize}{8\,975\,643}
\newcommand{\scorethreshold}{0.6}

\title[Ellipsoidal Binaries in TESS]{15000 Ellipsoidal Binary Candidates in \textit{TESS}: Orbital Periods, Binary Fraction, and Tertiary Companions}

\author[M. J. Green]{
Matthew J. Green\tlv$^,$\warwick\thanks{E-mail: mjgreenastro@gmail.com},
Dan Maoz\tlv\thanks{E-mail: danmaoz1@gmail.com},
Tsevi Mazeh\tlv,
Simchon Faigler\tlv,
\newauthor
Sahar Shahaf\weissman$^,$\tlv,
Roy Gomel\tlv,
Kareem El-Badry\harvard$^,$\mpia,
and Hans-Walter Rix\mpia.
\\
\tlv School of Physics and Astronomy, Tel-Aviv University, Tel-Aviv 6997801, Israel
\\
\warwick Astronomy and Astrophysics Group, Department of Physics, University of Warwick, Coventry, CV4 7AL, United Kingdom
\\
\weissman Department of Particle Physics and Astrophysics, Weizmann Institute of Science, Rehovot 7610001, Israel
\\
\harvard Center for Astrophysics | Harvard \& Smithsonian, 60 Garden Street, Cambridge, MA 02138, USA
\\
\mpia Max-Planck Institute for Astronomy, Königstuhl 17, D-69117 Heidelberg, Germany
}

\date{Accepted XXX. Received YYY; in original form ZZZ}

\pubyear{2023}

\begin{document}
\label{firstpage}
\pagerange{\pageref{firstpage}--\pageref{lastpage}}
\maketitle

\begin{abstract}
We present a homogeneously-selected sample of \samplesize\ candidate binary systems with main sequence primary stars and orbital periods shorter than 5 days.
The targets were selected from \textit{TESS} full-frame image lightcurves on the basis of their tidally-induced ellipsoidal modulation.
Spectroscopic follow-up suggests a sample purity of $83 \pm 13$ per cent.
Injection-recovery tests allow us to estimate our overall completeness as $28 \pm 3$ per cent with \porb\,$< 3$\,days and to quantify our selection effects.
$39 \pm 4$ per cent of our sample are contact binary systems, and we disentangle the period distributions of the contact and detached binaries.
We derive the orbital period distribution of the main sequence binary population at short orbital periods, finding a distribution continuous with the log-normal distribution previously found for solar-type stars at longer periods, but with a significant steepening  at \porb\,$\lesssim 3$\,days, and a pile-up of contact binaries at \porb\,$\approx 0.4$\,days.
Companions in the period range 1--5 days are an order of magnitude more frequent around stars hotter than $\approx 6250\,K$ (the Kraft break) when compared to cooler stars, suggesting that magnetic braking shortens the lifetime of cooler binary systems.
However, the period distribution in the range 1--10 days is independent of temperature.
We detect resolved tertiary companions to $9.0 \pm 0.2$ per cent of our binaries with a median separation of 3200\,AU.
The frequency of tertiary companions rises to $29 \pm 5$ per cent among the systems with the shortest ellipsoidal periods. 
This large binary sample with quantified selection effects will be a powerful resource for future studies of detached and contact binary systems with \porb$<5$\,days.
\cmnt{word count: 246/250}
\end{abstract}


%%%%%%%%%%%%%%%%%%%%%%%%%%%%%%%%%%%%%%%%%%%%%%%%%%

%%%%%%%%%%%%%%%%% BODY OF PAPER %%%%%%%%%%%%%%%%%%

\section{Introduction}
\label{sec:introduction}

Binary star systems, particularly those with short orbital periods, play numerous important roles in astrophysics, such as as the progenitors of merged stars, cataclysmic variables, several types of supernova explosions, and gravitational-wave-detected merging neutron stars and black holes, to name just a few. 
However, the complex evolutionary processes that lead from an initial main-sequence pair to those final stages remain poorly understood \citep{Ivanova2013,Khouri2022}.
A necessary first step for understanding the physics and evolution of binaries in their many astrophysical roles is to characterize the observed binary population. 

\subsection{Characterizing binary populations}

The review by \citet{Moe2017} drew upon a number of samples of binary stars, the largest of which was a study of 454 FGK-type stars with a total of 168 confirmed companions \citep{Raghavan2010}.
Combining these samples, and correcting each to account for their selection effects, \citet{Moe2017} characterised the binary population in terms of companion frequency, orbital period, mass ratio, and eccentricity.
They found that the companion frequency increases significantly with primary star mass, from $0.50 \pm 0.04$ for solar-type stars to $2.1 \pm 0.3$ for O-type primary stars.
The orbital period distribution of FGK binary systems can be approximated by a log-normal distribution centred on a period of $\approx 10^5$ days.
A follow-up work by \citet{Moe2019} also showed that the frequency of binary companions typically increases with decreasing metallicity.

More recently than the \citet{Raghavan2010} sample, larger samples of binary systems have been selected using large-scale surveys. 
For binary systems with long orbital periods and wide orbital separations, a sample of 1.1\,million resolved binary systems with $a \gtrsim 50$\,AU was selected from \textit{Gaia}, and subsets characterised, by \citet{El-Badry2018,El-Badry2019a} and \citet{El-Badry2019,El-Badry2021a}.
A sample of 392 resolved triple systems was also selected from \textit{Gaia} by \citet{Tokovinin2022}.
\citet{Hartman2022} examined a sample of resolved K+K binaries, and found at least 40 per cent to be hierarchical triple systems with an unresolved inner binary.

Large samples of unresolved binary systems, with a diversity of orbital period ranges, have been selected from photometric, spectroscopic, and astrometric surveys.
Data Release 3 of the \textit{Gaia} survey presented orbital solutions for approximately 87\,000 eclipsing binaries with orbital periods $\lesssim 200$\,days, 220\,000 spectroscopic binaries with orbital periods $\lesssim 2000$\,days, and 170\,000 astrometric binaries with orbital periods $\lesssim 10\,000$\,days, with some overlap between the samples \citep{GaiaCollaboration2022,Halbwachs2022,Bashi2022,Shahaf2022}.
\citet{Hwang2022} searched for resolved tertiary companions (with separations of $10^3$--$10^4$ AU) to \textit{Gaia} binary systems and found that the frequency of companions was significantly enhanced for eclipsing inner binaries, and reduced for astrometric inner binaries, relative to the general stellar population.

Spectroscopically, the binary population has been explored with the Apache Point Observatory Galactic Evolution Experiment (APOGEE), a part of the Sloan Digital Sky Survey (SDSS).
The frequency of binary companions was explored as a function of the location of the primary star on the Hertzsprung-Russel diagram \citep{Badenes2018}, and a sample of 20\,000 binaries was compiled with orbital periods $\lesssim$ 20\,000 days \citep{Price-Whelan2020}.
Binarity was found to decrease as the primary star evolves off the main sequence, in a way that is consistent with binary systems ceasing to appear as binaries at the point of contact between the component stars.
APOGEE data have also been used to explore the dependence of binarity on both metallicity and the abundence of $\alpha$-elements \citep{Mazzola2020}, and to characterise the subset of binaries and higher-order systems which are double-lined in the APOGEE spectra \citep{El-Badry2018a,Kounkel2021}.

Samples of eclipsing binary systems include 2878 eclipsing binaries from the \textit{Kepler} survey \citep{Prsa2011,Kirk2016}, 450\,000 eclipsing binaries from the Optical Gravitational Lensing Experiment \citep[OGLE;][]{Soszynski2016}, 4584 eclipsing binaries from the Transiting Exoplanet Survey Satellite \citep[\textit{TESS};][]{Prsa2022}, 35\,000 eclipsing binaries from the All Sky Automated Search for SuperNovae \citep[ASAS-SN;][]{Rowan2022}, and 3879 eclipsing binaries from the Zwicky Transient Facility \citep[ZTF;][]{El-Badry2022b}.

If the completeness of a sample is low (either overall, or in particular regions of parameter space), then uncertainties in the selection function will dominate our understanding of the underlying population.
Quantifying the selection effects of a sample is rarely a simple matter.
In particular, machine learning methods or by-eye inspection are widely used during the classification of variable stars.
For both of these methods, it is not trivial to quantify the selection efficiency or to verify whether any parameter-dependent selection effects were introduced in the process.

For most of the samples discussed above, there have so far been only a few attempts to quantify or correct for the selection functions. 
Alongside the ZTF sample of eclipsing binary systems, \citet{El-Badry2022b} presented injection-recovery tests which allowed them to reconstruct the underlying period distribution.
\citet{Kirk2016} and \citet{Moe2019} approximated the selection function of the \textit{Kepler} sample of eclipsing binary systems at short periods as proportional to $(R_1 + R_2) / a$, where $R_1$ and $R_2$ refer to the component stellar radii and $a$ to the orbital separation; however, they acknowledged that at short periods, the presence of ellipsoidal modulation may introduce further selection effects that were not modelled.
There is thus a considerable advantage to selection methods that are relatively complete and whose selection functions are easy to model.

\subsection{Ellipsoidal modulation}

A relatively under-utilised method by which close binary systems may be identified is to search for their ellipsoidal modulation.
Binary systems with small orbital separations ($a \lesssim 15 R_\odot$ for two Sun-like stars) show a number of photometric signatures that reveal their binary nature. 
Assuming that the binary does not eclipse, the strongest of these signatures is usually the ellipsoidal modulation, which results from the tidal deformation of each of the two stars by the gravity of the other \citep{Kopal1959,Morris1985}. 
Other photometric signatures will include the reflection of each star's light from the surface of the other, and the relativistic Doppler beaming of the light due to the velocities of the stars \citep{Rybicki1979,Morris1993}.\footnote{The ellipsoidal effect is typically the most significant for binaries consisting of main sequence or giant stars, but note that, in the case that one of the stars is a luminous, compact object such as a white dwarf or hot subdwarf, the reflection effect will often be stronger than the ellipsoidal.}

These photometric signals -- BEaming, Ellipsoidal modulation, and Reflection (collectively, BEER) -- can be predicted with reasonable accuracy based on the component stellar properties and a relatively small number of assumptions \citep{Morris1993, Zucker2007, Faigler2011, Gomel2021a}.
In combination, these signatures present a useful tool by which close binaries can be selected from a photometric survey.
Because the selection criteria are relatively simple and well understood, it is not too challenging to recover the selection efficiency as a function of binary parameters, and hence reconstruct the underlying population from which the sample was selected. 

There are several differences between the characteristics of an ellipsoidal selection strategy and those of a search for eclipses. 
Firstly, only a small fraction of binaries eclipse (e.g.~$\approx$25\% for two sun-like stars with orbital period \porb$=2$\,days). 
A much greater fraction of close binaries show a detectable ellipsoidal signals ($\approx$85\% for a \textit{TESS} lightcurve of the same binary with magnitude $G=12$).
Secondly, the selection biases are somewhat different to an eclipsing binary search.
For a binary system with primary radius $R_1$, the probability of an eclipse is approximately\footnote{\review{Note that a number of simplifying assumptions are present in this approximation. Note also that, for contact binary systems, the true relation is quite different.}} proportional to $R_1$\porb$^{-2/3} \sin i$. Ellipsoidal modulation, in contrast, {\it always} occurs (probability of one), but the amplitude of the ellipsoidal signal is approximately proportional to $R_1^3 M_1^{-1} $\porb$^{-2} \sin^2 i$.
An ellipsoidal sample can therefore achieve a significantly greater completeness than that of an eclipsing sample at short orbital periods ($\lesssim 3$\,days), but its sensitivity will rapidly decrease towards longer orbital periods.

The greatest disadvantage of an ellipsoidal survey in comparison to an eclipsing survey is that it is more susceptible to contamination from photometrically similar variables, such as certain pulsators or configurations of star spots; however, with careful selection it is possible to create a sample with a high purity, as we demonstrate in Section~\ref{sec:purity}.

There have been several previous photometric searches for ellipsoidal binaries.
\citet{Faigler2013,Faigler2015b}, \citet{Faigler2015a}, and \citet{Tal-Or2015} analysed lightcurves from \textit{Kepler} and \textit{CoRoT}, from which they identified 70 binaries with luminous companions, primarily main sequence--main sequence (MS--MS) binaries; four main sequence--white dwarf (MS--WD) binaries; and one planet. 
\citet{Rowan2021} analysed lightcurves from ASAS-SN and selected a sample of 369 candidate ellipsoidal binaries with a priority on purity.
\citet{Soszynski2016} identified 25\,405 candidate ellipsoidal binaries from the OGLE survey, from which \citet{Gomel2021c} selected 136 candidate main sequence--black hole (MS--BH) and main sequence--neutron star (MS--NS) binaries.
These searches have typically focused on the prospect of discovering non-accreting MS--BH and MS--NS binaries. 
Such binaries are expected to exist, but only a few discoveries have been claimed, and several of those claims are disputed \citep{Thompson2019,Liu2019,Rivinius2020,El-Badry2020,El-Badry2021,Jayasinghe2021,El-Badry2022,El-Badry2022a,Shenar2022,Mazeh2022}.
In this work, we concentrate on what can be learned about the overall binary population characteristics from the BEER signatures of a large sample of MS--MS ellipsoidal binaries.

\subsection{Contact binary systems}

At orbital periods shorter than a day, many binary systems exist in a configuration where either one or both stars overfill their Roche lobe.
Contact binaries\footnote{In this work, we adopt the terminology `contact' binaries for systems where both stars fill their Roche lobes, `semi-detached' for systems where one star fills its Roche lobe, and `detached' for systems where neither star fills its Roche lobe. We consider all of these systems to be subtypes of ellipsoidal binaries.} typically have FGK stellar types and orbital periods close to eight hours. 
They come into contact at short periods and evolve to longer periods as a result of phases of mass transfer in alternating directions.
This period change is slow and the systems have a long ($\sim$ Myr -- Gyr) lifetime, while the temperatures of the two stars equalize on a relatively short timescale \citep[see the reviews in][]{Eggleton2012,Kobulnicky2022}. 
There are a number open questions in the study of contact binary systems, including their orbital period cut-off \citep{Zhang2020}, the relations between orbital period, temperature, and mass \citep{Pawlak2016,Jayasinghe2020,Poro2022}, and the frequency of tertiary companions \citep{Kobulnicky2022}.

In all of the short-period binary samples cited above, contact binary systems were either removed from the sample using a by-eye inspection \citep[e.g.][]{El-Badry2022b}, classified separately by eye \citep[e.g.][]{Kirk2016,Prsa2022}, or diverted to a separate sample using a machine-learning classification \citep[e.g.][]{Soszynski2016}.
However, given the quite limited success rates of previous studies in unambiguously discerning between contact and detached binaries based on photometric data alone, any classification method that attempts to separate contact and detached binaries may introduce additional selection effects into the final sample.
There is therefore value in a homogeneously-selected sample that includes both contact and detached binary systems.

We take an approach in which both contact and detached binary systems are included in the sample as efficiently as possible, while we handle the two classes differently on a statistical level when correcting for their respective selection effects.
We propose a probabilistic classification for each target based on simulated light curves, but leave to future users of the sample the option to perform their own classifications, if preferred.

\subsection{Ellipsoidal sample}

We present a sample of \samplesize\ candidate, ellipsoidal, MS--MS binaries which show BEER-like signatures in their \textit{TESS} lightcurves.
The sample is made publicly available alongside this paper.
In this work we also perform a preliminary analysis in which we attempt to disentangle contact and detached binary systems, and estimate the orbital period distribution, the frequency of companions as a function primary stellar mass, and the frequency of tertiary companions.

The layout of the paper is as follows. 
In Section~\ref{sec:data}, we describe the origin and quality filtering of our input \textit{TESS} lightcurves.
Section~\ref{sec:beer} presents the \software{beer} algorithm used to select our candidates.
Section~\ref{sec:purity} describes spectroscopic follow-up obtained for a subset of targets to ensure the validity of the sample.
In Section~\ref{sec:simulations} we perform injection-recovery tests to explore the completeness and selection functions of the sample.
Section~\ref{sec:contacts} discusses the handling and classification of contact binary systems in our sample.
Section~\ref{sec:sample-properties} derives the physical property distributions of the sample and the implied distributions of the underlying population, after correcting for selection effects.
In Section~\ref{sec:discussion} we discuss our sample and results and compare them to the existing literature. We summarise our conclusions in Section~\ref{sec:conclusion}.

\section{\textit{TESS} Data}
\label{sec:data}

\begin{figure*}
\includegraphics[width=\columnwidth]{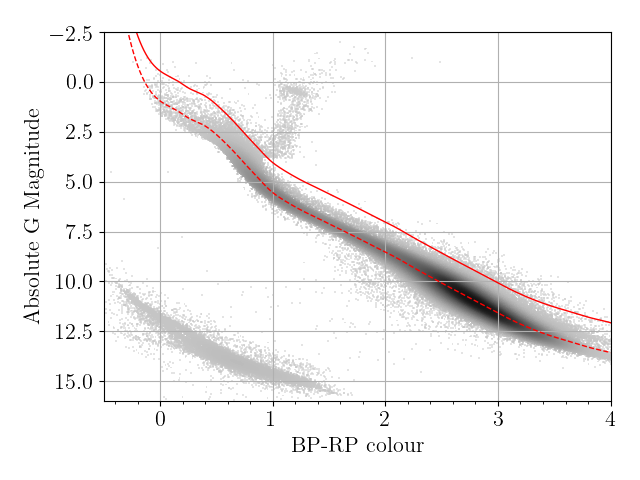}
\includegraphics[width=\columnwidth]{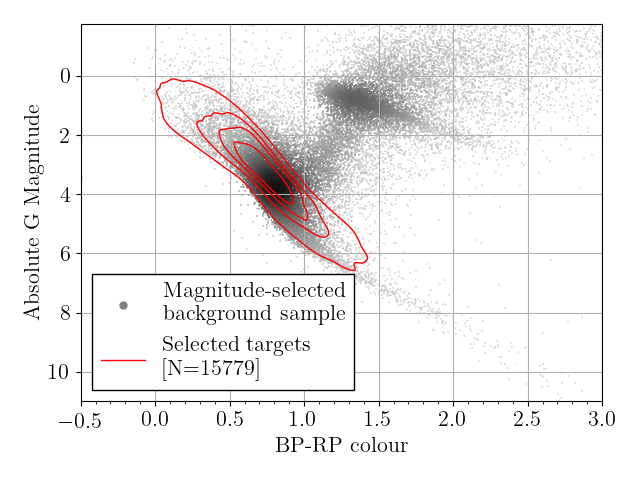}
\caption{\textit{Left:} A colour-magnitude diagram of stars within 100\,pc, from \textit{Gaia} \citep{GaiaCollaboration2022}.
We plot a spline fit to the main sequence (red dashed line) and the cut applied to remove subgiants and giants (red solid line), which is offset by 1.5 magnitudes, as described in Section~\ref{sec:candidate-selection}.
\textit{Right:} Target density in our sample (contours) compared to a magnitude-selected background sample of stars from \textit{Gaia} (grey dots).
The upper side of the envelope of stars in our sample is caused by the subgiant cut shown in the left panel.
The locus of the sample is somewhat vertically offset above that of the the main sequence -- see discussion in Section~\ref{sec:magnitude}.
}
\label{fig:hrd}
\end{figure*}

The Transiting Exoplanet Survey Satellite \citep[\textit{TESS},][]{Ricker2014} recorded, in the first two years of its operation, full-frame images with a cadence of thirty minutes, in addition to higher-cadence lightcurves for a small subset of high-priority targets.
The \textit{TESS} footprint is divided into sectors, each of which is observed for 27.4 days (continously except for a 16-hour downlink gap halfway through the observations).
Approximately 60\% of the sky was observed in at least one \textit{TESS} sector during the first two years of observations, with the sectors having significant overlap towards the ecliptic poles.
\textit{TESS} sectors 1--13 are in the southern ecliptic hemisphere, and sectors 14--26 are in the northern ecliptic hemisphere.

We used separate reduction pipelines to generate lightcurves from the full-frame images for the northern and southern ecliptic hemispheres.
In the southern hemisphere we used the \software{eleanor} package \citep{Feinstein2019} with default parameters to generate lightcurves for all targets with \textit{TESS} magnitude $T < 13.5$.
In the northern ecliptic hemisphere, the absence of downloadable \software{eleanor} "postage stamps" made large-scale reduction using that method difficult. 
Instead, we used the publicly available lightcurves generated by the Quick-Look Pipeline group \citep[QLP,][]{Huang2020a,Huang2020b}.
The QLP targets also have a limiting magnitude of $T < 13.5$. 
The detrending performed during the QLP reduction process filters all periodic signals longer than 0.3\,days, which removes most BEER signals from the lightcurves. Therefore, we use the raw, un-detrended lightcurves which are also publicly available.
As a result there are more systematic effects present in these lightcurves than in our \software{eleanor} reductions, but these are satisfactarily removed by the detrending that we apply during our selection process (described in section~\ref{sec:candidate-selection}).
Across the whole sky this provided us with lightcurves for \longinputsize\ targets, of which approximately one third are visible in multiple sectors.

To the above targets we applied several additional filters. 
Giants and sub-giants cannot be contained within a binary system at this period range \citep[e.g.][]{Eggleton1983} but, as single-star pulsators, they can be a major source of contamination.
In order to ensure that giants and sub-giants were excluded from the sample, we applied a cut in the \textit{Gaia} colour-magnitude diagram, found by fitting a spline to the main sequence, and removing all sample stars $>1.5$ magnitudes above the spline (see Fig.~\ref{fig:hrd}).
Note that binary systems will be systematically more luminous than isolated main sequence stars by up to a factor of two, and we therefore ensured that systems shifted upwards by this factor will not be removed by our cut.
As \textit{Gaia} data are important for some of our analysis, we also applied the following \textit{Gaia} quality checks \citep{Babusiaux2018}:
\begin{multline}
\texttt{parallax\_over\_error} > 10,\\
\texttt{phot\_g\_mean\_flux} / \texttt{phot\_g\_mean\_flux\_error} > 50,\\
\texttt{phot\_rp\_mean\_mag} / \texttt{phot\_rp\_mean\_mag\_error} > 20,\\
\texttt{phot\_bp\_mean\_flux} / \texttt{phot\_bp\_mean\_flux\_error} > 20,\\
\texttt{phot\_bp\_rp\_excess\_factor} < 1.3 + 0.06 \times \texttt{bp\_rp}^2),\\
\texttt{phot\_bp\_rp\_excess\_factor} > 1.0 + 0.015 \times \texttt{bp\_rp}^2),\\
\texttt{visibility\_periods\_used} > 8,\\
\texttt{astrometric\_chi2\_al} / (\texttt{astrometric\_n\_good\_obs\_al} - 5) \\ \qquad < 1.44 \times \max(1, \exp(-0.4 \times (\texttt{phot\_g\_mean\_mag} - 19.5))).
\label{eq:gaia-quality}
\end{multline}

For each target, the contamination by nearby stars was estimated by retrieving all \textit{Gaia} sources within five arcminutes, estimating their \textit{TESS} magnitudes from their \textit{Gaia} magnitudes and colours (after correcting for extinction), and then using an assumed Gaussian point-spread function with a FWHM of 23 arcseconds for each star.
The contamination ratio was estimated only at the centre of light for the target in question, and no attempt was made to account for discrete pixels.
Estimates of contamination made in this way are generally lower limits, because (a) the contamination fraction will typically be larger when summing over a several-pixel aperture than at the point location of the target, (b) the \textit{TESS} point-spread function is typically broader near the edges of the field of view than at the centre, and (c) our estimates do not account for light from unresolved outer companions, which are typically not detected by \textit{Gaia} at separations less than one arcsec \citep{El-Badry2019}.
For targets that also appeared in the \textit{TESS} Candidate Target List (CTL), we verified that the contamination fractions estimated in this manner were generally consistent with those estimated by \citet{Stassun2019}.
Any target with an estimated contamination greater than 50\% was removed.

After these selection filters, we were left with lightcurves for \inputsize\ stars (of which approximately 30 per cent spanned multiple sectors of observations).
In Table~\ref{tab:long-table} we list all \longinputsize\ initial \textit{TESS} targets and highlight the subset of \inputsize\ that passed the selection filters listed above.
The user may repeat our process with stricter or less-strict filters if desired.

For luminous binary systems, the light from the secondary star can bias photometric estimates of the primary mass and radius.
We therefore did not use the mass and radius estimates available from the TESS Input Catalogue \citep[TIC;][]{Stassun2019}.
We considered, instead, the TIC temperature estimates, which are based on the colour of the target, to be somewhat more reliable, as the colour of the binary star should be less affected by the secondary light than the overall magnitude, and we adopted the TIC temperature as the primary star effective temperature.
Primary star mass and radius were estimated from the temperature by interpolation of the solar-metallicity main sequence models tables of \citet{Pecaut2013}.\footnote{http://www.pas.rochester.edu/$\sim$emamajek, version number 2021/03/02.}
There are considerable uncertainties in the derivation of these masses and radii (resulting from unknown photometric contamination and metallicity), and we adopt for them a large fractional uncertainty of 20 per cent.

\section{BEER Algorithm}
\label{sec:beer}

We processed each lightcurve using the \software{beer} algorithm \citep{Mazeh2010,Mazeh2012,Faigler2011,Faigler2015a,Faigler2013,Faigler2015b,Tal-Or2013,Tal-Or2015,Engel2020,Gomel2021a,Gomel2021b,Gomel2021c}.
In this section, we first set out the equations by which an ellipsoidal binary can be described, and then describe the process of selecting ellipsoidal binary candidates using the algorithm.

\subsection{The BEER Equations}

\begin{figure}
\includegraphics[width=\columnwidth]{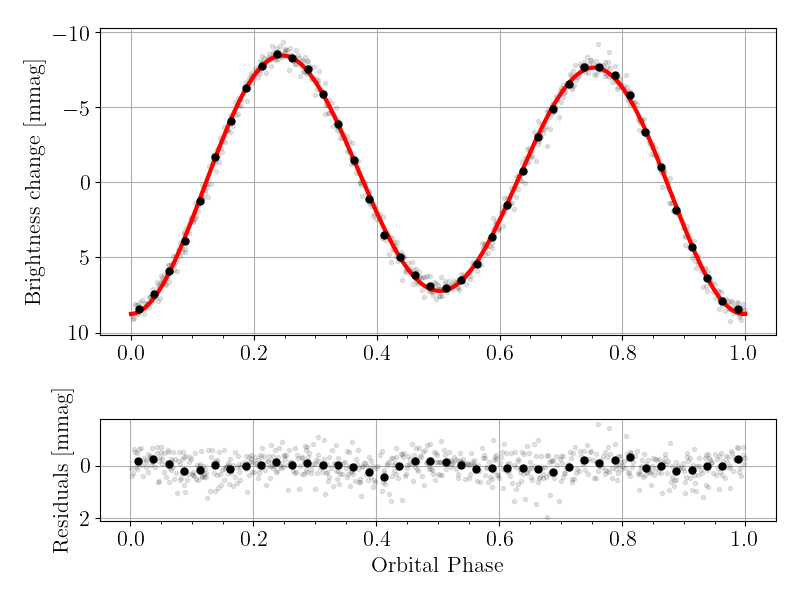}
\caption{Phase-folded \textit{TESS} lightcurve of TIC\,115471382, a typical ellipsoidal binary system in our sample. 
Light grey points show the individual data, black points show the phase-binned mean data, and the red curve is the best-fit BEER model.
The lightcurve is dominated by the second-order ellipsoidal modulation (Equation~\ref{eq:ell2}), with a difference in the height of the maxima due to Doppler beaming (Equation~\ref{eq:beam}) and a difference in the depths of the minima caused by the reflection effect and the first-order ellipsoidal modulation (Equations~\ref{eq:refl} and \ref{eq:ell1}).
TIC\,115471382 was confirmed to be a binary system by spectroscopic follow-up observations detailed in Section~\ref{sec:purity}. 
}
\label{fig:lightcurve}
\end{figure}

For a binary system in a circular orbit\footnote{Eccentric orbits are somewhat more complicated, see \citet{Engel2020}.} in which both stars are luminous, the combined BEER signal may be approximated with the form
\begin{equation}
\begin{split}
B(\phi) =& \left( a_{1,\text{ell}} + a_{\text{ref}} \right) \cos(2 \pi \phi) + a_{2, \text{ell}} \cos(4 \pi \phi)\\ 
+& a_{3, \text{ell}} \cos(6 \pi \phi) + a_{\text{beam}} \sin(2 \pi \phi),
\label{eq:beer}
\end{split}
\end{equation}
where $\phi$ is the orbital phase (defined such that the photometric primary is at its furthest distance from Earth at integer values of $\phi$) and each value $a$ represents the amplitude of one component of the BEER signal.

The reflection amplitude is related to the physical properties of the system by 
\begin{equation}
\begin{split}
a_\text{ref} =& - \left( f_1 \alpha_\mathrm{ref,2} \left(\frac{R_2}{a}\right)^2 - f_2 \alpha_\mathrm{ref,1} \left(\frac{R_1}{a}\right)^2  \right) \sin i \\
\approx& - 56514 \sin i \left( \frac{M_1 + M_2}{M_\odot} \right)^{-2/3} \left( \frac{P_\text{orb}}{\text{day}}\right)^{-4/3} \\
\times& \left( f_1 \alpha_{\text{ref},2} \left(\frac{R_2}{R_\odot}\right)^2 - f_2 \alpha_{\text{ref},1} \left(\frac{R_1}{R_\odot}\right)^2 \right) \text{ppm},
\label{eq:refl}
\end{split}
\end{equation}
where $f_1$ and $f_2$ are the relative flux contribution of the primary and secondary stars in the observed band such that $f_1 + f_2 = 1$; $R_1$ and $R_2$ are the radii of the two stars; $M_1$ and $M_2$ are their masses; $\alpha_{\text{ref},1}$ and $\alpha_{\text{ref},2}$ are their albedos; \porb\ is the orbital period; $i$ is the orbital inclination with respect to the observer; and ppm stands for parts per million \citep{Morris1993,Mazeh2010}.
While it is possible for the reflection effect to be offset in phase due to rotation, for instance in hot Jupiters \citep{Faigler2013}, the component stars in our binary systems are likely to be tidally locked and therefore we neglected this possible offset.

The Doppler beaming amplitude can be approximated as
\begin{equation}
\begin{split}
a_\text{beam} =& - 4 \left( f_1 \alpha_{\text{beam},1} \frac{K_1}{c} - f_2 \alpha_{\text{beam},2} \frac{K_2}{c} \right) \\
\approx& - 2830 \sin i \left( \frac{M_1 + M_2}{M_\odot}\right)^{-2/3} \left(\frac{P_\text{orb}}{\text{day}}\right)^{-1/3} \\ 
\times& \left(f_1 \alpha_{\text{beam},1} \left(\frac{M_2}{M_\odot}\right) - f_2 \alpha_{\text{beam},2} \left(\frac{M_1}{M_\odot} \right) \right) \text{ppm},
\label{eq:beam}
\end{split}
\end{equation}
where $\alpha_{\text{beam},1}$ and $\alpha_{\text{beam},2}$ characterise the efficiency of Doppler beaming from the two stars, and are a function of stellar temperature and surface gravity with values of order unity; $K_1$ and $K_2$ are the radial velocity semi-amplitudes of the two stars; and $c$ is the speed of light \citep{Rybicki1979,Loeb2003,Zucker2007,Mazeh2010}.

The ellipsoidal modulation contributes a signal at several harmonics of the orbital period, the strongest of which is the second harmonic. The first harmonic is approximated by
\begin{equation}
\begin{split}
a_\text{1,ell} \approx& \left( f_1 \alpha_\text{e1,1} \frac{M_2}{M_1} \left(\frac{R_1}{a}\right)^4 - f_2 \alpha_\text{e1,2} \frac{M_1}{M_2} \left(\frac{R_2}{a}\right)^4 \right) \\
\times& \left(4 \sin i - 5 \sin^3 i \right)
\\ \approx& 3194 \left(4 \sin i - 5 \sin^3 i\right) \left(\frac{M_1 + M_2}{M_\odot}\right)^{-4/3} \left(\frac{P_\text{orb}}{\text{day}}\right)^{-8/3} \\
\times& \left( f_1 \alpha_\text{e1,1} \frac{M_2}{M_1} \left(\frac{R_1}{R_\odot}\right)^4  - f_2 \alpha_\text{e1,2} \frac{M_1}{M_2} \left(\frac{R_2}{R_\odot}\right)^4      \right) \text{ppm},
\label{eq:ell1}
\end{split}
\end{equation}
while the second harmonic is approximated by
\begin{equation}
\begin{split}
a_\text{2,ell} \approx& - \left( f_1 \alpha_\text{e2,1} C_{1} \frac{M_2}{M_1} \left(\frac{R_1}{a}\right)^3 + f_2 \alpha_\text{e2,2} C_{2} \frac{M_1}{M_2} \left(\frac{R_2}{a}\right)^3 \right) \sin^2 i \\
\approx& - 13400 \sin^2 i \left(\frac{M_1 + M_2}{M_\odot}\right)^{-1} \left(\frac{P_\text{orb}}{\text{day}}\right)^{-2} \\
\times& \left(f_1 \alpha_\text{e2,1} C_{1} \frac{M_2}{M_1} \left(\frac{R_1}{R_\odot}\right)^3 + f_2 \alpha_\text{e2,2} C_{2} \frac{M_1}{M_2} \left(\frac{R_2}{R_\odot}\right)^3  \right) \text{ppm},
\label{eq:ell2}
\end{split}
\end{equation}
and the third harmonic amplitude is approximately
\begin{equation}
\begin{split}
a_\text{3,ell} \approx& - \left( f_1 \alpha_\text{e3,1} \frac{M_2}{M_1} \left(\frac{R_1}{a}\right)^4 - f_2 \alpha_\text{e3,2} \frac{M_1}{M_2} \left(\frac{R_2}{a}\right)^4  \right) \sin^3 i
\\ \approx& - 3194 \sin^3 i \left(\frac{M_1 + M_2}{M_\odot}\right)^{-4/3} \left(\frac{P_\text{orb}}{\text{day}}\right)^{-8/3} \\
\times& \left( f_1 \alpha_\text{e3,1} \frac{M_2}{M_1} \left(\frac{R_1}{R_\odot}\right)^4 - f_2 \alpha_\text{e3,2} \frac{M_1}{M_2} \left(\frac{R_2}{R_\odot}\right)^4      \right) \text{ppm}.
\label{eq:ell3}
\end{split}
\end{equation}
These approximations are the low-order terms of the expansion in \citet{Morris1993}.
In each of the above, $\alpha_{e1,1}$ and similar constants are functions of the effective temperature and surface gravity of the relevant star with values of order unity, and $C_{1}$ and  $C_{2}$  are correction factors that depend on the Roche lobe filling factor of the relevant star, as calculated by \citet{Gomel2021a}.

An example \textit{TESS} lightcurve of a BEER binary is shown in Fig.~\ref{fig:lightcurve}, compared to a best-fit BEER model described by Equations~\ref{eq:beer}--\ref{eq:ell3}.

\subsection{Candidate selection}
\label{sec:candidate-selection}

The candidate selection algorithm consisted of three stages, as further detailed below. First, each lightcurve was detrended and fit with a three-harmonic sinusoidal model. Second, an attempt was made to find a set of physical binary parameters that were consistent with the amplitudes of the harmonic model. Finally, the quality of the candidate was assessed against a range of criteria, and an overall score was assigned.

\subsubsection{Three-harmonic model}

In the first stage, it was necessary to estimate the orbital period. A Lomb-Scargle periodogram \citep{Lomb1976,Scargle1982} was derived from the lightcurve across a grid of frequencies from 0.1 to 12 cycles per day with a step size of 0.001 cycles per day. 
The five most prominent periods were selected, while insisting that selected peaks be separated by at least 0.02 cycles per day. 
For a BEER binary, either the orbital period or the period of the ellipsoidal modulation (half the orbital period) may be more significant; therefore, for each of the five selected periods, we considered both that period itself and twice that period to be candidate orbital periods, for a total of ten orbital periods considered per binary.

At each of these trial orbital periods, the detrending and the harmonic fit were simultaneously applied, as in \citet{Mazeh2010}.
The detrending removed any linear trend, as well as filtering a comb of frequencies with a step size of $1 / 2 T$ where $T$ is the time-span of the data, up to a maximum frequency equal to one-quarter the proposed orbital frequency.  
The harmonic fit consisted of three sine and three cosine functions, at the proposed orbital period and the first two harmonics.
Each half-sector of the lightcurve was treated separately, and the uncertainties on the measured amplitude of each harmonic were inflated until the measurements between half-sectors were consistent with each other.
In this way, variable amplitudes were given larger uncertainties.
For targets which have lightcurves from multiple \textit{TESS} sectors available, any proposed orbital frequency which was not consistent to within 0.008 cycles per day across all sectors was discarded.

A phase offset was applied to the harmonic fit such that the amplitude of the second-harmonic sine term became zero \citep{Faigler2011}. 
There will always be two phase offsets which are suitable, and we considered both, resulting in twenty sets of three-harmonic models for each candidate (two phase offsets for each of the ten trial orbital periods).
The resulting three-harmonic models had the form
\begin{equation}
\begin{split}
B(\phi) =& a_\text{1c} \cos (2 \pi \phi) + a_\text{2c} \cos (4 \pi \phi) + a_\text{3c} \cos (6 \pi \phi) \\ +&
a_\text{1s} \sin (2 \pi \phi) + a_\text{3s} \sin (6 \pi \phi),
\end{split}
\label{eq:harmonics}
\end{equation}
where each $a$ is a measured amplitude. This matches the form of Equation~\ref{eq:beer}.\footnote{Strictly, a BEER binary as described by Equation~\ref{eq:beer} should have $a_\text{3s} = 0$. We do not enforce this, but in practise $a_\text{3s}$ is negligible in almost all cases.}

\subsubsection{Physical model}

In the second stage of candidate selection, we attempted to find a physical set of binary parameters that can describe the measured amplitudes of $a_\text{1c}$, $a_\text{2c}$, and $a_\text{1s}$.
Note that these physical parameters were not adopted as the definitive description of the target, even if the target was selected as an ellipsoidal binary candidate; this stage of the process was used only to select against targets for which no reasonable physical description was possible.

A Levenberg-Marquardt algorithm was used to converge our physical BEER model (Equations~\ref{eq:beer}--\ref{eq:ell3}) on the measured amplitudes.
The parameters that we allowed to vary were $M_1$, $R_1$, $M_2$, $\sin i$, and the albedos of the two stars ($\alpha_\text{ref,1}$ and $\alpha_\text{ref,2}$).
The effective temperature of the primary, $T_\text{eff,1}$, was taken from the TIC.
The parameters $R_2$ and $T_\text{eff,2}$ were calculated under the assumption that the secondary was a main sequence star.
Other $\alpha$ constants in the BEER equations were interpolated from a grid of theoretical values using $T_\text{eff}$, $M$ and $R$ for the star in question.

For each trial set of parameters, a $\chi^2_\text{\sc beer}$ statistic that includes several priors was calculated by
\begin{equation}
\begin{split}
\chi^2_\text{BEER} =& \left(\frac{ a_\text{ref} + a_\text{1,ell} - a_\text{1c}}{\sigma_{a\text{1c}}}\right)^2 + 
\left(\frac{a_\text{2,ell} - a_\text{2c}}{\sigma_{a\text{2c}}}\right)^2 \\ +& \left(\frac{a_\text{beam} - a_\text{1s}}{\sigma_{a\text{1s}}}\right)^2 + 
\left(\frac{\exp(\lvert\log( M_1 / M_\text{1,can} )\rvert) - 1}{\sigma_{M_\text{1,can}} / M_\text{1,can}}\right)^2 \\
+& \left(\frac{\exp(\lvert\log( R_1 / R_\text{1,can} )\rvert) - 1}{\sigma_{R_\text{1,can}} / R_\text{1,can}}\right)^2 \\
+& \left(\exp(\lvert\log( \sin i / (\pi / 4) )\rvert) - 1\right)^2 + \left(\frac{\max(\alpha_\text{ref,1}, \alpha_\text{ref,2})}{1.5}\right)^2 \\
+& \left(\frac{R_1}{R_\text{RL,1}}\right)^{20} + 
\left(\frac{R_2}{R_\text{RL,2}}\right)^{20},
\end{split}
\end{equation}
where $a$ constants represent the various amplitudes defined in Equations~\ref{eq:beer}--\ref{eq:harmonics}, $\sigma_{a\text{1c}}$ and similar are the uncertainties on the corresponding measurements, $M_\text{1,can}$ and $R_\text{1,can}$ are the canonical ($T_\text{eff}$-derived) mass and radius of the primary star, and $R_\text{RL,1}$ and $R_\text{RL,2}$ are the Roche lobe radii of the two stars.
The terms involving $M_1$, $R_1$, and $\sin i$ are symmetric about the canonical value in logarithmic space.
The $\sin i$ and $\alpha_\mathrm{ref}$ terms pull the solution towards mean values ($\pi / 4$ and 0, respectively) with loose uncertainties.
The final two terms with high powers were intended to strongly select against models in which either star extends significantly beyond its Roche lobe.

During this fitting process we placed lower limits on the uncertainties of the measured amplitudes, to account for the possibility of these uncertainties being underestimated. 
Uncertainties on $a_\text{1c}$ and $a_\text{2c}$ were forced to be at least 10\% of the measured amplitude. 
We have previously found that empirical measurements of Doppler beaming are less reliable than measurements of reflection or ellipsoidal modulation, perhaps due to the typically lower amplitude of the Doppler beaming signal, and confusion with other physical effects such as the O'Connell effect \citep{OConnell1951,Knote2022}. 
We therefore forced the uncertainty on $a_\text{1s}$ to be at least 50\% of the measured amplitude.
As a result, Doppler beaming placed only a very weak constraint on the physical values assigned to each binary.

For each target, the fitting process described above was applied to each of the twenty sets of $a_\text{1c}$, $a_\text{2c}$, and $a_\text{1s}$ values that were derived from each lightcurve.
The trial orbital period and phase offset which had the lowest value of $\chi^2_\text{BEER}$ were selected and others were discarded.

\subsubsection{Scoring}

\begin{table}
\caption{The parameters $p_j$ used to calculate the overall score of a binary, and the scaling constants $k_j$ associated with each parameter. Each parameter and constant should be applied to one of the scoring functions shown in Equations~\ref{eq:selection-function-positive} and~\ref{eq:selection-function-negative}. The final column of this table describes whether to use the function from Equation~\ref{eq:selection-function-positive} or~\ref{eq:selection-function-negative}.
}
\begin{tabular}{lccc}
\label{tab:score-functions}
Name of score & $p_j$ & $k_j$ & Eq.\\
\hline
Ellipsoidal SNR			 & $a_\text{2c} / \sigma_{a\text{2c}}$  		  & 4     & 10 \\
$a_1$ or $a_3$ SNR        & max($a_1 / \sigma_{a1}, a_3 / \sigma_{a3}$)   & 4     & 10 \\
Model 	        		& $\chi^2_\text{BEER}$		  					  & 20	  & 11 \\
Power spectrum          & max(PS) / mean(PS)		 					  & 200   & 10 \\
Alarm 					& Alarm statistic								  & 1000  & 11 \\
High ellipsoidal		& $a_\text{2c}$									  & 7     & 11 \\
\end{tabular}
\end{table}

\begin{table*}
\caption{The results of the \software{beer} analysis of all input \textit{TESS} lightcurves, including the assigned score, the measured amplitudes, and the $\chi^2$ of the best-fit model. We include the first ten targets as an example, and for brevity we exclude the uncertainties; the full table (including uncertainties) may be found via the Centre de Données astronomiques de Strasbourg (CDS) and on Zenodo (DOI: 10.5281/zenodo.7750121).
}
\begin{tabular}{lccccccccc}
\label{tab:long-table}
TIC ID & RA [deg] & Dec [deg] & Score & $a_{1c}$ & $a_{2c}$ & $a_{3c}$ & $a_{1s}$ & $a_{3s}$ & $\chi^2$ \\
\hline
671 & 218.750233466797 & -28.8705020752408 & 0.0 & -0.00526 & -0.00187 & 0.00159 & -0.00086 & 0.00193 & 0.37 \\
813 & 218.771943117848 & -28.6426153121805 & 0.0 & 0.00026 & -0.00029 & 0.00019 & 0.00074 & 0.00027 & 0.88 \\
826 & 218.777186363537 & -28.6178633065608 & 0.0 & 0.00001 & -0.00007 & 0.00004 & 0.00012 & -0.00006 & 0.22 \\
834 & 218.828665786114 & -28.6060830157426 & 0.0 & -0.00115 & -0.00190 & 0.00013 & 0.00070 & 0.00006 & 0.24 \\
846 & 218.786672155155 & -28.5800240972002 & 0.0 & 0.00105 & -0.00104 & -0.00022 & 0.00127 & -0.00044 & 1.78 \\
873 & 218.749900122974 & -28.5201929634215 & 0.0 & -0.00011 & -0.00035 & -0.00028 & 0.00011 & -0.00005 & 0.06 \\
972 & 218.842356181595 & -28.3787741345186 & 0.0 & -0.00084 & -0.00040 & 0.00014 & 0.00085 & 0.00006 & 0.81 \\
1129 & 218.734585493562 & -28.1547448331321 & 0.0 & -0.00083 & -0.00095 & -0.00026 & -0.00055 & -0.00020 & 0.35 \\
1170 & 218.732389366934 & -28.1043097546334 & 0.0 & -0.00034 & -0.00003 & -0.00004 & -0.00023 & -0.00008 & 0.64 \\
1178 & 218.736595772979 & -28.0938861637931 & 0.0 & -0.00012 & -0.00027 & -0.00006 & 0.00009 & 0.00013 & 0.45 \\
\textit{continued...}\\
\hline
\end{tabular}
\end{table*}

\begin{table*}
\caption{The ellipsoidal binary candidates selected for our sample, including the measured orbital periods. 
$T_\mathrm{eff}$ is obtained from the TIC. 
We also include the estimated probability that the target is a contact binary, calculated in the manner described in Section~\ref{sec:contacts}.
We include the first ten targets as an example, and for brevity we exclude certain columns; the full table (including uncertainties) may be found in the online material of this journal, via CDS, or on Zenodo (DOI: 10.5281/zenodo.7750121).
}
\begin{tabular}{lcccccccc}
\label{tab:short-table}
TIC ID & RA [deg] & Dec [deg] & Score & $T_\mathrm{eff}$ [K] & $T$ [mag] & $P_\mathrm{orb}$ [days] & $t_0$ [JD] & P(contact) \\
\hline
364768528 & 299.581489449 & 52.5925656957 & 1.0 & 7587 & 9.96 & 1.219320 & 2458722.2188 & 0.0 \\
293064485 & 35.2282187466 & 46.7774506004 & 1.0 & 7556 & 11.22 & 1.094631 & 2458804.3839 & 0.0 \\
120000024 & 334.480075709 & 36.0811779216 & 1.0 & 5403 & 11.83 & 0.656783 & 2458751.1638 & 0.0 \\
384999832 & 137.241007929698 & -59.0433494717591 & 1.0 & 7581 & 12.41 & 1.249790 & 2458569.3510 & 0.0 \\
372446253 & 150.893429551377 & -66.4014609259132 & 1.0 & 8225 & 11.25 & 1.340175 & 2458597.6791 & 0.0 \\
388183492 & 57.1335831977116 & -78.422825032229 & 1.0 & 6510 & 12.57 & 0.711983 & 2458461.5786 & 0.0 \\
408041054 & 77.5181566060557 & 8.93139469888493 & 1.0 & 6752 & 11.96 & 1.238186 & 2458451.6889 & 0.0 \\
279283191 & 313.900656501 & 47.5481513602 & 1.0 & 7676 & 10.34 & 0.957464 & 2458739.2164 & 0.0 \\
351899015 & 346.763084691 & 41.7430121839 & 1.0 & 5850 & 12.10 & 0.693679 & 2458750.0374 & 0.0 \\
470123809 & 330.313374443 & 79.7406960591 & 1.0 & 7270 & 9.70 & 0.982388 & 2458925.6142 & 0.0 \\
\textit{continued...}\\
\hline
\end{tabular}
\end{table*}

In the third stage of candidate selection, each lightcurve was assigned a score between 0 and 1 to describe its quality as a candidate.
The overall score was the product of several component scores, each for a particular criterion, similar to the method described by \citet{Tal-Or2015}.
Each component score takes one of the following forms:
\begin{align}
S_j &= \exp \left(-p_j / k_j\right), \label{eq:selection-function-positive} \\
S_j &= 1 - \exp \left(-p_j / k_j\right) \label{eq:selection-function-negative},
\end{align}
for a parameter $p_j$ and a calibration constant $k_j$ which we adjusted manually until the desired selection was achieved.
This function was chosen due to its simple form (requiring only one calibration constant per parameter) and because the function varies smoothly as a function of $p_j$. 

In Table~\ref{tab:score-functions} we show the parameters $p_j$ which were used to calculate scores and the values of $k_j$ associated with each.
Note that $a_1^2 = a_\text{1c}^2 + a_\text{1s}^2$, and similarly for the other scores.
The SNR scores demand that there be a significant signal at the ellipsoidal period and at either the orbital period or the third harmonic.
The model score demands that there exist a physical model able to describe the lightcurve.
The power spectrum score insists that the ellipsoidal signature be strong relative to the rest of the power spectrum (PS), by comparing the peak height to the mean of the power spectrum (where the mean is calculated while excluding the orbital period and its harmonics).
The alarm statistic penalises any correlation among the residuals of the lightcurve after subtracting the harmonic model \citep{Tamuz2006}.
The high-ellipsoidal score selects against any target with an unphysically high amplitude ($\gtrsim 1$\,mag), in order to exclude some classes of pulsators.
In addition to these score calculations, any target with $a_1 > a_2$ or $a_3 > 0.3 a_2$ was given a score of zero, as these regions of parameter space are difficult to fill with true ellipsoidal binaries and therefore such targets are likely to be contaminants.
In Table~\ref{tab:long-table} we list the score assigned to each target analysed, as well as the measured amplitudes and $\chi^2$.

Once each target was given a score, it was necessary to set a score threshold with which to select candidates.
After examining a large number of lightcurves by eye, and spectroscopically following up a number of targets (Section~\ref{sec:purity}), we found that a threshold of \scorethreshold\ provides a reasonably pure sample, and we adopted this threshold for the analysis performed in the later sections of this work.
If a user prefers a higher completeness at the expense of a lower purity, they may wish to select their own sample from Table~\ref{tab:long-table} using a lower score threshold.

In Section~\ref{sec:purity}, we justify a further cut in terms of orbital period and ellipsoidal amplitude, to remove a region of parameter space which is dominated by contaminants which are not spectroscopically variable.
At the end of this selection process, using \scorethreshold\ as the score threshold, there were \samplesize\ remaining targets in the sample.
These targets are listed in Table~\ref{tab:short-table}.
We describe the properties of these targets in Sections~\ref{sec:contacts} and \ref{sec:sample-properties}.

\section{Sample Validation and Contaminants}
\label{sec:purity}

We validated the purity of our sample by measuring radial velocities (RVs) for a subset of targets. 
RVs were obtained using spectroscopic follow-up observations with the Las Cumbres Observatory Global Telescope network \citep[LCOGT;][]{Brown2013}. 

\subsection{LCOGT spectroscopy}

\begin{figure*}
\includegraphics[width=500pt]{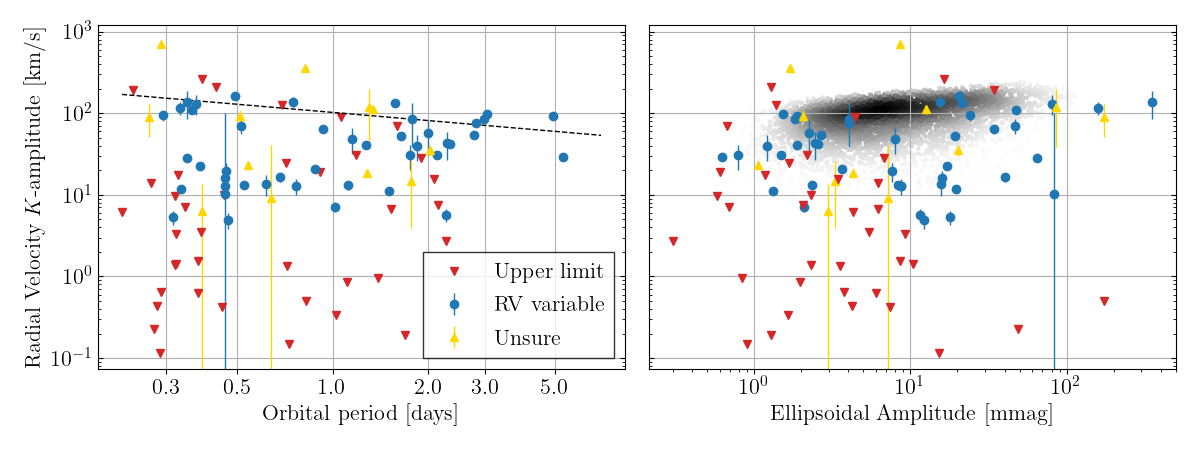}
\caption{\textit{Left:} The measured radial velocity $K$-amplitude of each target observed using LCOGT, or $3 \sigma$ upper limits for targets classed as not variable. The dashed line shows the expected $K$-amplitude for an example binary system with a 1.4\,$M_\odot$ primary star, $q = 0.5$, and $\sin i = 0.86$.
\textit{Right:} Comparison of the RV $K$-amplitudes and the photometric ellipsoidal amplitudes of all LCOGT targets, with $3 \sigma$ upper limits for non-RV variable targets.
Grey points show the amplitude distribution of a simulated set of binary systems.
Many of the upper limits on non-variable targets are significantly lower than expected for their photometric ellipsoidal amplitudes.
}
\label{fig:lco-amps-period}
\end{figure*}

\begin{figure}
\includegraphics[width=\columnwidth]{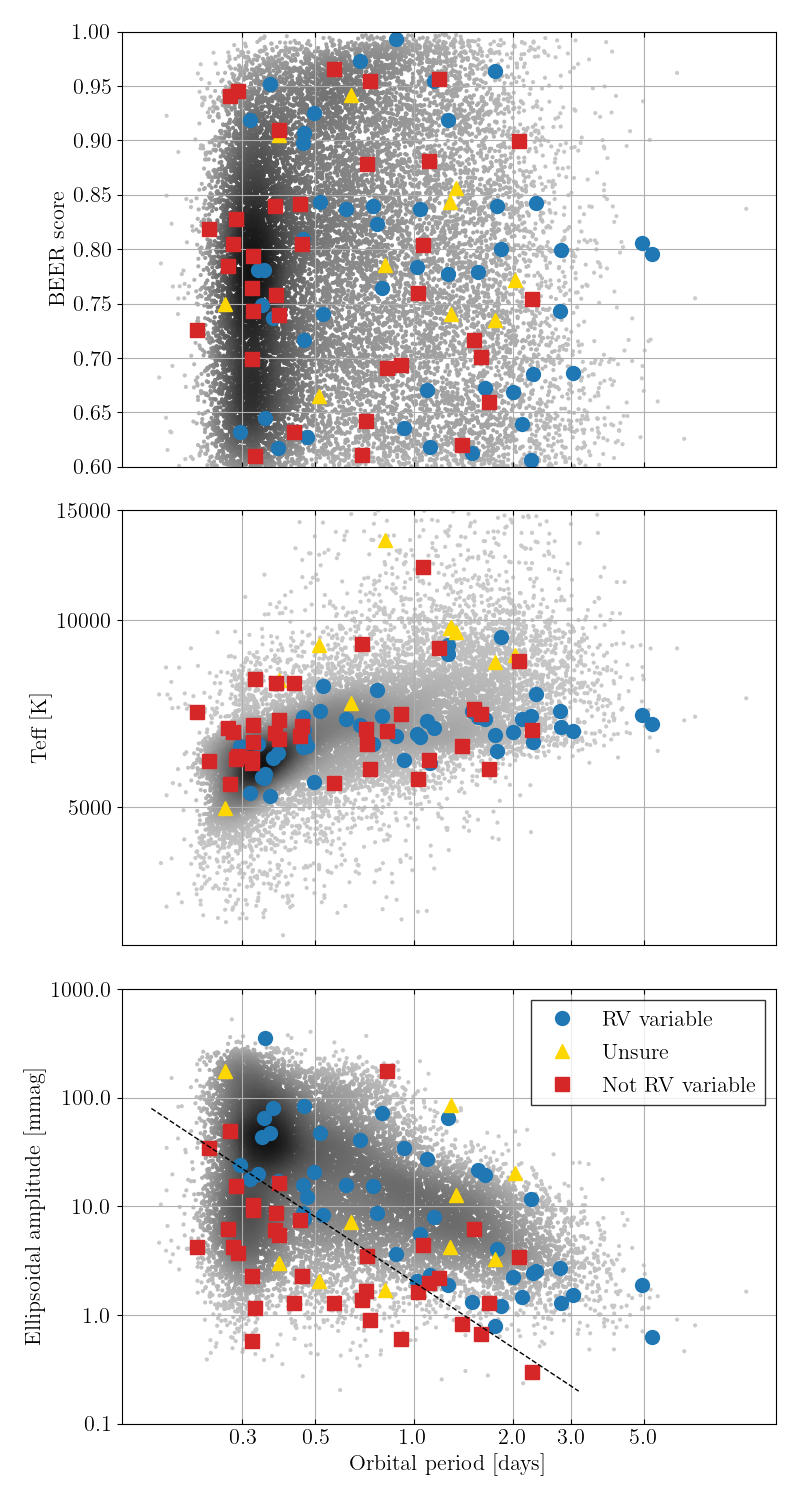}
\caption{Properties of 93 targets with score $> 0.6$ that were observed with LCO spectroscopic follow-up. Nine targets with \software{beer} scores $< 0.6$ have not been plotted.
The upper panel plots \software{beer} score, the middle panel shows primary $T_\mathrm{eff}$ from TIC, and lower panel shows the second-order ellipsoidal amplitude, all as a function of orbital period.
Coloured points indicate targets observed with LCOGT, coloured according to whether or not they showed RV variability, while grey points show the other targets in our sample (before applying the cut in Equation~\ref{eq:lco-cut}).
The cut in Equation~\ref{eq:lco-cut} is shown as a dashed line in the lower panel, which separates the lower-left region populated primarily by targets that are not RV variable.
}
\label{fig:lco-targets}
\end{figure}

\begin{figure}
\includegraphics[width=\columnwidth]{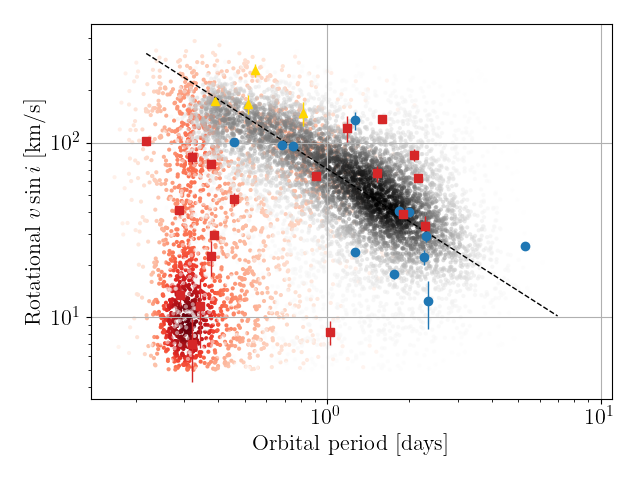}
\caption{The rotational velocity ($v_\mathrm{rot} \sin i$) of our sample from \textit{Gaia} Data Release 3.
Only targets with \textit{Gaia} broadening measurements are shown (8500 targets). 
Red targets are below the amplitude cut in Equation ~\ref{eq:lco-cut}, and grey targets are above. 
LCOGT targets are also plotted, using the same colour scheme as Fig.~\ref{fig:lco-targets}. 
The dashed line shows the expected rotational velocity for a 1.4\,$M_\odot$ star that is tidally locked and viewed edge-on.
There is a separation between the two populations, with the majority of those below the cut showing significantly lower rotational broadening.
}
\label{fig:lco-vsini}
\end{figure}

\begin{figure*}
\includegraphics[width=500pt]{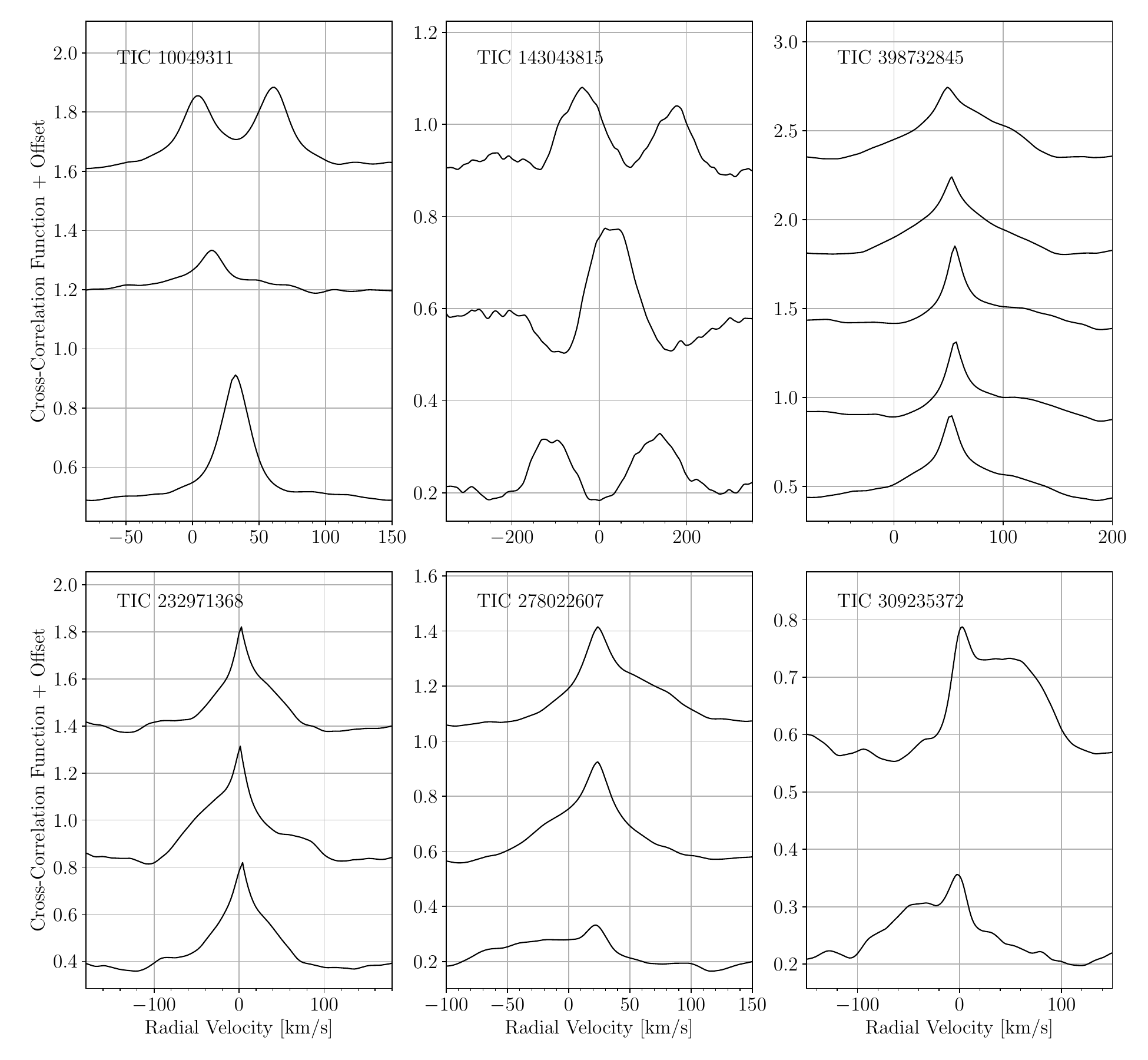}
\caption{Example cross-sections along the $v_2$ axis of TODCOR cross-correlation maps of selected targets, each at several epochs.
The first two targets are double-lined binary systems with varying degrees of rotational broadening. 
The other four targets exhibit a broadened, moving, spectral component and a narrow, stationary, spectral component.
As discussed in the text, we suggest these as candidate unresolved triple systems.
}
\label{fig:todcor-examples}
\end{figure*}

\begin{figure}
\includegraphics[width=\columnwidth]{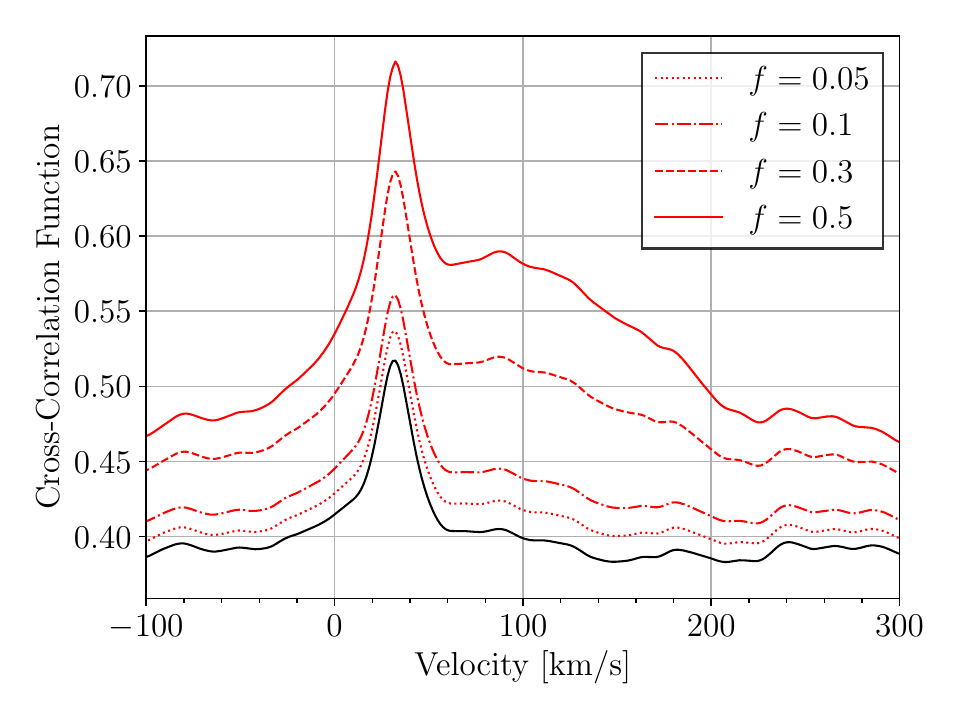}
\includegraphics[width=\columnwidth]{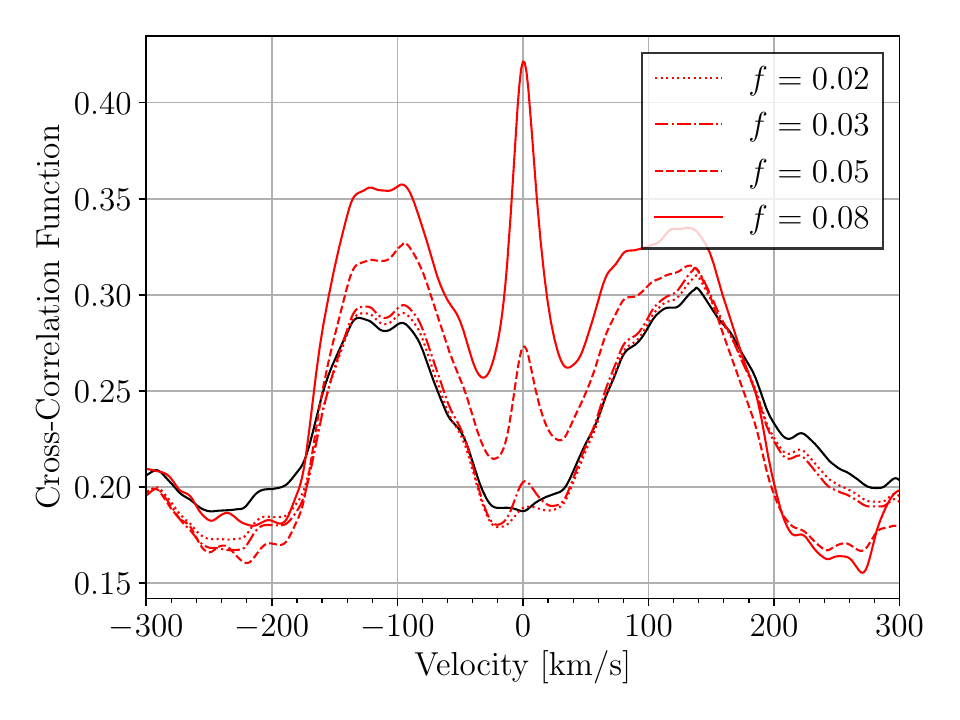}
\caption{Cross-sections along the $v_2$ axis of the TODCOR cross-correlation map of TIC\,154067797 (above) and TIC\,143043815 (below).
In both panels, the black lines show the cross-correlation function of the raw data alone.
The red lines show the cross-correlation function after the addition of a broadened third body with $v_\mathrm{rot} \sin i = 100$\,km\,s$^{-1}$ and a flux ratio $f$ relative to the original flux of the target.
The panel above replicates a triple system with a dominant outer body and a less luminous, broadened inner binary.
The panel below replicates a triple system with a dominant inner binary and a less luminous, unbroadened outer body.
In this way we ascertain the approximate flux ratio range in which a triple system can be detected spectroscopically.
}
\label{fig:todcor-inject}
\end{figure}

We obtained spectroscopic follow-up of 107 ellipsoidal binary candidates using the Network of Robotic Echelle Spectrographs (NRES). 
NRES instruments are fibre-fed, high-resolution ($R \approx 53000$) echelle spectrographs with a wavelength coverage of 3800--8600\,\AA. 
They are mounted on a number of 1\,m telescopes at multiple observing sites: the Cerro Tololo Observatory in Chile, the Wise Observatory in Israel, the McDonald Observatory in USA, and the Sutherland Observatory in South Africa (though note that due to technical issues, the Sutherland NRES instrument was not operational during our observations).

The 107 targets were selected to cover a range of values of orbital period, ellipsoidal amplitude, effective temperature, and \software{beer} score.
Our goal was to observe each candidate over at least three epochs. 
In some cases further epochs were obtained for targets where a clear classification was not initially possible (e.g.~due to poor phase coverage).
In a minority of cases, only two epochs were observed, either due to poor weather conditions or scheduling constraints. 

Each spectrum was reduced using the automated NRES-BANZAI pipeline.\footnote{DOI 10.5281/zenodo.1257560}
We made use of the NRES orders covering wavelengths 5150--5280\,\AA, 5580--5660\,\AA, 6100--6180\,\AA, and 6360--6440\,\AA, which were selected due to their inclusion of strong, narrow stellar lines and the absence of telluric lines.
The RV of the target at each epoch was measured using the SPectroscopic vARiabiliTy Analysis package  \citep[\software{sparta};][]{Shahaf2020},\footnote{github.com/SPARTA-dev/SPARTA} by cross-correlation with a PHOENIX template spectrum \citep{Husser2013}.
The $T_\mathrm{eff}$, surface gravity ($\log g$), and metallicity used for the template spectrum were taken from the TIC, with metallicity assumed to be solar if no TIC value was available. 
A rotational broadening was applied to the template spectrum. 
In some cases an initial guess for the rotational broadening of $v_r \sin i = 0.86 (2 \pi R_1) / P_\mathrm{orb}$ was sufficient (implying tidally induced corotation, as is generally expected in such short-period systems). 
In other cases $v_r$ was adjusted manually until the broadening was similar to that of the primary star.
A single template was always used, although we note that in some cases a second luminous component of the target was visible as a second peak in the cross-correlation function (CCF).
The RV was taken to be the peak of the CCF for each epoch. 

Once RV measurements were obtained, an orbital solution was found using a Markov Chain Monte Carlo (MCMC) method \citep{Foreman-Mackey2013}. 
The photometric orbital period and phase were used as priors.
The best-fit RV amplitude $K_1$ and its uncertainties were taken to be the median of the distribution and half the difference of the 16th and 84th percentiles.

Each target was examined to ensure that the cross-correlation and best-fit $K$ were of good quality.
Each target was then classified as either RV variable or not RV variable, based on whether a significant RV shift between epochs was detected.
A number of targets were difficult to classify, mostly in cases where rotational broadening of the spectral lines prevented a clean RV measurement. 
These targets were classified into a separate category, `unsure'.
Five further targets were not classified as insufficient data were obtained (these are not included in the `unsure' category).

The full list of spectroscopic targets is presented in Table~\ref{tab:lco-table}.
Of the 93 classified targets which have a \software{beer} score greater than 0.6, 47 are RV variable, 36 are not, and 10 are unclear.
Of the RV variable targets, 12 have two visible spectroscopic components.
A further nine targets with scores in the range 0.4--0.6 were observed, of which three were classed as RV variable (two of those were double-lined), four were classed as not RV variable, and two were unclear.

In Fig.~\ref{fig:lco-amps-period}, we plot the measured radial velocity amplitudes of each LCOGT target.
In the right panel, we plot the $K$-amplitude as a function of photometric ellipsoidal amplitude for LCOGT targets and for a set of simulated binary systems, assuming a reasonable distribution of $\sin i$ and $q$. 
The simulation is described in Section~\ref{sec:simulations}.
For the majority of LCOGT targets classed as not RV variable, the upper limit on the $K$-amplitude is 1--2 orders of magnitude lower than would be expected for their ellipsoidal amplitude.
It is therefore difficult to understand the non-detection of radial velocity variations as resulting from a small $\sin i$ in these systems.

Additional measured properties of the LCOGT targets, and of the sample as a whole, are plotted in Fig.~\ref{fig:lco-targets}.
As the bottom panel of Fig.~\ref{fig:lco-targets} shows, the majority of observed targets which were not RV variable can be isolated to one region of parameter space: having short periods and low ellipsoidal amplitudes.
We conjecture that these targets are some form of non-binary contaminant, whose nature we discuss in the following section.
The cut 
\begin{equation}
\label{eq:lco-cut}
\log(a_{2} \mathrm{[mmag]}) > -2 \log (P_\mathrm{orb} \mathrm{[days]}) + 0.3,
\end{equation}
shown in Fig.~\ref{fig:lco-targets}, removes 33 targets, of which 3 were classed as variable, 27 as not variable and 3 as unclear.

Throughout the rest of this work, the cut in Equation~\ref{eq:lco-cut} was applied to our sample.
Following this cut, \samplesize\ ellipsoidal binary candidates remained in the sample.
Of the 60 LCOGT targets which passed this cut and had a \software{beer} score greater than 0.6, 44 were classed as RV variable, 9 were not RV variable, and 7 were unclear.
Excluding targets which were classified as unclear, we therefore estimate a sample purity of $83 \pm 13$ per cent, assuming Poisson uncertainties.

\subsection{The nature of the contaminants}

The nature of the contaminants discussed in the previous section, those targets falling below the cut in Equation~\ref{eq:lco-cut}, is unclear.
Although the amplitudes of photometric variability are lower than those of the confirmed ellipsoidal binaries, the lightcurves of these contaminants are otherwise remarkably similar to true ellipsoidal binaries, with similar harmonic ratios $a_1/a_2$ and $a_2/a_3$.
Their spectral types are FGK, similar to the ellipsoidal binary candidates that lie above the proposed cut at the same period range. 
Note, however, that the candidates above the cut show a correlation between orbital period and temperature (as will be discussed in Section~\ref{sec:teff}), while the candidates below do not.

The targets removed by this cut can be separated into two populations on the basis of their spectroscopic broadening.
In Fig.~\ref{fig:lco-vsini} we plot the velocity broadening values published in \textit{Gaia} Data Release 3 \citep{GaiaCollaboration2022} for targets above and below the cut in Equation~\ref{eq:lco-cut}.
LCOGT targets are highlighted where such a measurement was available from the catalogue.
While targets above the cut have approximately the rotational broadening expected for a tidally-locked star, the majority of targets below the cut show a lower rotational broadening.

We discuss the targets removed by the cut in Equation~\ref{eq:lco-cut} as two categories: those with the expected rotational broadening and those without.
The broadened stars possibly include some genuine ellipsoidal binaries whose photometric amplitude was reduced by an unusually face-on inclination.
However, note that such an inclination would also reduce $v \sin i$.
More likely, the broadened stars may be rotating  stars with star spot configurations that happen to mimic an ellipsoidal signature.

The majority of targets removed by the cut in Equation~\ref{eq:lco-cut} have low values of rotational broadening (we refer to these stars as `unbroadened' for brevity, although we note that many do still have a statistically-significant measurement of broadening).
A possible interpretation of the unbroadened, non-RV variable targets below the cut is that they are triple systems, in which the photometric ellipsoidal signal is diluted by a brighter third body. 
The stationary third body would also dominate the spectroscopy of the system, masking any RV signature.
An assumed mean dilution factor of $\sim 10$ (based on the vertical offset of the red and blue distributions in Fig.~\ref{fig:lco-targets}) would require that the third body is 2.5 magnitudes brighter in the $T$ band than the combined flux of the inner binary. 
If the third body is F-type then the inner binary may consist of K-type stars, while if the dominant star is of G- or K-type, the inner binary must be of M-type or later.
Note that even in the case that all \textit{unbroadened} targets below the amplitude cut are triple systems, the amplitude cut is still justified on the grounds of purity, as a significant fraction of the targets below the line are strongly rotationally broadened and therefore more likely to be rotating single star contaminants.

An alternate suggestion for the non-broadened contaminants may be that the photometric signature arises from some form of low-amplitude pulsation.
As we discuss in the following section, it is difficult to distinguish between these two propositions using the spectroscopic data in hand.

\subsection{Searching for triple systems among LCOGT data}
\label{sec:todcor}

We used a two-dimensional cross-correlation method \citep[TODCOR;][]{Mazeh1994, Zucker1994,Zucker2003}, as implemented by the software package \software{saphires}\footnote{https://github.com/tofflemire/saphires/}, to search for spectroscopic signatures of triple systems in the LCOGT spectra from our sample, among both the RV-variable and the non-RV-variable targets.
The TODCOR method produces a two-dimensional map of the CCF between the target spectrum and two template spectra, in which the axes are the velocities of the primary and secondary template ($v_1$ and $v_2$).
Our primary PHOENIX template matched the template used in the one-dimensional analysis, while a secondary template of $T_\mathrm{eff} = 5000 K$, $\log g = 4$ (a K3-type star), and solar metallicity was used for all targets.
The overall flux ratio, $f_2 / f_1$, between the secondary and primary templates at a central wavelength of 6500\,\AA\ was allowed to vary between zero and one to produce the best cross-correlation value, while between spectral orders the flux ratio was adjusted according to the wavelength of each order, proportional to the ratio between two blackbodies with the temperatures of the templates used.
The same spectral orders were used as in the one-dimensional analysis.

In Fig.~\ref{fig:todcor-examples}, we show several examples of cross-sections through the two-dimensional TODCOR cross-correlation maps along the $v_2$ axis, at the optimal value of $v_1$.
The first two cases show typical examples of double-lined binary systems, in which the two RV-variable components with similar amounts of Doppler broadening are resolved.

The latter four cases show a somewhat different configuration: one component is narrow and stationary, while the second is broadened and RV-variable.
It is difficult to understand these two components as belonging to main sequence stars in the same binary system: the similar strength of their contributions to the CCF implies their luminosity must be similar, while the very different radial velocity amplitudes of the two components would imply an extreme mass ratio if the stars are in orbit around each other at the ellipsoidal period.

We suggest that these four targets are examples of unresolved triple systems, in which the broad component originates from one star of the inner binary and the narrow component originates from the third body.
We also note that three of these four targets (excluding TIC\,309235372) have unusually small ellipsoidal amplitudes, lying either below or very close to the amplitude cut applied in Equation~\ref{eq:lco-cut}.
These reduced amplitudes are consistent with the presence of flux dilution by a third body.
We inspected the TODCOR maps and cross-sections of all 107 targets observed using LCOGT, but found only these four candidates with this appearance.

We investigated the range of flux ratios in which triple systems with this appearance would be detectable. 
In the top panel of Fig.~\ref{fig:todcor-inject}, we show the CCF for an example unbroadened, non-RV-variable target.
To this, we add hypothetical templates representing one star within the inner binary, with a range of flux ratios.
Each template had $T_\mathrm{eff} = 4500 K$, was velocity broadened by 100\,km\,s$^{-1}$, and was redshifted by 100\,km\,s$^{-1}$.
The flux ratio $f_1/f_3$ was again defined at a central wavelength of 6500\,\AA, and adjusted between spectral orders as before.
As our inspection of the TODCOR maps was performed manually with no quantifiable criteria, it is difficult to precisely quantify the critical detectability threshold.
However, we can approximately say that a triple with a flux ratio of 0.3 between the broadened inner binary and the stationary third component would be likely detectable, while a flux ratio of 0.1 would not be detectable.

On this basis, we performed a first-order Monte Carlo estimate of the fraction of unresolved, tertiary-dominated triple systems in which the inner binary would be detected spectroscopically. 
The mass of the tertiary object was drawn from the distribution of masses in the input set of \textit{TESS} lightcurves.
The mass ratio between the most massive star of the inner binary and the tertiary ($M_1 / M_3$) was drawn from a uniform distribution between 0.1 and 1. 
The $G$-band magnitudes of both stars were then estimated by the \citet{Pecaut2013} tables, and the flux ratio calculated.
If the detection threshold is somewhere between $f_1/f_3 = 0.1$ and 0.3, we might expect $\approx 25$--40 per cent of tertiary-dominated triples to be detected as such by our LCOGT spectroscopy.

Among our LCOGT targets were four detected triple systems, and 10 unbroadened, non-RV variable targets which were suggested in the previous section as possible tertiary-dominated triples.
We therefore conclude that an interpretation in which some, most, or all of the \textit{unbroadened}, low-amplitude contaminants (which are themselves approximately one third of all the low-amplitude contaminants) in Fig.~\ref{fig:lco-targets} are tertiary-dominated triple systems is consistent with the current data.
However, with current data we cannot confirm this hypothesis.
Neither can we ascertain what fraction may be some other form of contaminant.
With the relatively small sample size of our LCOGT targets, it is difficult to say more.

With TODCOR, we also investigated the possibility that among our LCOGT targets are a different form of triple system: triples in which the inner binary is photometrically dominant.
These systems are much easier to detect spectroscopically, because the hypothetical third spectral component is narrow relative to the broadened inner binary.
In the bottom panel of Fig.~\ref{fig:todcor-inject}, we show the TODCOR CCF for an example double-lined binary system.
To this, we added hypothetical templates representing the third object, with a range of flux ratios.
Each template had $T_\mathrm{eff} = 4500 K$, and was not velocity broadened or redshifted.
We find that even a third body which contributes five per cent of the total light should be marginally detectable, and one which contributes 10 per cent will be easily detectable.

We repeated the previous Monte Carlo estimation, this time drawing $M_1$ from the mass distribution of our sample, and assuming a uniform mass ratio distribution of $M_3 / M_1$ between 0.1 and 1.
Taking the critical flux ratio to be between $f=0.05$ and $f=0.1$, we estimate that $\approx 40$--50 per cent of all binary-dominated triples should be detectable as such spectroscopically. Given again that only four triple systems were spectroscopically identified as triples, out of the total 47 RV-variable targets, we conclude that only a small fraction of our sample can conceal low-mass, unresolved third bodies.
The implications of this will be discussed in Section~\ref{sec:discussion}.

\section{Completeness of the sample}
\label{sec:simulations}

\begin{figure*}
\includegraphics[width=500pt]{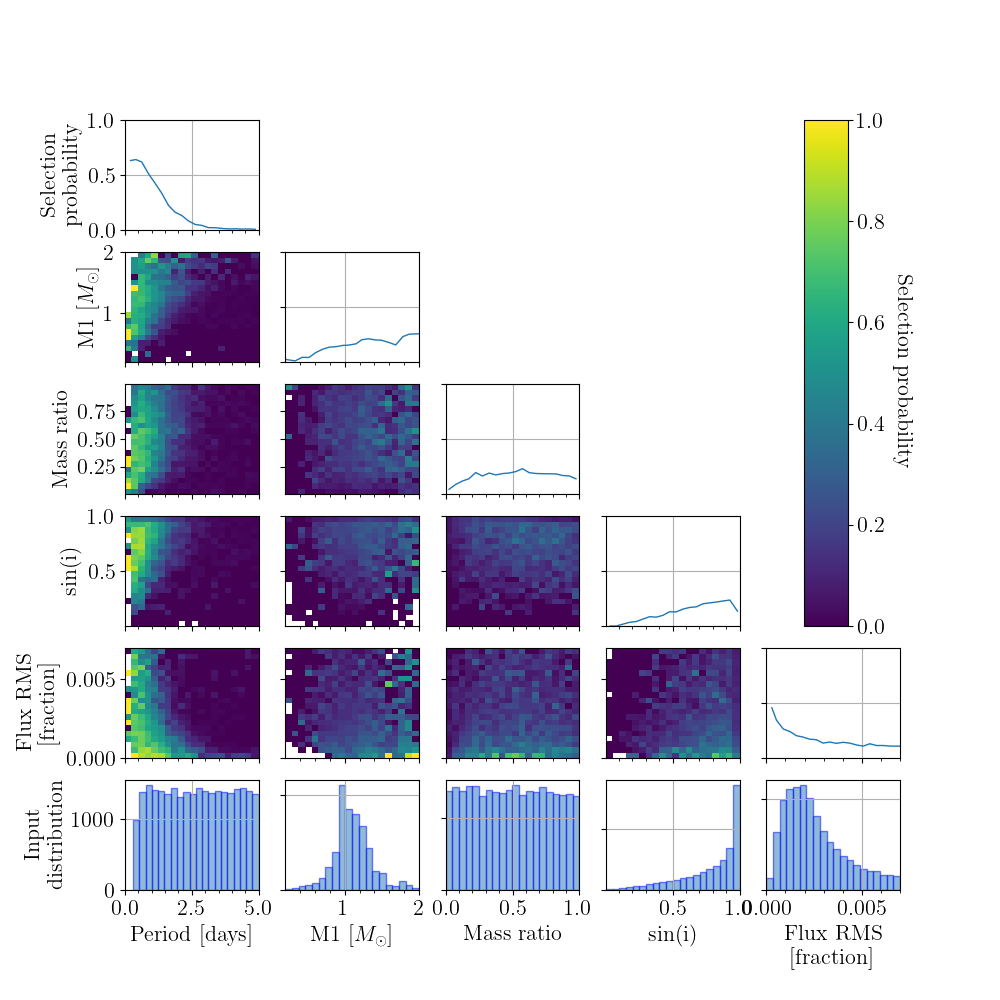}
\caption{The fraction of simulated binaries which were selected as candidates by our algorithm, as a function of binary system parameters.
The plots along the diagonal show the selection function for each parameter, with other parameters marginalised over.
The two-dimensional histograms show the selection efficiency in the phase space of each pair of parameters.
Lastly, the one-dimensional histograms along the bottom row show the initial distribution of each parameter in the simulated population (in other words, before selection effects have been applied).
}
\label{fig:cornerplot}
\end{figure*}

\begin{figure}
\includegraphics[width=\columnwidth]{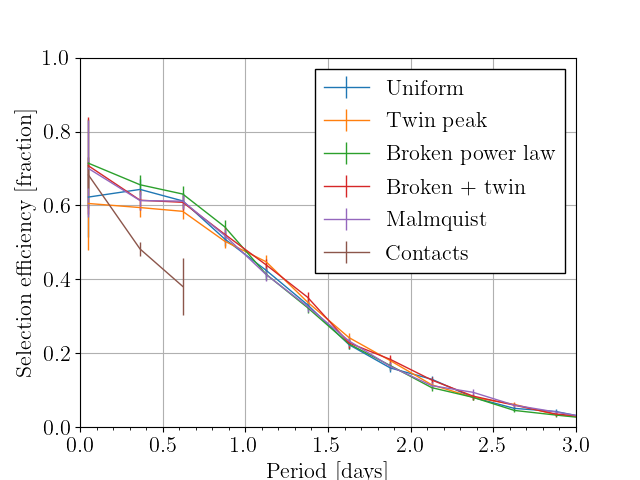}
\caption{Selection efficiency of the \software{beer} algorithm as a function of orbital period, measured using injection-recovery tests with a variety of mass-ratio distributions. 
The distributions are described in the text.
}
\label{fig:selection-functions}
\end{figure}

In order to infer from our sample the properties of the underlying population of short-period binary systems, it is necessary to understand how the selection efficiency of our sample varies as a function of the various input parameters.
We explored this using a series of injection-recovery simulations.
For each injection-recovery test, a sample of simulated binary lightcurves was generated by drawing properties from preset distributions.

First, for each simulated binary, a random base lightcurve was selected from the list of all input lightcurves.
By using a true \textit{TESS} lightcurve as the base for each simulated lightcurve, we replicate the noise profile and any instrumental effects induced by \textit{TESS}, as well as any correlated noise which may be present due to underlying stellar variability.
However, note that we do not allow for the possibility that stellar variability may correlate with binarity.

The properties (\teff, $M_1$, $R_1$) of the star which was the source of the base lightcurve were adopted as those of the primary star in the simulated binary.
The orbital inclination was chosen from a geometric distribution.
Two parameters whose true distributions are not known are \porb\ and the mass ratio ($q = M_2 / M_1$).
A value of \porb\ was drawn from a uniform distribution between 0 and 5\,days.
Several distributions of $q$ were tested, as detailed later in this section, with the simplest being a uniform distribution.
Primary and secondary stellar albedos were drawn from a uniform distribution between 0 and 0.3, but were not found to have a significant impact on detectability.
Using these adopted values, a BEER signature was calculated using Equations~\ref{eq:beer}--\ref{eq:ell3} and added to the selected base lightcurve.

We also wanted to account for selection effects due to any eclipses that a binary might show. Deep eclipses will be selected against by our algorithm, introducing selection effects that depend on parameters of interest such as \porb\ and $q$.
Therefore, we combined each simulated lightcurve with the predicted eclipse profile for the simulated binary, if any, generated using the package \software{ellc} \citep{Maxted2016}.

This process was repeated until 30\,000 simulated lightcurves were generated.
Any simulated binary for which either primary or secondary star would be Roche-lobe filling was thrown out.
The set of lightcurves generated in this manner were passed through the selection algorithm described in Section~\ref{sec:candidate-selection}.
Through this process, we replicate any selection effects introduced as a result of our procedure.

For a uniform input distribution of $q$, the fraction of simulated binaries that were selected (with a \software{beer} score greater than \scorethreshold) is shown in the corner-plot in Fig.~\ref{fig:cornerplot}.
The strongest dependences of selection efficiency are on \porb\ and $M_1$, with shorter-period systems and higher-mass primaries being preferentially selected (as expected from Equation~\ref{eq:ell2}). 
There is also a dependence on mass ratio and orbital inclination, again as expected.
The parameter `flux RMS' in Fig.~\ref{fig:cornerplot} shows the RMS of the lightcurve before the simulated binary was added, and includes shot noise, \textit{TESS} systematic lightcurve artefacts, and any variability from stellar activity.
The overall completeness of our sample under these assumptions, respectively for periods shorter than  1, 2, 3, and 5 days, respectively, is $58.1 \pm 1.2$, $40.4 \pm 0.6$, $28.2 \pm 0.4$, and $16.6 \pm 0.2$ per cent.

In order to test the impact of the unknown $q$ distribution on the completeness, further injection-recovery tests were performed with various plausible distributions of $q$, including: 
a uniform distribution of $q$; 
a distribution which was uniform for $q < 0.95$, with a `twin peak' probability enhancement for $q > 0.95$ parametrised by $\mathcal{F}_{twin} = 0.29$ \citep{Moe2017}; 
a broken power-law distribution with $q_\mathrm{break} = 0.3$, $\gamma_\mathrm{small} = 0.3$, and $\gamma_\mathrm{large} = -0.6$ \citep{Moe2017}; 
a similar power-law distribution, combined with a twin peak of $\mathcal{F}_{twin} = 0.29$; 
and a uniform distribution multiplied by a Malmquist correction to replicate a selection bias in favour of high-mass companions. 
The latter can be justified as follows: binaries with high-mass companions are intrinsically brighter due to the flux contribution from the secondary star, and will therefore be over-represented in a magnitude-limited sample (the Malmquist bias).
The Malmquist correction that we used was
\begin{equation}
P_\mathrm{Malm}(q) \propto 1 + q^{3.5}
\end{equation}
which was derived from an assumption that the secondary star has luminosity $\propto M_2^{3.5}$ and that the number of visible targets goes as the square of the distance within which they are visible (the latter is a reasonable assumption for distances $\gtrsim 300$\,pc).

In Fig.~\ref{fig:selection-functions} we show the one-dimensional selection efficiency as a function of period for each of the $q$ distributions that were simulated.
We find that the $q$ distribution does not have a strong effect on selection efficiency.
The scatter of selection efficiency between different input distributions is of order a few per cent.
We adopt a relatively conservative uncertainty of ten per cent on all estimates of the selection efficiency, to account for the unknown distribution of mass ratios. 
We therefore revise the previous completeness estimates to $58 \pm 6$, $40 \pm 4$, $28 \pm 3$, and $17 \pm 2$ per cent for periods less than  1, 2, 3, and 5 days, respectively.

Our sample also contains a significant number of contact binaries (see Section~\ref{sec:contacts} for further discussion of these targets.)
As Equations~\ref{eq:beer}--\ref{eq:ell3} are not valid for binary systems that are beyond Roche-lobe filling, we repeat the completeness simulations for the 
case of contact binaries.
We generate contact binary lightcurves using the software package \software{phoebe} \citep{Prsa2005,Prsa2016}.\footnote{We first confirmed that, for detached binaries, \software{phoebe} and Equations~\ref{eq:beer}--\ref{eq:ell3} produce consistent amplitudes.} 
Because the parameter distributions for contact binaries are somewhat different, we calculate reasonable orbital periods and mass ratios using equations 1, 2, and 9 of \citet{Gazeas2009}, using the primary mass of the randomly-selected input lightcurve, and setting the temperature of both stars equal to the main-sequence temperature of the primary star.
We then confirm that the resulting parameters describe a physically-realistic contact binary using the internal checks of \software{phoebe}.
1\,500 lightcurves were generated in this way.
The resulting selection efficiency as a function of period is also shown in Fig.~\ref{fig:selection-functions}.
The lower selection efficiency of simulated contact binary when compared to detached binaries arises 
\review{
from a combination of factors.
Firstly, Equations~\ref{eq:beer}--\ref{eq:ell3} are not valid for contact binary systems, and therefore it is less likely that a physical solution to those equations exists to describe the observed lightcurve.
Secondly, contact binary systems are more likely to display eclipses, which affect selection in a complex manner (described in more detail below).
Thirdly, for a large fraction of contact binaries, the \software{beer} algorithm fails to accurately measure the orbital period}
(the equal eclipse depth leads to \porb\,$/ 2$ being selected instead); those targets are then removed by the $a_1 > a_2$ cut described in Section~\ref{sec:candidate-selection}.

\review{
The selection efficiency of a binary system is also affected by the potential presence of, and morphology of, eclipses.
To investigate this, we used the package \software{ellc} to estimate the depth of eclipses in our previously-described simulated populations of detached and contact binary systems.
}

\review{
For the detached binary systems, approximately 20 per cent of the simulated binaries show eclipses, and 7 per cent show eclipses which are deeper than $0.2 \times$ the total light.
We find that eclipses shallower than $\approx 0.2 \times$ the total light do not notably decrease the selection efficiency. 
Indeed, shallow eclipses may increase the selection efficiency, as they effectively increase the apparent amplitude and hence make the photometric signature easier to detect.
Across the population, 17 per cent of the non-eclipsing targets, 30 per cent of targets with eclipses shallower than $0.2 \times$ the total light, and 8 per cent of targets with eclipses deeper than $0.2 \times$ the total light were selected.
Overall, we can estimate that the fraction of detached binary systems that were removed from our sample due to their eclipses is $\lesssim 5$ per cent.
}

\review{
Of the contact binary systems in the simulated population, 70 per cent show eclipses. 
For contact binaries, we do not find that an eclipse of any depth increases the selection efficiency.
The overall selection efficiency of non-eclipsing contact binary systems is 53 per cent, and that of eclipsing contact binary systems is 40 per cent. 
Overall, we can estimate that $\approx 20$ per cent of all contact binary systems that would otherwise have been detected were missed due to the presence of eclipses.
}

\subsection{Comparison to \textit{Gaia} spectroscopic binaries}

For a further test of the completeness of our sample, we cross-match our candidates with the \textit{Gaia} Data Release 3 catalogue of single-lined spectroscopic binaries.
We use the high-quality catalogue of \citet{Bashi2022}, who used a comparison with publicly-available spectra to remove many spurious \textit{Gaia} binary candidates.
We adopt their suggested quality score threshold of 0.587.

We cross-matched the input \inputsize\ \textit{TESS} targets from which our sample was drawn with the \citet{Bashi2022} catalogue, in order to establish which binary systems were available to be detected by our algorithm.
Out of the detectable systems, we recovered $64 \pm 16$ per cent, $36 \pm 4$ per cent, $19 \pm 2$ per cent, and $8.3 \pm 0.7$ per cent for orbital periods less than 1, 2, 3, and 5 days, respectively.
These numbers are of a similar order to the completeness estimates based on simulated binary systems.

Note that the recovery rate relative to another sample is not the same as completeness: it also depends on the completeness of the comparison sample.
Our recovery rate of the \textit{Gaia} spectroscopic binaries can be understood as the ratio between our completeness function (approximately $\propto$\,$P_\mathrm{orb}^{-2}$) and that of the spectroscopic sample (roughly $\propto$\,$P_\mathrm{orb}^{-1/3}$).
As a result, our recovery rate relative to the spectroscopic sample decreases with increasing orbital period at a steeper rate than does our completeness estimate based on our simulations.

\section{Contact Binaries}
\label{sec:contacts}

\begin{figure*}
\includegraphics[width=500pt]{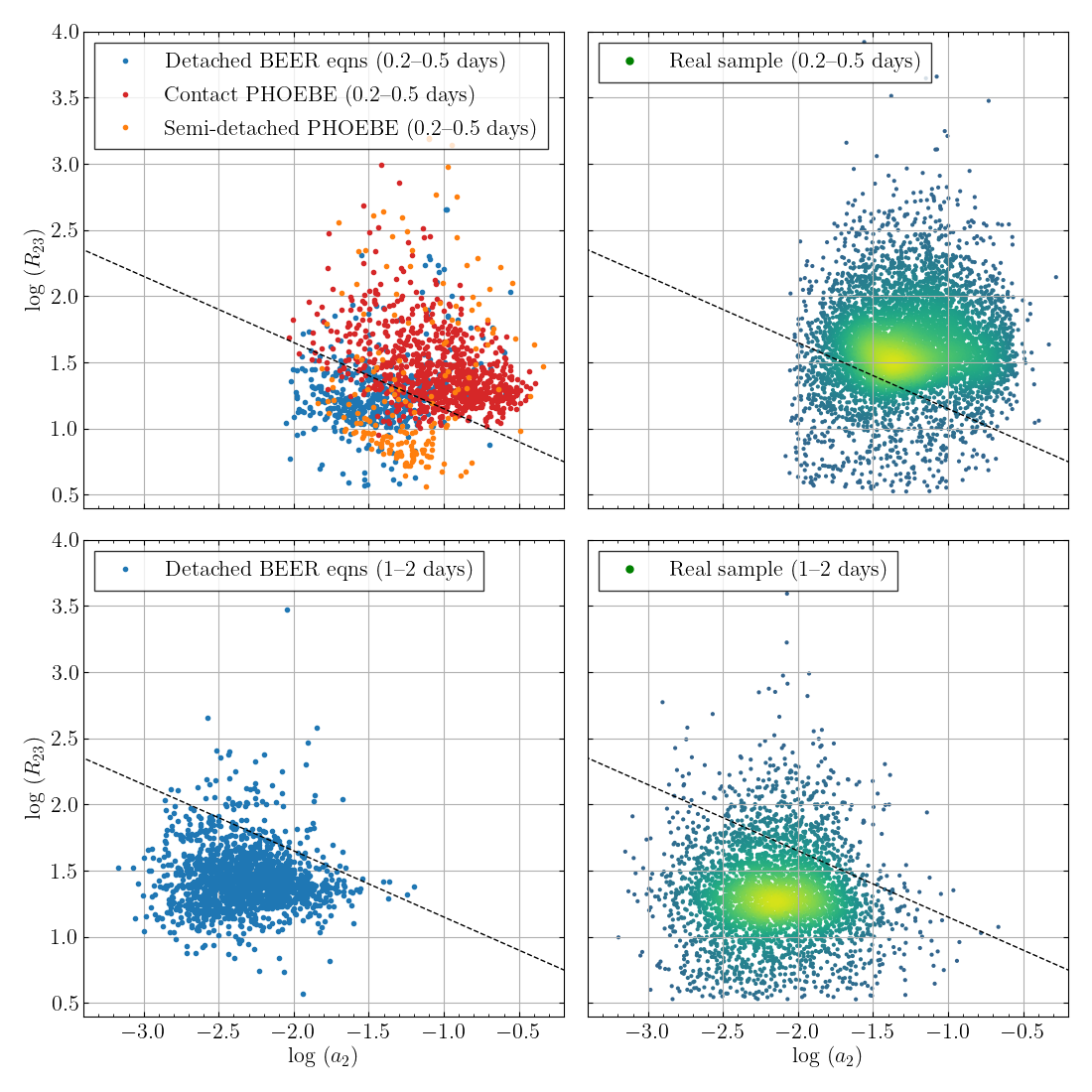}
\caption{$R_{23} = a_2 / a_3$ as a function of $a_2$ for several sets of simulated lightcurves (left) and for the real sample (right).
The simulated lightcurves were generated as described in Section~\ref{sec:simulations}.
The top row shows short-period systems, while the bottom row shows longer orbital periods.
The dashed line shows the cut in Equation~\ref{eq:contacts}, which approximately separates the detached and contact populations from each other.
}
\label{fig:contacts}
\end{figure*}

\begin{figure}
\includegraphics[width=\columnwidth]{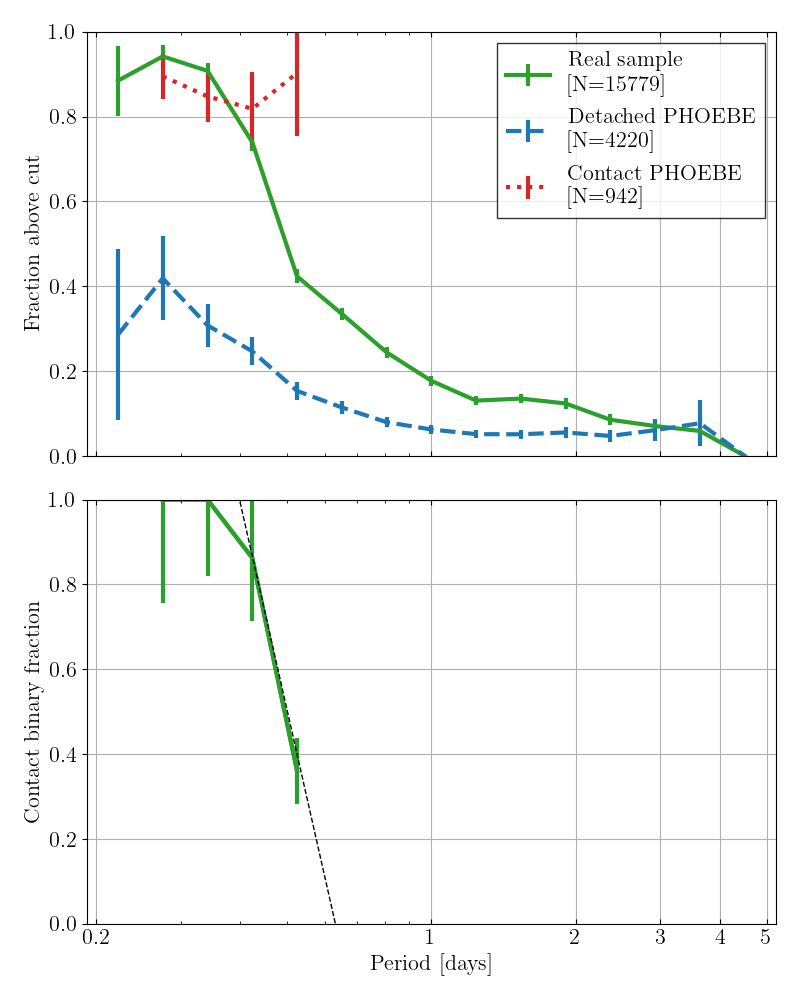}
\caption{\textit{Top:} The fraction of lightcurves with $R_{23}$ greater than the cut in \ref{eq:contacts} as a function of orbital period, for our real sample and for simulated lightcurves of detached and contact binary systems.
\textit{Below:} The fraction of the real sample which must be contact binary systems as a function of orbital period, in order to reproduce the curve of the real sample in the top figure.
The dotted grey line shows an approximation to the contact binary fraction described in Equation~\ref{eq:contact-fraction}.
}
\label{fig:contact-fraction}
\end{figure}

As already noted above, Equations~\ref{eq:beer}--\ref{eq:ell3} can be applied to detached and semi-detached binaries, but not to contact binaries, and as such contact binaries must be treated somewhat  differently.
It is therefore important to determine the fraction of contact binary systems at each orbital period.
Several previous works \citep[e.g.][]{Rucinski1997a,Rucinski1997,Rucinski1998,Rucinski2002,Pojmanski2002,Paczynski2006} have distinguished between contact and detached \textit{eclipsing} binaries by comparison of their measured harmonic amplitudes.
While we note that not all of the binaries in our sample are eclipsing, we nevertheless find a similar method to be effective using $a_2$ and the third harmonic amplitude $a_3$.

In the left panels of Fig.~\ref{fig:contacts}, we plot the $a_2$ amplitude and the ratio $R_{23} = a_2 / a_3$ of simulated lightcurves generated for detached, semi-detached, and contact binaries, as described in Section~\ref{sec:purity}. 
For each simulated population, we consider only simulated binary systems that were selected as candidates by our algorithm for consistency with the true sample.
Detached and contact binaries tend to concentrate in different regions of amplitude space, separated by the line
\begin{equation}
\log(R_{23}) = 0.65 - 0.5 \log(a_2),
\label{eq:contacts}
\end{equation}
although each population has some members that cross that line.

Because both the simulated populations of detached and contact binaries overlap the line in Equation~\ref{eq:contacts}, we cannot confidently determine for each target whether it is a contact or detached binary. 
However, in any given period range it is possible to estimate what fraction of targets are in contact, by comparison of the fraction of targets that lie above the line in Equation~\ref{eq:contacts} to the expected fraction for a purely detached or a purely contact population.
The right panels of Fig.~\ref{fig:contacts} show the measured amplitudes of our real sample in comparison to the same line.
At short periods, a significant fraction of our sample lies above the line, suggesting that a high fraction of targets are contact binaries.
At long periods, the amplitude distribution of our sample matches closely with that of the simulated detached sample.
The top panel of Fig.~\ref{fig:contact-fraction} shows, for the real sample and both simulated datasets, the fraction of targets that lie above the line of Equation~\ref{eq:contacts} as a function of orbital period. 
Within each period bin, the fraction of targets that are contact binaries can be found by solving the equation
\begin{equation}
F_\mathrm{real} = f_\mathrm{cb} F_\mathrm{contact} + (1 - f_\mathrm{cb}) F_\mathrm{detached}
\end{equation}
where $f_\mathrm{cb}$ is the fraction of targets in the sample that are contact binaries, and $F_\mathrm{real}$, $F_\mathrm{contact}$, and $F_\mathrm{detached}$ are respectively the fractions of the real targets, the simulated contact binaries, and the simulated detached binaries that lie above the line in Equation~\ref{eq:contacts}.
The uncertainty on $f_\mathrm{cb}$ in each bin is estimated using Poisson statistics on the numbers of simulated and real binary systems in that bin.

The resulting contact binary fraction decreases with increasing orbital period, as expected (Fig.~\ref{fig:contact-fraction}).
The fraction as a function of orbital period can be approximated by the function
\begin{equation}
f_\mathrm{cb}(P_\mathrm{orb}) = -5.0 \log (P_\mathrm{orb}) - 1.0,
\label{eq:contact-fraction}
\end{equation}
in the period range in which this function returns a value between 0 and 1.
Note that, as a result of the \citet{Gazeas2009} parameter relationships used, our simulated contact binary sample contains very few systems with orbital periods longer than 0.6\,days, and as such we are unable to estimate the contact binary fraction at orbital periods longer than this. 
Fig.~\ref{fig:contact-fraction} shows that the real sample does not quite behave like the simulated, detached binary sample even at longer orbital periods. 
This might be explained by some contact binary contribution (or other contamination) on the level of 10 per cent at longer orbital periods.
Indeed, some $\sim 20$ per cent of contact binary systems are expected to have orbital periods of 0.6--1.0\,days \citep{Jayasinghe2020}.

By multiplying the estimated contact binary fraction by the orbital period distribution of our sample, we estimate that our sample contains $5200 \pm 500$ contact binaries, or $39 \pm 4$ per cent of the sample.

On an individual level, a probability that a particular target is in contact or detached can be estimated based on its position in amplitude and period space, by comparison with the simulated lightcurves.
Around each target, we drew a spheroid with logarithmic radii of 0.1\,dex in the $a_2$ and $R_{23}$ planes and 0.025\,dex in the \porb\ plane.
If fewer than 20 total simulated lightcurves were included in that spheroid, the radii were scaled up in integer multiple steps, up to a factor of five, until at least 20 simulated lightcurves were included; if there were still fewer than 20 simulated lightcurves included after the factor of five increase in radius, then no probability was calculated.
For each target, the probability that it is a contact system was then estimated by 
\begin{equation}
P(\mathrm{contact}) = \frac{ \frac{n(\mathrm{contact})}{N(\mathrm{contact})}} {\frac{n(\mathrm{contact})}{N(\mathrm{contact})} + \frac{n(\mathrm{detached})}{N(\mathrm{detached})}},
\end{equation}
where $N$ and $n$ respectively refer to the total number of simulated lightcurves, and the number of simulated lightcurves inside the described spheroid, for contact and detached simulations as indicated.
This procedure was repeated for all targets in the selected sample, and the estimated probabilities are included in a column in Table~\ref{tab:short-table}.

%%%%%%%%%%%%%%%%%%%%%%%%%%%%%%%%%%%%%5

\section{Sample Properties}
\label{sec:sample-properties}

%%%%%%%%%%%%%%%%%%%%%%%%%%%%%%%%%%%%%%
With the ellipsoidal binary sample we have constructed, and the selection effects that we have estimated based on followup spectroscopy and simulations, we now proceed
to evaluate the statistical properties of the short-period binary population, both the observed
one and the underlying one, after accounting for the observational effects.

\subsection{Companion frequency}
\label{sec:frequency}

Our sample contains a total of \samplesize\ ellipsoidal binary systems, out of an initial \inputsize\ \textit{TESS} targets that were analysed by our \software{beer} code.
The frequency of confirmed ellipsoidal companions among these input targets is therefore $0.0037 \pm 0.0006$, where an uncertainty of 17 per cent is used because of the estimated purity of 83 per cent.

The overall frequency of companions to main-sequence stars in the period range \porb~$< 3$\,days can be estimated by correcting our frequency of detected companions for the selection effects that were measured in Section~\ref{sec:simulations}.
As described in that section, the strongest selection effect is relative to orbital period.
We divide our sample into period bins (the same bins used in Section~\ref{sec:period}), and within each bin correct the number of systems by the period-dependent selection efficiency.
In this manner, we estimate that the overall frequency of companions to main-sequence stars in the period range \porb~$< 3$\,days is $0.010 \pm 0.002$, where the uncertainty is a combination of the estimated 17 per cent impurity and the 10 per cent uncertainty on the correction for selection efficiency.

If we divide the sample at \porb$= 0.6$\,days so as to separate the contact binaries from the rest of the sample, we find a corrected companion frequency of $0.0036 \pm 0.007$ in the period range 0.2--0.6\,days and $0.007 \pm 0.002$ in the period range 0.6--3.0\,days.

It is common to consider companion frequency per logarithmic period interval, $f_{\log P} = d N / d \log P$ \citep[see for example,][]{Moe2017,El-Badry2022b}.
Expressed in this manner, our sample overall has $f_{-0.7 < \log P < 0.5} = 0.0079 \pm 0.0015$.
Dividing this frequency at a period of 0.6\,days gives $f_{-0.7 < \log P < -0.2} = 0.0076 \pm 0.0015$ for periods of 0.2--0.6\,days, and $f_{-0.2 < \log P < 0.5} = 0.0080 \pm 0.0015$  for periods of 0.6--5.0\,days.

%%%%%%%%%%%%%%%%%%%%%%%%%%%%%%%%%%%%%%

\subsection{Amplitudes}
\label{sec:amplitudes}

\begin{figure*}
\includegraphics[width=500pt]{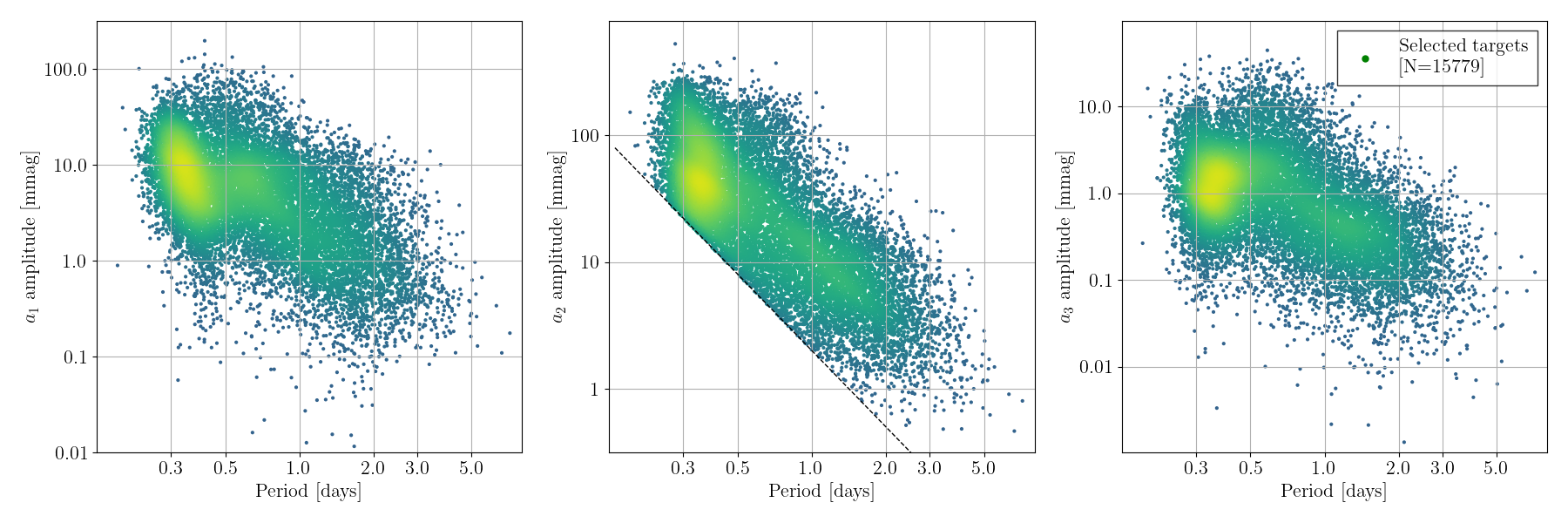}
\caption{The distribution of measured amplitudes as a function of orbital period for our selected targets.
The dashed line in the second panel shows the cut applied in Equation~\ref{eq:lco-cut}.
A separation in amplitude-period space can be seen between contact and detached binary systems.
}
\label{fig:amplitudes}
\end{figure*}

Fig.~\ref{fig:amplitudes} shows the harmonic amplitudes $a_1$, $a_2$, and $a_3$ versus \porb\ for our sample.
The diagonal line shows the cut that was applied in Section~\ref{sec:purity} to remove a significant number of (likely) non-binary contaminants.
A discontinuity in amplitudes (and a bimodal period distribution) can be seen between the contact binary systems at short orbital periods and the detached binary systems at longer orbital periods.
Targets with $a_2 \gtrsim 100$\,mmag are likely to be eclipsing binary systems \citep{Gomel2021b}.

%%%%%%%%%%%%%%%%%%%%%%%%%%%%%%%%%%%%%%%%

\subsection{Magnitude}
\label{sec:magnitude}

\begin{figure}
\includegraphics[width=\columnwidth]{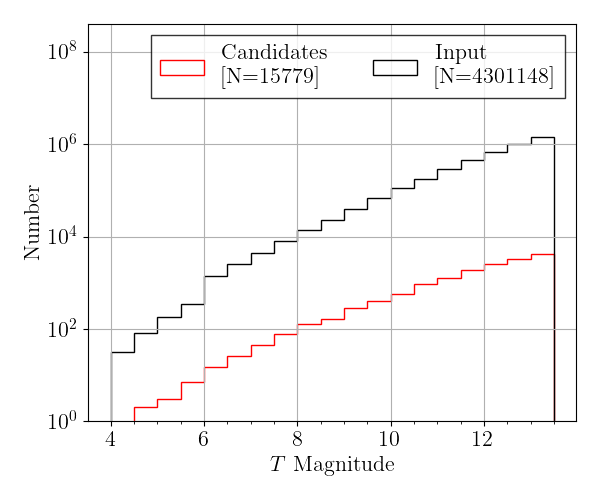}
\caption{The distribution of magnitudes in our sample, compared to the magnitude distribution of the \textit{TESS} lightcurves from which our sample was selected.
Some magnitude dependence can be seen in the selection process.
}
\label{fig:magnitudes}
\end{figure}

The \textit{TESS} $T$-band apparent magnitude distributions of our sample are shown in Fig.~\ref{fig:magnitudes}. 
93 per cent of targets in our sample have magnitudes in the range $10 < T < 13.5$.

Our sample shows a bias towards brighter magnitudes when compared to the input target list, most likely due to the lower signal-to-noise ratio (SNR) of fainter targets.
Our sample has a factor of 10 more stars in the $T=$13--13.5\,mag bin than the $T=$10--10.5\,mag bin, while for the input sample the same ratio is 15.

In the right-hand panel of Fig.~\ref{fig:hrd}, above, we showed a \textit{Gaia}-based colour-magnitude diagram of our BEER sample, compared to a volume-selected sample of background stars.
Our selected sample has an approximately similar colour distribution to a magnitude selected background sample (once giants and sub-giants are removed). 

Our sample is offset towards brighter magnitudes relative to a magnitude-selected background sample. 
At most colours, the offset of the centre of the sample is approximately 0.8 magnitudes.
By comparison, a twin-mass binary system is expected to be 0.75 magnitudes brighter than a single main-sequence star.
A companion of lower mass will have a smaller effect on the magnitude, but will push the binary system towards redder colours, contributing to an apparent vertical offset.
For a minority of targets in our sample, the apparent vertical offset relative to the main sequence is as much as 1.4 magnitudes, perhaps suggestive of the presence of a triple or higher-order system, or perhaps evidence that a minority of primary stars have begun to evolve off the main sequence.
We note that selection effects towards higher SNR and larger radii will also contribute to this offset.

%%%%%%%%%%%%%%%%%%%%%%%%%%%%%%%%%%%%%%%%

\subsection{Distance}
\label{sec:distance}

\begin{figure}
\includegraphics[width=\columnwidth]{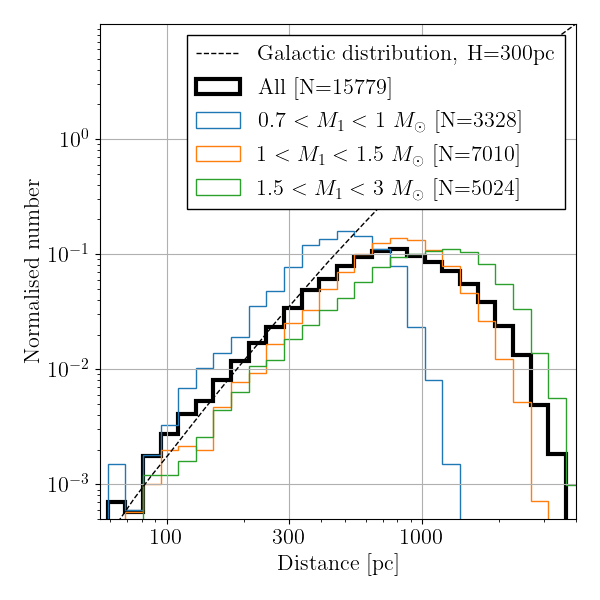}
\caption{Distribution of distances in our sample (thick black line), and for several primary mass bins (coloured lines), showing that brighter targets are detected to larger distances. 
The dashed line shows an approximate Galactic distribution of stars, as described in the text.
Incompleteness effects are present among low-mass stars at distances $\gtrsim 300$\,pc.
}
\label{fig:distances}
\end{figure}

The distribution of distances to the targets in our sample, as calculated from their \textit{Gaia} Data Release 3 parallaxes, is shown in Fig.~\ref{fig:distances}.
For comparison, we also calculate the expected distribution following the prescription of \citet{Pretorius2007}, approximating the Galaxy as an axisymmetric disc with a scale height of 300\,pc, without accounting for halo, bulge, spiral structure or thick disc.

Distance selection effects are present in our sample for distances $\gtrsim 300$\,pc. 
Targets with lower-mass primary stars are more vulnerable to these selection effects, as demonstrated in Fig.~\ref{fig:distances}.

%%%%%%%%%%%%%%%%%%%%%%%%%%%%%%%%%%%%%%

\subsection{Orbital period}
\label{sec:period}

\begin{figure*}
\includegraphics[width=\columnwidth]{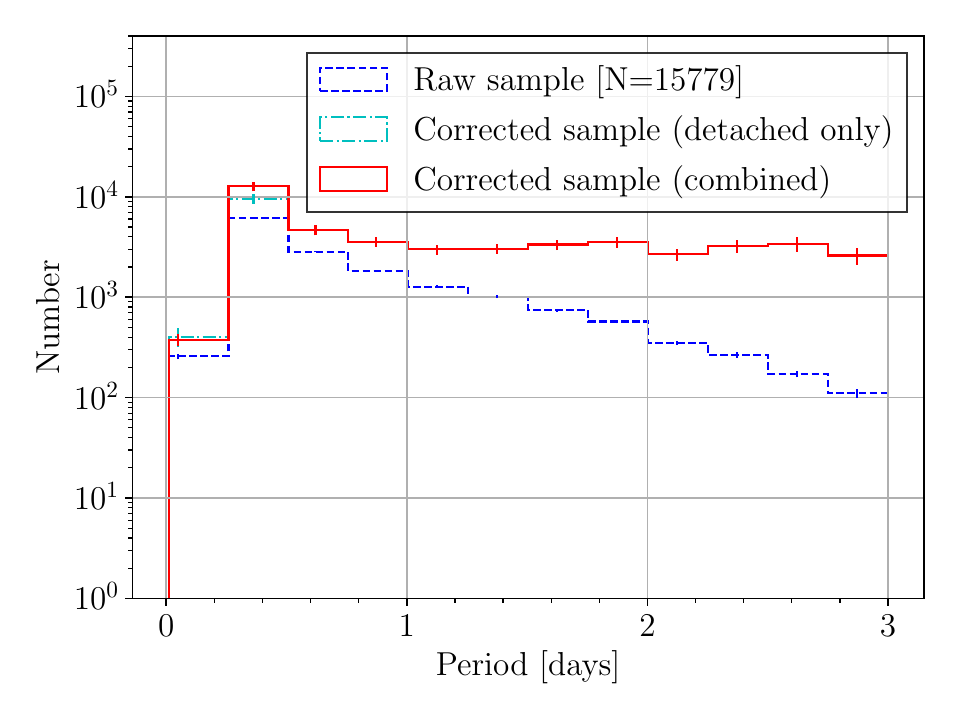}
\includegraphics[width=\columnwidth]{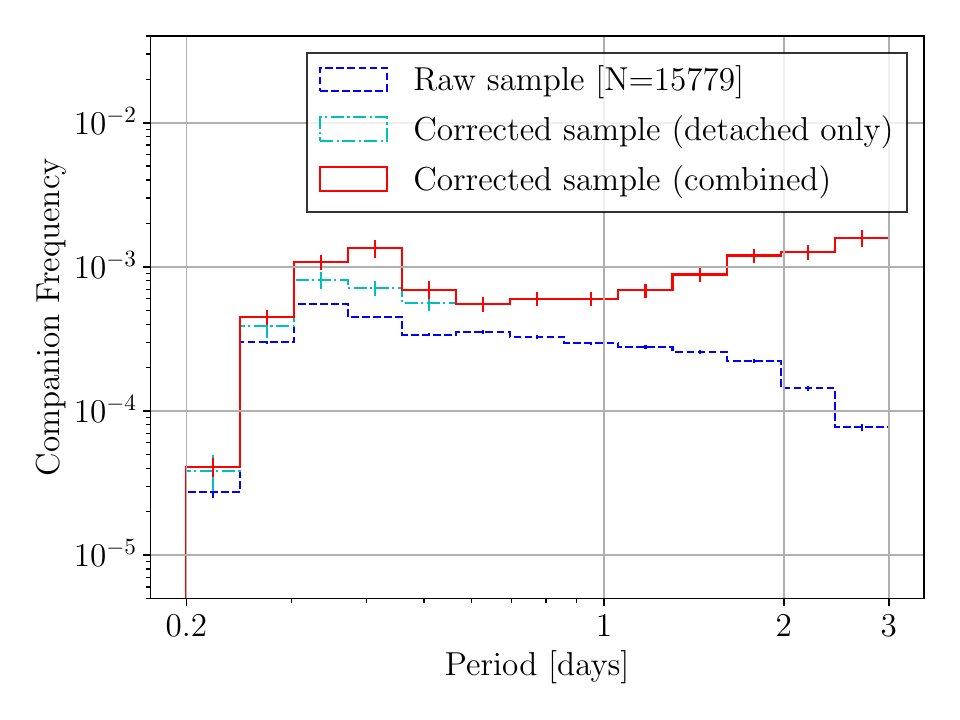}
\caption{The orbital period distribution of our sample, shown with linear (left) and logarithmic (right) period bins and axes.
The vertical axes show the number of targets detected (left) and the frequency of companions within the analyzed \textit{TESS} lightcurves (right).
The dashed line shows the raw distribution of our sample, while the solid line shows the distribution corrected for selection effects (taking into account the estimated contact binary fraction), and the dash-dotted line shows the correction assuming all targets are detached binary systems.
The orbital period distribution is approximately uniform on a linear scale between orbital periods of 1 and 3\,days.
}
\label{fig:period}
\end{figure*}

\begin{figure}
\includegraphics[width=\columnwidth]{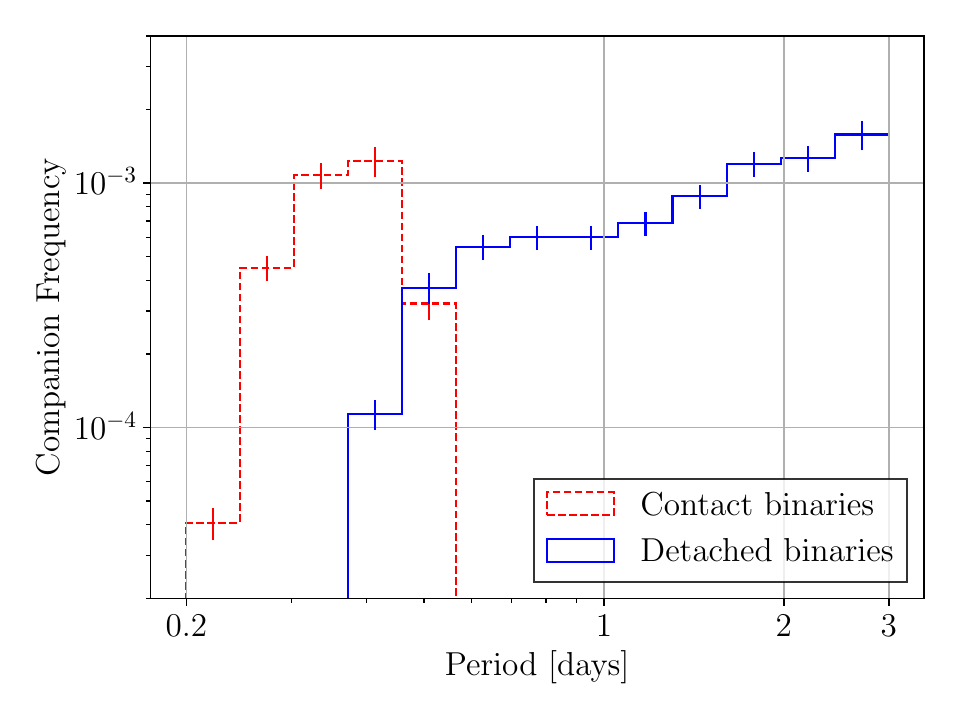}
\caption{The separated orbital period distributions of contact and detached binary systems, after correcting for selection effects in the manner described in the text. The period-dependent contact binary fraction is estimated using Equation~\ref{eq:contact-fraction}.
}
\label{fig:period-cbs}
\end{figure}

In Fig.~\ref{fig:period}, we show the orbital period distribution for our sample. 
We correct the period distribution by the selection function derived in Section~\ref{sec:simulations}, to approximate the true period distribution of the underlying population.
At short periods, we apply two forms of correction for the selection efficiency.
The first, shown by the dash-dotted line in Fig.~\ref{fig:period}, was calculated under the assumption that all target binary systems are detached.
The second, shown by the solid line, combines our estimated selection efficiencies for detached and contact binary systems, assuming that the fraction of binary systems that are in contact varies with period in the way calculated in Section~\ref{sec:contacts} (specifically, using the approximation in Equation~\ref{eq:contact-fraction}).

The overall period distribution of the underlying population has a peak centred on \porb$\approx 0.3-0.4$\,days, presumably a ``pile-up'' of contact binary systems, and is approximately flat at periods \porb$>1$\,day.

We also attempt to separate the period distributions of contact and detached binary populations by multiplying our approximation of the period distribution of the underlying population (Fig.~\ref{fig:period}) by the approximation of the contact binary fraction (Equation~\ref{eq:contact-fraction}). 
The separated period distributions of the two populations are shown in Fig.~\ref{fig:period-cbs}.
The peak at \porb$<1$\,day is entirely dominated by contact binaries.
Following the approximation in Equation~\ref{eq:contact-fraction}, we do not find any contact binaries with orbital periods longer than 0.6\,days; however, as discussed in Section~\ref{sec:contacts}, there are likely to be some contact binaries present at orbital periods 0.6--1.0\,days that are not replicated by our simulations, perhaps at a level of a few per cent of the detached population.

\subsection{Primary stellar temperature}
\label{sec:teff}

\begin{figure}
\includegraphics[width=\columnwidth]{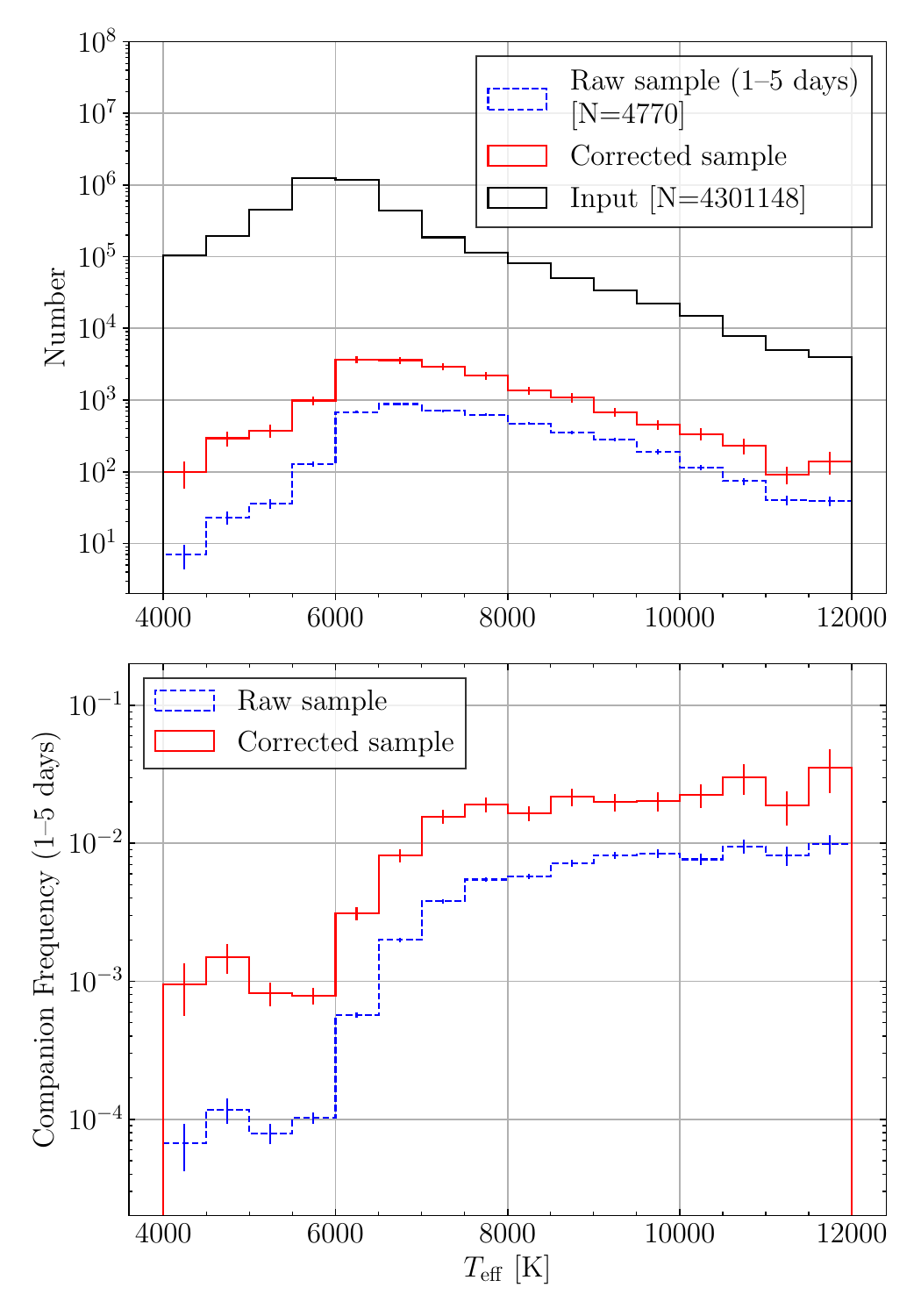}
\caption{\textit{Top:} The TIC \teff\ distribution of our binary sample (\porb\,$=1-5$ days) compared to the temperatures of all \textit{TESS} targets from which the sample was selected.
We plot both the distribution of our sample itself (red, dashed) and the distribution corrected for selection effects (red, solid).
Five targets from the sample with temperatures outside of the plotted range have been excluded.
\textit{Bottom:} Companion frequency among the analyzed \textit{TESS} lightcurves, as a function of TIC \teff.
Again, the dashed line shows the companion frequency calculated from our raw sample, and the solid line includes a correction for selection effects.
There is a significant step in the companion frequency at \teff\,$\sim 6000-7000K$.
}
\label{fig:teff}
\end{figure}

\begin{figure}
\includegraphics[width=\columnwidth]{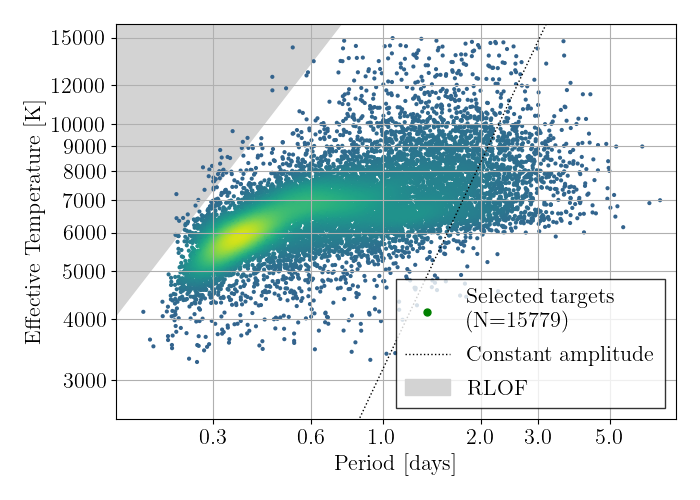}
\caption{The distribution of TIC effective temperatures in our sample as a function of orbital period.
The shaded grey region shows the approximate parameter space in which the primary star will experience Roche lobe overflow (RLOF) in an equal-mass binary system.
The dotted line shows an example diagonal on which any equal-mass main sequence binary system will have the same ellipsoidal amplitude. Selection effects dependent on amplitude will be parallel to this line.
At orbital periods longer than 0.6\,days, there is little visible correlation between period and \teff.
}
\label{fig:teff-period}
\end{figure}

In Fig.~\ref{fig:teff}, we show the distribution of temperatures of the photometric primary stars of our sample at \porb$> 1$\,day.
At these orbital periods the sample is dominated by detached binary systems.
In the lower panel of Fig.~\ref{fig:teff}, we show the frequency of companions in the period range 1--5\,days as a function of primary star temperature.
As in the previous section, we correct the primary star temperature distribution of our sample by the selection efficiency of our algorithm as a function of temperature (Section~\ref{sec:simulations}), so as to approximate the primary star temperature distribution of the underlying population of ellipsoidal binaries.
We show both corrected and uncorrected distributions, as well as the temperature distribution of the set of input \textit{TESS} targets from which our sample was selected.

We find that companions are more than an order of magnitude more common among stars with $T_\mathrm{eff} > 8000$ than with $T_\mathrm{eff} < 6000$, even after correcting for selection effects.
We discuss this in further detail in Section~\ref{sec:discussion}.

In Fig.~\ref{fig:teff-period} we plot the TIC effective temperature of the photometric primary star as a function of orbital period for all targets in our sample.
Two distinct populations can be seen.
Targets with orbital periods in the range 0.3--0.6\,days have approximately solar temperatures, and show a correlation between temperature and orbital period.
The slope of the correlation at short periods does not exactly replicate the expected slope for Roche-lobe filling primary stars.
Targets with orbital periods \porb$\gtrsim 0.6$\,days have a temperature distribution centred on 6000--8000\,K and show no correlation with orbital period.

For demonstrative purposes, on the same figure we also plot an example line of constant ellipsoidal amplitude, calculated from Equation~\ref{eq:ell2} with the approximations $M \propto R^{0.8}$ and $T_\mathrm{eff} \propto M$.
Any artificial trends introduced by a selection bias towards binaries with stronger ellipsoidal amplitudes should follow this diagonal.
Since the observed correlation at short periods has a significantly different gradient, we conclude that it is not artificially introduced.

A similar period-luminosity relationship is known for contact binaries \citep{Pawlak2016,Jayasinghe2020}.
We also examined the samples of ellipsoidal binaries selected from ASAS-SN \citep{Rowan2021} and OGLE \citep{Soszynski2016,Gomel2021c}, cross-matching both samples with the TIC to find primary temperatures.
Although both samples have a relatively small number of systems in this period range, a similar distribution was found in temperature-period space.

\section{Discussion}
\label{sec:discussion}

\subsection{Comparison to TESS eclipsing binary sample}
\label{sec:prsa}

\begin{figure}
\includegraphics[width=\columnwidth]{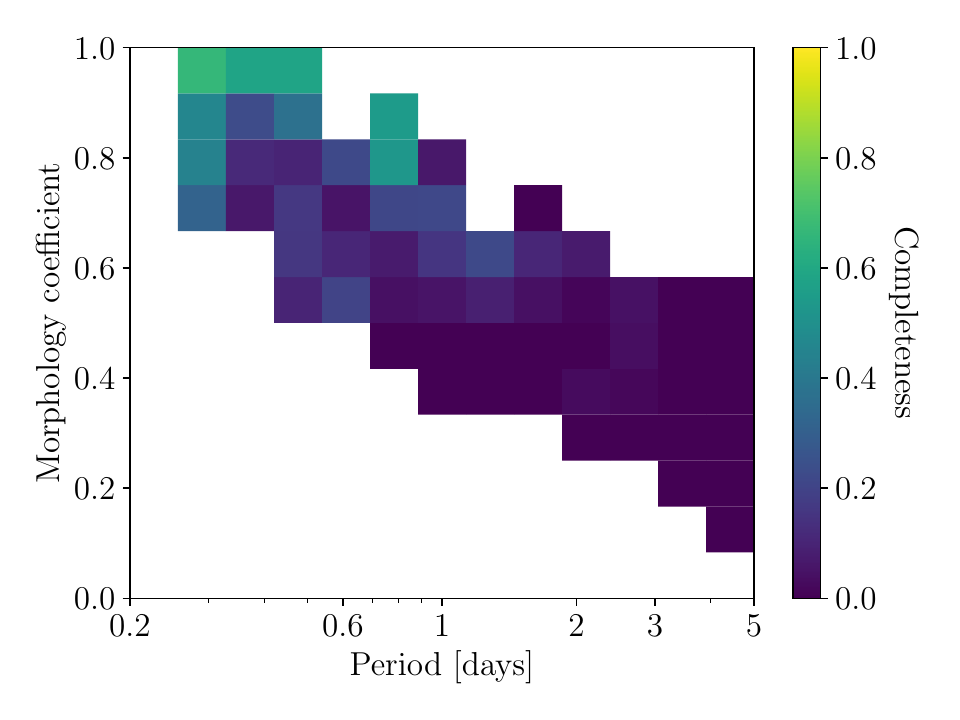}
\caption{The completeness of our ellipsoidal sample relative to the \textit{TESS} eclipsing binary sample of \citet{Prsa2022}.
Our completeness depends strongly on the orbital period and the morphological coefficient, which parametrizes the shape of the lightcurve (high values implying that ellipsoidal modulation dominates over eclipses).
}
\label{fig:prsa_morph}
\end{figure}

\review{
\citet{Prsa2022} have also presented a sample of 4584 photometrically-selected binary systems from the first two years of \textit{TESS}.
Although focused on eclipsing binary systems, that sample also included a number of ellipsoidal binary systems, and so it is reasonable to expect some overlap between that sample and the present sample.
This can be used to explore the completeness and selection effects of the two samples.
}

\review{
The \citet{Prsa2022} sample was drawn from only the targets which were observed by \textit{TESS} with short cadence (during the first two years, a cadence of two minutes), not from the full-frame images used for our sample.
Short-cadence data from the first two years of \textit{TESS} is available for some 230\,000 targets.
This gives their sample an occurrence rate of 0.021, as opposed to our occurrence rate of 0.0037 (Section~\ref{sec:frequency}).
The difference in occurrence rate can be partially explained by the larger period range of the \cite{Prsa2022} sample, but even at short periods the occurrence rate of the \citet{Prsa2022} sample is significantly higher.
}

\review{
A number of cuts were applied during our sample selection process, including on magnitude, \textit{Gaia} data quality, contamination, and colour-magnitude position (Section~\ref{sec:candidate-selection}).
Of the 230\,000 targets short-cadence \textit{TESS} targets from which the \citet{Prsa2022} sample was drawn, some 150\,000 (64 per cent) survive our cuts.
Of these targets, 605 ($0.41 \pm 0.1$ per cent) were selected for our ellipsoidal sample, comparable to the $0.39 \pm 0.01$ per cent selection rate of our full ellipsoidal sample from the entire \textit{TESS} dataset.
Of the 4854 targets in the \citet{Prsa2022} eclipsing binary sample, 2881 survive the same quality cuts, chiefly removed by the colour-magnitude cut.}

\review{
There is an overlap of 264 targets between our ellipsoidal sample and that of \citet{Prsa2022}.
Therefore, the overall completeness of our sample against the \textit{TESS} eclipsing binary sample is $9.2 \pm 0.6$ per cent, and the completeness of that sample against ours is $43.6 \pm 2.7$ per cent.
As was previously observed, the completeness of our ellipsoidal binary sample is a strong function of orbital period. 
Relative to the eclipsing binary sample, our completeness is $21 \pm 2$, $17 \pm 1$, $15 \pm 1$, and $12 \pm 1$ per cent for periods shorter than 1, 2, 3, and 5 days, respectively.
}

\review{
\citet{Prsa2022} use a `morphological coefficient' parameter, originally defined by \citet{Matijevic2012}, to classify the morphology of a lightcurve.
The parameter continuously varies between (approximately) 0 and 1, where values close to zero imply widely detached stars with little ellipsoidal modulation and values close to one imply contact binaries or purely ellipsoidal lightcurves.
}

\review{
As might be expected, the completeness of our sample relative to the sample of \citet{Prsa2022} shows a strong dependence on the morphological parameter, as shown in Fig.~\ref{fig:prsa_morph}.
If we consider only targets with a morphological parameter greater than 0.9, our completeness relative to \citet{Prsa2022} becomes $57 \pm 7$ per cent.
}

\review{ 
It is worth noting that, for morphological values close to 1, the lightcurves of targets in \citet{Prsa2022} may be close to purely sinusoidal. 
Among purely sinusoidal lightcurves, we have previously found that contamination by pulsating variables is relatively common \citep{Tal-Or2015,Gomel2019}, and so we applied a filter to remove such targets from our sample (as described in Section~\ref{sec:candidate-selection}).
By cross-matching the sample of \citet{Prsa2022} with our amplitude measurements listed in Table~\ref{tab:long-table} we find that, for 20 per cent of their targets with morphological parameters greater than $0.8$, we only measure a significant ($>3\sigma$) signal at the ellipsoidal frequency ($a_2$), not at $a_1$ or $a_3$, suggesting a lightcurve close to purely sinusoidal.
A further 10 per cent of \citet{Prsa2022} targets with morphological parameters greater than $0.8$ fall below the cut in amplitude-period space that we describe in Equation~\ref{eq:lco-cut}, below which we found that the majority of targets are not RV-variable.
}

\subsection{Period distribution compared to other samples}
\label{sec:comparisons}

\begin{figure}
\includegraphics[width=\columnwidth]{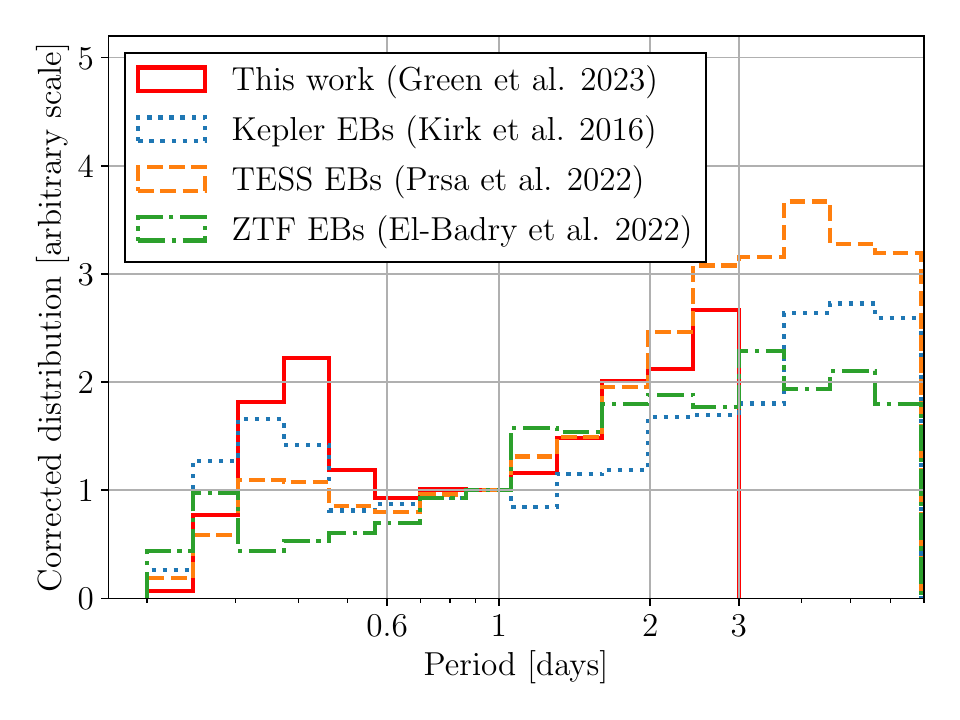}
\caption{A comparison of the orbital period distribution of our sample with several eclipsing binary samples, as summarized in Table~\ref{tab:ebs-samples}.
Between one and three days our period distribution is similar to those of other samples.
}
\label{fig:ebs_comparison}
\end{figure}

\begin{figure}
\includegraphics[width=\columnwidth]{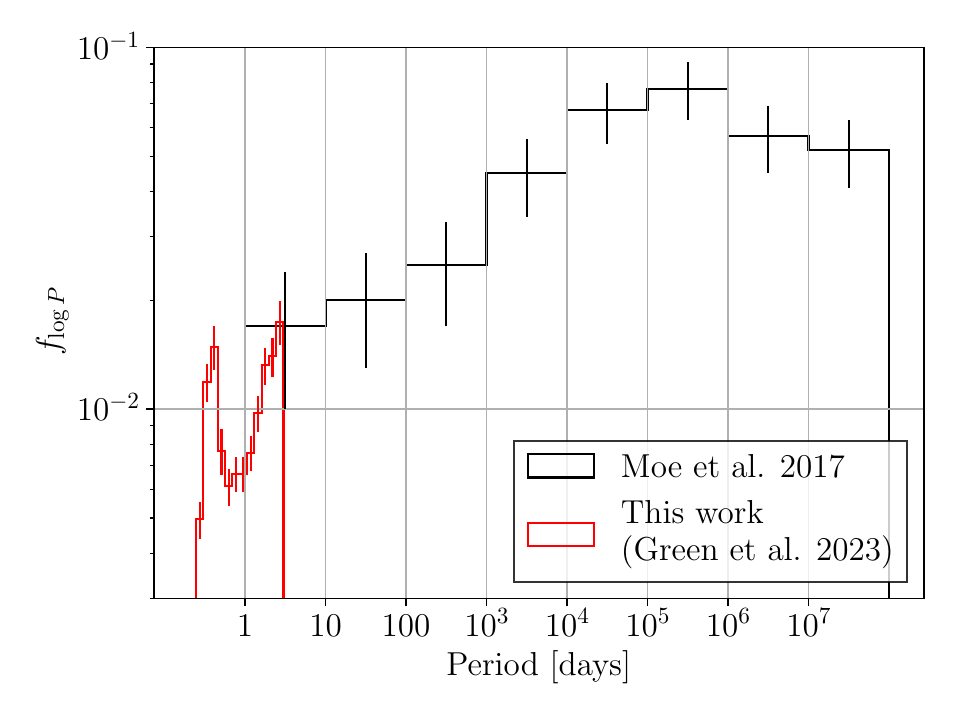}
\caption{The orbital period distribution (in terms of companion frequency per logarithmic bin) of the binary star population, as estimated from our sample in red, and from the \citet{Raghavan2010} sample by \citet{Moe2017} in black.
}
\label{fig:moe_comparison}
\end{figure}

\begin{table}
\caption{The eclipsing binary samples shown in Fig.~\ref{fig:ebs_comparison}, where $N$ is the number of systems in the sample. The temperature range shown is the 5--95 per cent range.
}
\begin{tabular}{lccc}
\label{tab:ebs-samples}
Survey & $N$ & $T_\mathrm{eff}$ [K] & Ref.\\
\hline
\textit{Kepler} EBs & 2876 & 4600--7900 & \citet{Kirk2016}\\
\textit{TESS}   EBs & 4584 & 3400--9500 & \citet{Prsa2022}\\
ZTF             EBs & 3879 & 4000--5400 & \citet{El-Badry2022b}\\
This work   & \samplesize\ & 4900--9400 & -- \\
\end{tabular}
\end{table}

In Fig.~\ref{fig:ebs_comparison}, we compare our corrected period distribution with the period distributions of eclipsing binary samples from \textit{Kepler}, \textit{TESS}, and ZTF \citep{Kirk2016,Prsa2022,El-Badry2022b}. 
Details of these samples are given in Table~\ref{tab:ebs-samples}; note that each sample has a somewhat different range of \teff.
We corrected the form of each eclipsing sample for completeness.
\citet{El-Badry2022b} found for the ZTF sample that the probability of detection is dominated by the probability of eclipse ($\propto P_\mathrm{orb}^{-2/3}$) at short periods, while at longer periods the eclipse duty cycle also becomes significant (also $\propto P_\mathrm{orb}^{-2/3}$, leading to a combined effect $\propto P_\mathrm{orb}^{-4/3}$).
We corrected the \citet{El-Badry2022b} sample using these two power laws, taking $P_\mathrm{orb} = 3$\,days to be the transition between the two.
For \textit{Kepler} and \textit{TESS}, as the window of observations in both surveys is significantly longer than the orbital periods considered, we assume a $\propto P_\mathrm{orb}^{-2/3}$ selection function 
\review{at $P_\mathrm{orb} > 1$\,day.
At shorter orbital periods, the \textit{Kepler} and \textit{TESS} samples are dominated by ellipsoidal and contact binary systems rather than eclipsing binaries, and the selection function becomes unclear.
For simplicity, we assume a constant selection efficiency for $P_\mathrm{orb} < 1$\,day for those two samples.}

As Fig.~\ref{fig:ebs_comparison} shows, in the period range 1--3\,days the shape of our distribution agrees reasonably well with those of all three eclipsing binary samples. 
(Note that we do not compare the absolute levels of the distributions, i.e., the binary frequency, which we have scaled arbitrarily in the figure.)
At shorter periods, the presence of the contact binary peak distinguishes our sample from the eclipsing binary samples (in which contact binary systems were deliberately selected against).
Our sample does not extend to longer orbital periods, but we note that for \porb\,$> 3$\,days the three eclipsing binary samples diverge somewhat from each other, perhaps due to further selection effects that are uncorrected (or perhaps a true difference arising from their different primary mass ranges).

In Fig.~\ref{fig:moe_comparison}, we show the $f_{\log P}$ distribution measured by \citet{Moe2017} based on the \citet{Raghavan2010} sample, compared to the $f_{\log P}$ distribution of our sample.
In both cases, the distribution has been corrected by the estimated completeness.
It should be noted that the two samples have not been handled in a consistent way: in the evaluation by \citet{Moe2017}, white dwarf companions and low-mass companions (those with $q < 0.1$) were removed, something that is not possible from our sample with the current data.
Our sample also contains a number of A-type stars, which were excluded from the \citet{Raghavan2010} sample.
Despite these caveats, there is a good continuity between the two distributions, given that our period distribution has a binary frequency value at a period of three days that is similar to that of the 1--10 day bin of the \citet{Moe2017} distribution.

The gradient of orbital periods in our sample appears steeper than the slope toward longer periods of the \citet{Moe2017} distribution.
We consider several possible interpretations of this steeper gradient. 
If the steep gradient of our sample is real, it suggests the presence of some physical process that decreases the companion frequency at \porb\,$\approx 1$\,day relative to \porb\,$\approx 3$\,days.
A natural interpretation would be magnetic braking, which (as discussed in Section~\ref{sec:magnetic-braking}) is likely to be a significant source of angular momentum loss among binary systems with orbital periods of a few days.

Alternatively, we could hypothesize some physical mechanism that enhances the companion frequency at periods of \porb\,$\approx 3$\,days relative to  \porb\,$\approx 1$\,day.
It has been suggested that Kozai-Lidov tidal interactions with a tertiary companion may cause a pile-up in binary orbital periods peaked at 2--3\,days \citep{Fabrycky2007}.
We discuss in Section~\ref{sec:triples} that at least $9.5 \pm 0.2$ per cent of our sample, and possibly a much greater fraction, have tertiary companions.
In order to explain the steep gradient of our sample by tidal interactions, the majority of our sample would require third components that remain undetected.

We also consider plausible systematic errors in our completeness correction or sample selection. 
Given the agreement between our period distribution and the eclipsing samples plotted in Fig.~\ref{fig:ebs_comparison}, such systematic errors are unlikely, but still worth discussing.
The steep gradient could be induced by period-dependent systematic error in our completeness correction, or by a period-dependent contamination in our sample.
A systematic error in our completeness correction would be caused by the presence of a period-dependent selection effect that we do not account for; for instance, a significant increase in stellar variability as orbital period decreases.
In order to reconcile our gradient with that of \citet{Moe2017}, our selection function at \porb\,$=1$\,day must be underestimated by approximately a factor of two relative to the selection function at \porb\,$=3$\,days.
Given that stellar activity typically saturates at \porb\,$\lesssim 3$\,days \citep[e.g.][]{Wright2011}, such a steep drop in completeness is difficult to explain from stellar variability.
Similarly, period-dependent contamination can only explain the steep gradient if the contamination rate at \porb\,$\approx 3$\,days is close to 50 per cent, which is difficult to reconcile with the LCOGT radial velocity observations presented in Section~\ref{sec:purity}.

Overall, we conclude that the steep drop in systems with decreasing orbital period in the range 1--3\,days is most likely the result of magnetic braking.

\subsection{Primary temperature and magnetic braking}
\label{sec:magnetic-braking}

\begin{figure}
\includegraphics[width=\columnwidth]{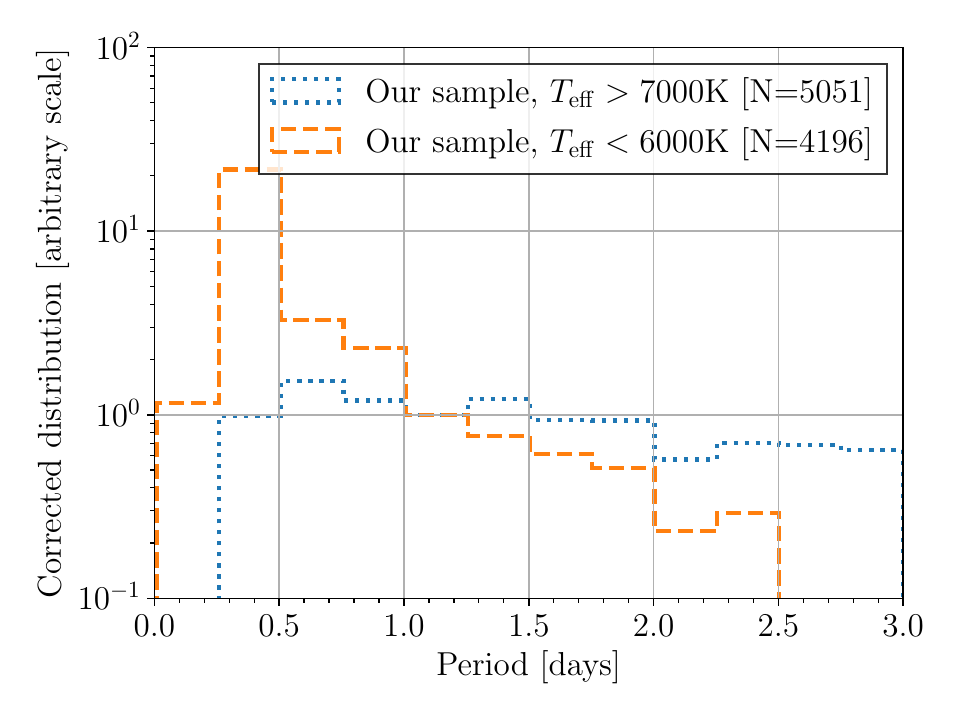}
\includegraphics[width=\columnwidth]{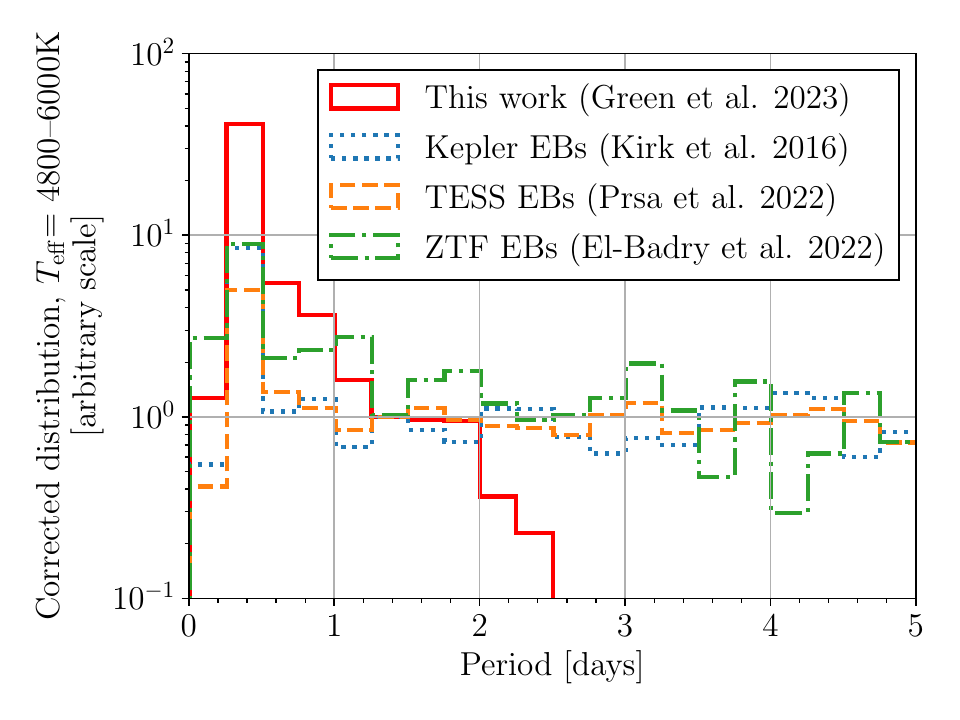}
\caption{\textit{Top:} The period distribution of binaries in our sample with primary $T_\mathrm{eff} > 7000$\,K or $T_\mathrm{eff} < 6000$\,K. They appear markedly different.
\textit{Bottom:} As Fig.~\ref{fig:ebs_comparison}, but limited to only binary systems with a primary temperature in the range 4800--6000\,K (note also that here the $x$-axis is linear). Apart from our sample, all agree that the period distribution is flat for \porb\,$\gtrsim 1$\,day. It therefore seems likely that the small number of systems per bin in our cool sub-sample has made it unreliable for \porb\,$\gtrsim 1.5$\,days.
}
\label{fig:period-hotcool}
\end{figure}

\begin{figure}
\includegraphics[width=\columnwidth]{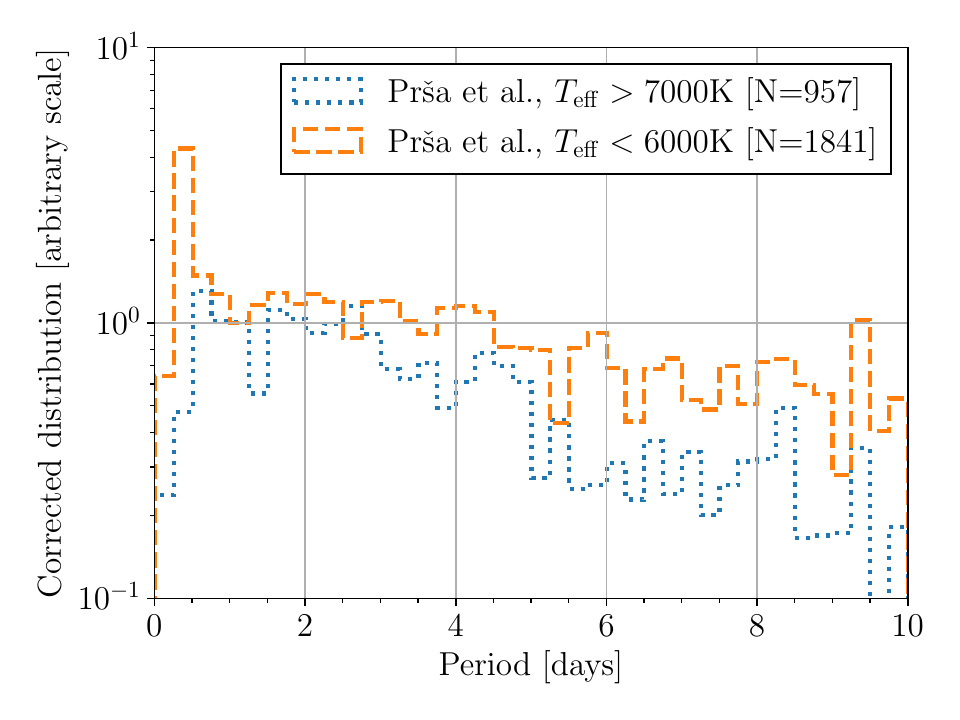}
\caption{As the upper panel of Fig.~\ref{fig:period-hotcool}, but applied to the \textit{TESS} eclipsing binary sample of \citet{Prsa2022}.
Their period distributions are approximately flat at short periods for both hot and cool binary systems, with a marginal suggestion of a divergence towards \porb\,$\gtrsim 3$\,days.
}
\label{fig:period-hotcool-prsa}
\end{figure}

In Section~\ref{sec:teff}, we showed an order-of-magnitude step change in the frequency of companions as a function of primary stellar temperature.
We note that \citet{Bashi2022} also commented on an apparent deficiency of binary systems in the \textit{Gaia} spectroscopic binary sample with \porb\,$\lesssim 5$\,days and $M_1 \lesssim 1.3 M_\odot$ ($T_\mathrm{eff} \lesssim 6300$ for main sequence primaries), although it should be considered that the completeness of the \textit{Gaia} spectroscopic sample is currently not well characterised. 
A correlation between multiplicity and primary star mass (and hence temperature) is a well-known feature of binary star populations \citep[e.g.][]{Moe2017}, but the scale of this step change is worthy of further discussion.

The dramatic increase in companion frequency occurs at approximately the Kraft break  \citep[$T_\mathrm{eff} \approx 6250K$,][]{Kraft1967,Jayasinghe2020,Avallone2022}. 
Stars hotter than the Kraft break have thin convective zones that cannot support a magnetized wind, while stars cooler than the Kraft break do exhibit a magnetized wind. 
Therefore, stars cooler than the Kraft break experience greater angular momentum loss due to magnetic braking.

It is reasonable to interpret this step change in binary frequency as a result of magnetic braking.
Cooler systems with more efficient magnetic braking will naturally lose angular momentum faster than hotter systems.
Angular momentum lost from the rotating component stars will be replaced from the orbit by tidal forces, causing the binary system to spiral inwards.
The result will be to shorten the lifetime of binary systems with primary stars cooler than 6250\,K which have orbital periods of a few days, leading to the observed deficiency of companions to cool primary stars within our period range.

\review{
\citet{El-Badry2022b} explored magnetic braking in their eclipsing binary sample and found that the period distribution does not follow that predicted by a canonical magnetic braking laws, but is instead more consistent with a model in which magnetic braking `saturates' at short rotation periods. 
Note that this does not necessarily imply that magnetic braking does not occur at all, only that its dependence on rotation period is weaker at short periods than at longer periods. 
The exact strength of the dependence at short periods is difficult to disentangle from the unknown birth period distribution of the population.
\citet{El-Badry2022b} searched also for any change in the orbital period distribution at the boundary between fully and partially convective M dwarfs ($\sim 3400$\,K), but did not find any significant change; previous works had suggested that magnetic braking may turn off below that boundary.
}

\review{
In Fig.~\ref{fig:period-hotcool}, we plot the period distribution of our sample, separated into systems with primary stars hotter and cooler than the Kraft break.
Any system with a detected, resolved third body (Section~\ref{sec:triples}) was removed, although this did not make any significant difference to the period distribution.
Completeness corrections were applied using the injection-recovery tests previously described, with the same temperature cuts applied.
The corrected period distributions are markedly different for the two populations, being approximately flat with linear orbital period for the hotter systems and going approximately as \porb$^{-2}$ for the cooler systems.
However, it should be noted that the cooler distribution is dominated by contact binaries at \porb\,$< 1$\,day, and has only a handful of systems per period bin for \porb\,$> 2$\,days, the latter making it especially vulnerable to any systematic issue with the correction for selection effects.
In the lower panel, we compare the period distribution of our cool sub-sample with the cool systems in the previously-discussed eclipsing binary samples\footnote{Note that we did not remove the triple systems from the eclipsing binary samples.}.
All samples apart from ours agree that the period distribution is approximately flat for \porb\,$> 1$\,days, suggesting indeed that our cool sub-sample is made unreliable by its small sample size at \porb\,$\gtrsim 2$\,days.
}

\review{
The \textit{TESS} eclipsing binary sample of \citet{Prsa2022} covers a similar temperature range to our sample, but has a greater representation of cool systems at long periods, making it ideal for the comparison of period distributions across the Kraft break.
In Fig.~\ref{fig:period-hotcool-prsa}, we plot the period distribution of that sample, likewise separated into hot and cool subsets.
A rough correction for completeness was applied, as described in Section~\ref{sec:comparisons}.
For both subsets of that sample, the period distribution is approximately consistent with flat for \porb\,$\gtrsim 1$\,day.
There is some marginal evidence that the period distributions diverge towards longer orbital periods, but given the crudeness of the correction for selection effects, it is unclear if this divergence is significant.
}

\review{
Overall, the sharp change in binary frequency coincident with the Kraft break is highly suggestive that magnetic breaking is shortening the lifetimes of short-period binary systems with primaries cooler than the Kraft break.
However, there is no strong evidence of any change in the period distribution between systems that are hotter and cooler than the Kraft break, with a number of independent samples all showing the same distribution that is constant with linear orbital period.
It may be that the weak dependence between the efficiency of saturated magnetic braking and orbital period, as shown by \citet{El-Badry2022b}, means that the mechanism reduces the overall number of cooler binaries without significantly changing their period distribution. 
Alternatively, it may be that magnetic braking still occurs in the hotter systems, with the same dependence on orbital period as for cooler systems, but a weaker overall strength.
}

\subsection{Triple systems}
\label{sec:triples}

\begin{figure}
\includegraphics[width=\columnwidth]{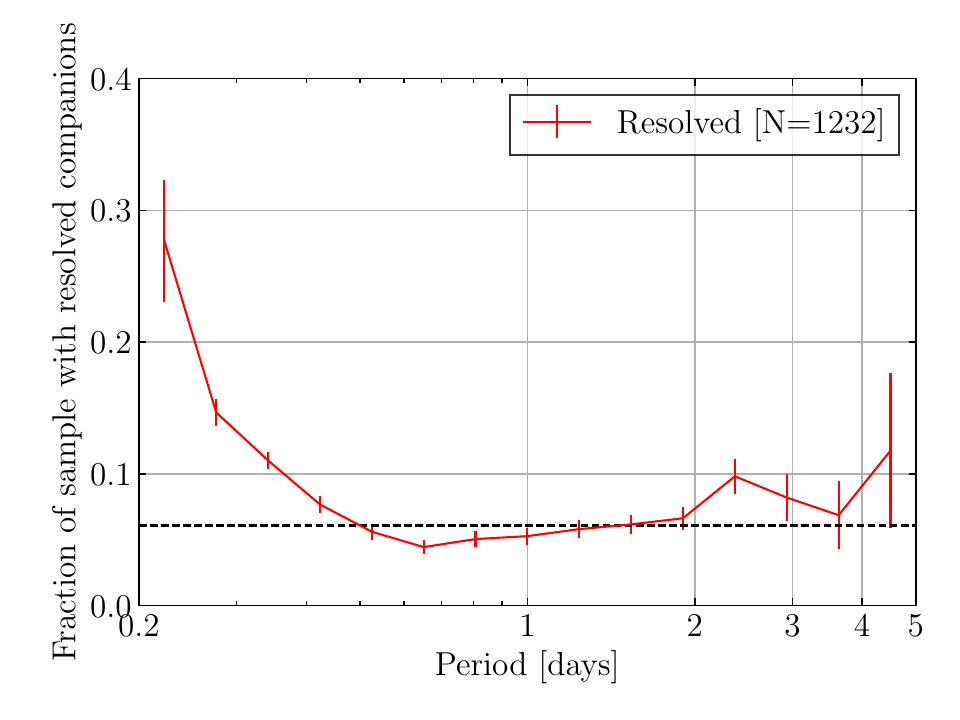}
\caption{Frequency of resolved tertiary companions to our binary candidate systems from \citet{El-Badry2021a} as a function of the orbital period of the inner binary.
The dashed line shows the frequency of resolved companions among the input \textit{TESS} targets from which our sample was selected. 
The frequency of companions is significantly elevated at short orbital periods, but not significantly greater than the value for single stars at inner orbital periods $\gtrsim 1$\,day.
}
\label{fig:tertiaries}
\end{figure}

\review{
A number of previous works have shown that most, perhaps all, short-period binary systems have tertiary companions, and that the probability of a third object being present is strongly anti-correlated with the binary orbital period \citep[e.g.][]{Tokovinin2006,Tokovinin2014,Laos2020,Hwang2022}.
Two explanations for this trend have been put forward. 
Firstly, tidal interactions with the tertiary object over the lifetime of the system may play a role in driving the inner binary towards shorter orbital periods, via a mechanism known as Kozai Cycles with Tidal Friction \citep[KCTF;][]{Lidov1962,Kozai1962,Mazeh1979,Kiseleva1998,Fabrycky2007,Naoz2014,Borkovits2022}.
Alternatively, it may be that dynamic interactions in the initially unstable triple during the pre-main sequence phase, combined with energy dissipation with the surrounding protostellar disc, are responsible for the short inner orbital periods \citep[often termed `dynamical unfolding';][]{Bate2002,Bate2009,Bate2012,Moe2018,Tokovinin2020,Borkovits2022}.
A population synthesis model by \citet{Moe2018} predicted that $\sim 60$ per cent of close binaries ($P_\mathrm{orb} < 10$\,days) form via the dynamical unfolding channel, 30 per cent via the KCTF channel (15 per cent during the pre-main sequence and 15 per cent during the main sequence), and 10 per cent by other mechanisms.
A review of these mechanisms and the relevant empirical constraints can be found in \citet{Offner2022}.
}

A full investigation into the presence of triple systems among our targets will be the subject of a follow-up paper, but we present a preliminary discussion here.
Triple systems may be divided into two categories: resolved (meaning that the third object is spatially resolved from the inner binary) and unresolved.
Unresolved triple systems may be further divided by whether the tertiary star is brighter or fainter than the combined light of the inner binary system.

We searched for resolved tertiary companions to our binary candidates by cross-matching our sample with the catalogue of resolved proper-motion companions in \textit{Gaia} \citep{El-Badry2021a}.
We found that 1416 of our targets have a resolved companion ($9.0 \pm 0.2$ per cent, assuming Poisson uncertainties).
In Fig.~\ref{fig:tertiaries} we show the tertiary frequency as a function of inner orbital period. %, and the separation distribution of resolved tertiary companions to our sample.
Binary systems with shorter inner orbital periods are more likely to have resolved tertiary companions, as high as $29.5 \pm 0.2$ for systems with \porb\,$\lesssim 0.3$\,days.

The resolved triples have a range of separations $>200$\,AU, with a median separation of 3200\,AU.
It should be noted that these separation distances are the projection on the sky of the separation vector at the epoch of \textit{Gaia} Data Release 3, not the orbital semi-major axis of the third body.
The majority of these separation distances imply a timescale of tidal interactions which is longer than the Hubble time, but it should be noted that this may not have been true at earlier times in the history of the binary.

\review{
It is interesting to make a comparison with the simulations of \citet{Moe2018}.
Of the formation channels of short-period binary systems modelled by \citet{Moe2018}, this cross-match is most sensitive to those which formed via KCTF interactions during the main sequence lifetime of the binary, which are predicted to have tertiary separations of 200--5000\,AU.
Our measured frequency of triple objects, $9.0 \pm 0.2$ per cent, is comparable to the $\sim 15$ per cent predicted by \citet{Moe2018}.
}

To understand the baseline frequency of resolved companions, we also cross-matched the input set of \inputsize\ \textit{TESS} targets from which  our sample was drawn with the catalogue of \citet{El-Badry2021a}.
Among those input targets, $6.08 \pm 0.01$ per cent have resolved companions. 
Therefore, the frequency of resolved tertiary companions to binary candidates in our sample is enhanced by a factor of $1.48 \pm 0.03$ relative to field \textit{TESS} stars.
Furthermore, as can be seen in Fig.~\ref{fig:tertiaries}, this excess is almost entirely driven by the targets with the shortest inner binary orbital periods.
For inner binaries with periods longer than $\approx 0.6$ days, the frequency of resolved companions is not significantly elevated above the baseline value for single stars.

This enhancement factor is significantly lower than the $2.28 \pm 0.10$ measured by \citet{Hwang2022}.
However, the comparison is not necessarily valid: note that \citet{Hwang2022} applied corrections for separation-dependent incompleteness which we have not applied.
In addition, the blending cut applied in Section~\ref{sec:data} may have introduced further selection effects, and resolved companions may have been missed if they are faint or if they fail the \textit{Gaia} parallax quality cuts applied by \citet{El-Badry2021a}.
We also note that the majority of the detected tertiary companions to the \citet{Tokovinin2006} sample have orbital separations $\lesssim 1000$\,AU, and therefore are in many cases below the resolution limit of \citet{El-Badry2021a} for our targets.
A full study of the prevalence of resolved tertiary companions to our targets will be the subject of a future paper.

In Section~\ref{sec:purity}, we used the LCOGT spectroscopic data obtained for a subset of targets to search for unresolved triples, and place constraints on their existence.
We showed that only a minority (4 out of 47) of the spectroscopically observed targets in the final sample showed evidence for both an RV-variable inner binary and a stationary third object.
We estimated that an unresolved, unbroadened triple star can only be hidden if its luminosity in the observed band is $\lesssim 5$ per cent that of the inner binary.
For an inner binary of 6000\,K stars, this implies that a hidden tertiary must be cooler than $\approx 4500$\,K; if the inner binary is of 5000\,K, the hidden tertiary object must be cooler than $\approx 3800$\,K.
With current data it is difficult to prove or disprove the presence of hidden tertiary objects of this nature.
We additionally note that many triple systems in which the outer body dominates may be entirely removed from the sample by the dilution of the ellipsoidal magnitude, but it is difficult to distinguish these triple systems from low-amplitude pulsating stars.

\section{Conclusions}
\label{sec:conclusion}

We have presented a comprehensive analysis of \samplesize\ ellipsoidal binary systems with orbital periods \porb$< 5$\,days.
Our sample was selected on the basis of their \textit{TESS} full-frame image lightcurves.
This sample offers some of the most detailed insight to date into the properties of short-period, main-sequence, binary systems.

The sample has an estimated purity of $83 \pm 13$ per cent on the basis of spectroscopic follow-up that was obtained for a subset of targets.
Using injection-recovery tests of synthetic lightcurves, we estimate an overall completeness of $58 \pm 6$, $40 \pm 4$, $28 \pm 3$, and $17 \pm 2$ per cent of all binary systems with main sequence primaries and orbital periods shorter than 1, 2, 3, and 5 days, respectively.
These injection-recovery tests also allow us to quantify, and correct for, the selection effects of the sample as a function of the physical properties of the binary systems.

30--50 per cent of the targets in our sample are contact binary systems, which are the dominant class of binary system in our sample for \porb$< 0.6$\,days.
The contact binaries in our sample appear to be typical, with FGK-type primary stars and a correlation between \porb\ and $T_\mathrm{eff}$.

We have explored the \porb\ and primary-$T_\mathrm{eff}$ distributions of our sample, in both cases correcting for the selection effects of the sample.
The detached binary systems in our sample have stellar types of AFGK, reflecting a combination of the magnitude-limited selection of our sample and the correlation between primary mass and companion frequency.
We find that companions with orbital periods in the range 1--5 days are an order of magnitude more common around stars with temperatures hotter than the Kraft break than around cooler stars.
The period distribution shows a contact binary pile-up at \porb\,$\lesssim 0.6$ days, and a uniform companion frequency as a function of orbital period in the range 1--3 days.
The period distribution and companion frequency of this sample are consistent with the the short-period end of the \citet{Moe2017} log-normal distribution at orbital periods of three days, while breaking from that distribution for shorter orbital periods.
On the basis of the orbital period and effective temperature distributions, we suggest that magnetic braking significantly shortens the lifetimes of detached binary systems with \porb\,$\lesssim 3$\,days and $T_\mathrm{eff} \lesssim 6250$\,K, \review{sending them to the contact binary pile-up at $\sim 0.4$\,days}.
\review{However, we note that there is no clear change in the distribution of orbital periods among systems with $T_\mathrm{eff} \lesssim 6250$\,K, only an overall reduction in their number.}

We detected probable third bodies bound to $9.0 \pm 0.2$ per cent of our sample, using the catalogue of resolved companions by \citet{El-Badry2021a}.
These have a median on-sky separation of 3200\,au.
Resolved tertiary companions are significantly more common ($29 \pm 5$ per cent) among the shortest-period binaries (\porb\,$< 0.3$\,days) than among the entire sample.
We also investigated the possibility of unresolved tertiary companions to our binary systems.
Third components were detected spectroscopically in only a minority of targets for which LCOGT follow-up is available.
Cooler third components may be hidden if the temperature difference is $\gtrsim 1000$\,K.
Dilution from a photometrically dominant third body may also remove some triples from our sample.
It is clear that third bodies are common around short-period binary systems, and that their occurrence has a connection to the orbital period of the inner binary.
However, with the current data it is difficult to place a strong constraint on their frequency.

The sample of ellipsoidal binary systems is publicly available in the online material associated with this paper.
This large, homogeneously-selected sample with quantified selection effects will be a powerful resource for future studies of detached and contact binary systems with \porb$<5$\,days.

\section*{Acknowledgements}

\review{We are grateful to the anonymous referee for their feedback, which has improved the quality of the manuscript.
We are grateful to Silvia Toonen, Carles Badenes, and Felipe Lagos for helpful discussions.}

This work was supported by the European Research Council (ERC) under the European Union's FP7 Programme, Grant No. 833031 (MJG, DM). 
We further acknowledge support by a grant from the GIF--the German Israeli Foundation for Science and Development (DM, TM, HWR). 
This research was supported in part by the National Science Foundation under Grant No.~NSF PHY-1748958 (Kavli Institute of Theoretical Physics).

This paper includes data collected by the \textit{TESS} mission. Funding for the \textit{TESS} mission is provided by the NASA's Science Mission Directorate.
This work makes use of observations from the Las Cumbres Observatory global telescope network.

This work made use of the software packages \software{beer}, \software{python}, \software{numpy}, \software{scipy}, \software{astropy}, \software{matplotlib}, \software{eleanor} \citep{Feinstein2019}, \software{sparta} \citep{Shahaf2020}, \software{ellc} \citep{Maxted2016}, \software{phoebe} \citep{Prsa2005,Prsa2016}, \software{topcat}, and \software{saphires}.
This work made use of the Astrophysics Data System (ADS) Abstract Service from NASA, and the Set of Identifications, Measurements and Bibliography for Astronomical Data (SIMBAD) database from the Centre de Donn\'{e}es astronomiques de Strasbourg (CDS).

\section*{Data Availability}

In the online material of this journal can be found the complete versions of Tables~\ref{tab:short-table} and \ref{tab:lco-table}.
Complete versions of Tables~\ref{tab:long-table} and \ref{tab:short-table} will be available on the associated CDS (Centre de données astronomiques de Strasbourg) tables, as well as on Zenodo (DOI: 10.5281/zenodo.7750121).
\textit{TESS} full-frame images are available from the Barbara A. Mikulski Archive for Space Telescopes (MAST),\footnote{https://archive.stsci.edu/missions-and-data/tess} as are the QLP extracted lightcurves used in the northern hemisphere.\footnote{https://archive.stsci.edu/hlsp/qlp}
LCOGT data are available via the LCOGT data archive.\footnote{https://archive.lco.global/}
Any other data can be made available upon reasonable request to the authors.
%The software implementation of the \software{beer} algorithm is proprietary.

%%%%%%%%%%%%%%%%%%%%%%%%%%%%%%%%%%%%%%%%%%%%%%%%%%



\bibliographystyle{mnras}
\bibliography{refs} % if your bibtex file is called example.bib

\begin{thebibliography}{}
\makeatletter
\relax
\def\mn@urlcharsother{\let\do\@makeother \do\$\do\&\do\#\do\^\do\_\do\%\do\~}
\def\mn@doi{\begingroup\mn@urlcharsother \@ifnextchar [ {\mn@doi@}
  {\mn@doi@[]}}
\def\mn@doi@[#1]#2{\def\@tempa{#1}\ifx\@tempa\@empty \href
  {http://dx.doi.org/#2} {doi:#2}\else \href {http://dx.doi.org/#2} {#1}\fi
  \endgroup}
\def\mn@eprint#1#2{\mn@eprint@#1:#2::\@nil}
\def\mn@eprint@arXiv#1{\href {http://arxiv.org/abs/#1} {{\tt arXiv:#1}}}
\def\mn@eprint@dblp#1{\href {http://dblp.uni-trier.de/rec/bibtex/#1.xml}
  {dblp:#1}}
\def\mn@eprint@#1:#2:#3:#4\@nil{\def\@tempa {#1}\def\@tempb {#2}\def\@tempc
  {#3}\ifx \@tempc \@empty \let \@tempc \@tempb \let \@tempb \@tempa \fi \ifx
  \@tempb \@empty \def\@tempb {arXiv}\fi \@ifundefined
  {mn@eprint@\@tempb}{\@tempb:\@tempc}{\expandafter \expandafter \csname
  mn@eprint@\@tempb\endcsname \expandafter{\@tempc}}}

\bibitem[\protect\citeauthoryear{Avallone et~al.,}{Avallone
  et~al.}{2022}]{Avallone2022}
Avallone E.~A.,  et~al., 2022, \mn@doi [The Astrophysical Journal]
  {10.3847/1538-4357/ac60a1}, 930, 7

\bibitem[\protect\citeauthoryear{Babusiaux et~al.,}{Babusiaux
  et~al.}{2018}]{Babusiaux2018}
Babusiaux C.,  et~al., 2018, \mn@doi [Astronomy & Astrophysics]
  {10.1051/0004-6361/201832843}, 616, A10

\bibitem[\protect\citeauthoryear{Badenes et~al.,}{Badenes
  et~al.}{2018}]{Badenes2018}
Badenes C.,  et~al., 2018, \mn@doi [The Astrophysical Journal]
  {10.3847/1538-4357/aaa765}, 854, 147

\bibitem[\protect\citeauthoryear{Bashi, Shahaf, Mazeh, Faigler, Dong, El-Badry,
  Rix  \& Jorissen}{Bashi et~al.}{2022}]{Bashi2022}
Bashi D.,  Shahaf S.,  Mazeh T.,  Faigler S.,  Dong S.,  El-Badry K.,  Rix
  H.-W.,   Jorissen A.,  2022, arXiv, 000, 2207.08832

\bibitem[\protect\citeauthoryear{Bate}{Bate}{2009}]{Bate2009}
Bate M.~R.,  2009, \mn@doi [Monthly Notices of the Royal Astronomical Society]
  {10.1111/j.1365-2966.2008.14106.x}, 392, 590

\bibitem[\protect\citeauthoryear{Bate}{Bate}{2012}]{Bate2012}
Bate M.~R.,  2012, \mn@doi [Monthly Notices of the Royal Astronomical Society]
  {10.1111/j.1365-2966.2011.19955.x}, 419, 3115

\bibitem[\protect\citeauthoryear{Bate, Bonnell  \& Bromm}{Bate
  et~al.}{2002}]{Bate2002}
Bate M.~R.,  Bonnell I.~A.,   Bromm V.,  2002, \mn@doi [Monthly Notices of the
  Royal Astronomical Society] {10.1046/j.1365-8711.2002.05775.x}, 336, 705

\bibitem[\protect\citeauthoryear{Borkovits, Rappaport, Toonen, Moe, Mitnyan  \&
  Cs{\'{a}}nyi}{Borkovits et~al.}{2022}]{Borkovits2022}
Borkovits T.,  Rappaport S.~A.,  Toonen S.,  Moe M.,  Mitnyan T.,
  Cs{\'{a}}nyi I.,  2022, \mn@doi [Monthly Notices of the Royal Astronomical
  Society] {10.1093/mnras/stac1983}, 515, 3773

\bibitem[\protect\citeauthoryear{Brown et~al.,}{Brown et~al.}{2013}]{Brown2013}
Brown T.~M.,  et~al., 2013, \mn@doi [Publications of the Astronomical Society
  of the Pacific] {10.1086/673168}, 125, 1031

\bibitem[\protect\citeauthoryear{Eggleton}{Eggleton}{1983}]{Eggleton1983}
Eggleton P.~P.,  1983, \mn@doi [The Astrophysical Journal] {10.1086/160960},
  268, 368

\bibitem[\protect\citeauthoryear{Eggleton}{Eggleton}{2012}]{Eggleton2012}
Eggleton P.,  2012, Journal of Astronomy and Space Science, 29, 145

\bibitem[\protect\citeauthoryear{El-Badry \& Quataert}{El-Badry \&
  Quataert}{2020}]{El-Badry2020}
El-Badry K.,  Quataert E.,  2020, Monthly Notices of the Royal Astronomical
  Society: Letters, 493, L22

\bibitem[\protect\citeauthoryear{El-Badry \& Quataert}{El-Badry \&
  Quataert}{2021}]{El-Badry2021}
El-Badry K.,  Quataert E.,  2021, \mn@doi [Monthly Notices of the Royal
  Astronomical Society] {10.1093/MNRAS/STAB285}, 502, 3436

\bibitem[\protect\citeauthoryear{El-Badry \& Rix}{El-Badry \&
  Rix}{2018}]{El-Badry2018}
El-Badry K.,  Rix H.-W.,  2018, \mn@doi [Monthly Notices of the Royal
  Astronomical Society] {10.1093/mnras/sty2186}, 480, 4884

\bibitem[\protect\citeauthoryear{El-Badry \& Rix}{El-Badry \&
  Rix}{2019}]{El-Badry2019a}
El-Badry K.,  Rix H.~W.,  2019, \mn@doi [Monthly Notices of the Royal
  Astronomical Society: Letters] {10.1093/mnrasl/sly206}, 482, L139

\bibitem[\protect\citeauthoryear{El-Badry et~al.,}{El-Badry
  et~al.}{2018}]{El-Badry2018a}
El-Badry K.,  et~al., 2018, \mn@doi [Monthly Notices of the Royal Astronomical
  Society] {10.1093/mnras/sty240}, 476, 528

\bibitem[\protect\citeauthoryear{El-Badry, Rix, Tian, Duch{\^{e}}ne  \&
  Moe}{El-Badry et~al.}{2019}]{El-Badry2019}
El-Badry K.,  Rix H.~W.,  Tian H.,  Duch{\^{e}}ne G.,   Moe M.,  2019, \mn@doi
  [Monthly Notices of the Royal Astronomical Society] {10.1093/mnras/stz2480},
  489, 5822

\bibitem[\protect\citeauthoryear{El-Badry, Rix  \& Heintz}{El-Badry
  et~al.}{2021}]{El-Badry2021a}
El-Badry K.,  Rix H.~W.,   Heintz T.~M.,  2021, \mn@doi [Monthly Notices of the
  Royal Astronomical Society] {10.1093/mnras/stab323}, 506, 2269

\bibitem[\protect\citeauthoryear{El-Badry et~al.,}{El-Badry
  et~al.}{2022a}]{El-Badry2022b}
El-Badry K.,  et~al., 2022a, arXiv, 000, arXiv:2208.05488

\bibitem[\protect\citeauthoryear{El-Badry et~al.,}{El-Badry
  et~al.}{2022b}]{El-Badry2022}
El-Badry K.,  et~al., 2022b, arXiv, 000, arXiv:2209.06833

\bibitem[\protect\citeauthoryear{El-Badry, Seeburger, Jayasinghe, Rix, Almada,
  Conroy, Price-Whelan  \& Burdge}{El-Badry et~al.}{2022c}]{El-Badry2022a}
El-Badry K.,  Seeburger R.,  Jayasinghe T.,  Rix H.-W.,  Almada S.,  Conroy C.,
   Price-Whelan A.~M.,   Burdge K.,  2022c, \mn@doi [Monthly Notices of the
  Royal Astronomical Society] {10.48550/arxiv.2203.06348}, 512, 5620

\bibitem[\protect\citeauthoryear{Engel, Faigler, Shahaf  \& Mazeh}{Engel
  et~al.}{2020}]{Engel2020}
Engel M.,  Faigler S.,  Shahaf S.,   Mazeh T.,  2020, \mn@doi [Monthly Notices
  of the Royal Astronomical Society] {10.1093/mnras/staa2182}, 497, 4884

\bibitem[\protect\citeauthoryear{Fabrycky \& Tremaine}{Fabrycky \&
  Tremaine}{2007}]{Fabrycky2007}
Fabrycky D.,  Tremaine S.,  2007, \mn@doi [The Astrophysical Journal]
  {10.1086/521702}, 669, 1298

\bibitem[\protect\citeauthoryear{Faigler \& Mazeh}{Faigler \&
  Mazeh}{2011}]{Faigler2011}
Faigler S.,  Mazeh T.,  2011, \mn@doi [Monthly Notices of the Royal
  Astronomical Society] {10.1111/j.1365-2966.2011.19011.x}, 415, 3921

\bibitem[\protect\citeauthoryear{Faigler \& Mazeh}{Faigler \&
  Mazeh}{2015}]{Faigler2015a}
Faigler S.,  Mazeh T.,  2015, \mn@doi [Astrophysical Journal]
  {10.1088/0004-637X/800/1/73}, 800, 73

\bibitem[\protect\citeauthoryear{Faigler, Tal-Or, Mazeh, Latham  \&
  Buchhave}{Faigler et~al.}{2013}]{Faigler2013}
Faigler S.,  Tal-Or L.,  Mazeh T.,  Latham D.~W.,   Buchhave L.~A.,  2013,
  \mn@doi [Astrophysical Journal] {10.1088/0004-637X/771/1/26}, 771, 26

\bibitem[\protect\citeauthoryear{Faigler et~al.,}{Faigler
  et~al.}{2015}]{Faigler2015b}
Faigler S.,  et~al., 2015, \mn@doi [Astrophysical Journal]
  {10.1088/0004-637X/815/1/26}, 815, 26

\bibitem[\protect\citeauthoryear{Feinstein et~al.,}{Feinstein
  et~al.}{2019}]{Feinstein2019}
Feinstein A.~D.,  et~al., 2019, \mn@doi [Publications of the Astronomical
  Society of the Pacific] {10.1088/1538-3873/AB291C}, 131, 094502

\bibitem[\protect\citeauthoryear{Foreman-Mackey, Hogg, Lang  \&
  Goodman}{Foreman-Mackey et~al.}{2013}]{Foreman-Mackey2013}
Foreman-Mackey D.,  Hogg D.~W.,  Lang D.,   Goodman J.,  2013, \mn@doi
  [Publications of the Astronomical Society of Pacific] {10.1086/670067}, 125,
  306

\bibitem[\protect\citeauthoryear{{Gaia Collaboration} et~al.,}{{Gaia
  Collaboration} et~al.}{2022}]{GaiaCollaboration2022}
{Gaia Collaboration} et~al., 2022, \mn@doi [Astronomy and Astrophysics]
  {10.1051/0004-6361/202039657}, 649, arXiv:2208.00211

\bibitem[\protect\citeauthoryear{Gazeas}{Gazeas}{2009}]{Gazeas2009}
Gazeas K.~D.,  2009, in {Sonja Schuh} {Gerald Handler} eds, Communications in
  Asteroseismology, Vol. 159: JENAM 2008 Symposium No 4: Asteroseismology and
  Stellar Evolution. Verlag der Osterreichischen Akademie der Wissenschaften,
  pp 129--130

\bibitem[\protect\citeauthoryear{Gomel et~al.,}{Gomel et~al.}{2019}]{Gomel2019}
Gomel R.,  et~al., 2019, \mn@doi [Monthly Notices of the Royal Astronomical
  Society] {10.1093/mnras/sty2828}, 482, 5327

\bibitem[\protect\citeauthoryear{Gomel, Faigler  \& Mazeh}{Gomel
  et~al.}{2021a}]{Gomel2021a}
Gomel R.,  Faigler S.,   Mazeh T.,  2021a, Monthly Notices of the Royal
  Astronomical Society, 501, 2822

\bibitem[\protect\citeauthoryear{Gomel, Faigler  \& Mazeh}{Gomel
  et~al.}{2021b}]{Gomel2021b}
Gomel R.,  Faigler S.,   Mazeh T.,  2021b, Monthly Notices of the Royal
  Astronomical Society, 504, 2115

\bibitem[\protect\citeauthoryear{Gomel, Faigler, Mazeh  \& Pawlak}{Gomel
  et~al.}{2021c}]{Gomel2021c}
Gomel R.,  Faigler S.,  Mazeh T.,   Pawlak M.,  2021c, \mn@doi [Monthly Notices
  of the Royal Astronomical Society] {10.1093/mnras/stab1235}, 504, 5907

\bibitem[\protect\citeauthoryear{Halbwachs et~al.,}{Halbwachs
  et~al.}{2022}]{Halbwachs2022}
Halbwachs J.-L.,  et~al., 2022, \mn@doi [arXiv] {10.1051/0004-6361/202243969},
  p. arXiv:2206.05726

\bibitem[\protect\citeauthoryear{Hartman, L{\'{e}}pine  \& Medan}{Hartman
  et~al.}{2022}]{Hartman2022}
Hartman Z.~D.,  L{\'{e}}pine S.,   Medan I.,  2022, \mn@doi [Astrophysical
  Journal] {10.48550/arxiv.2205.13715}, 934, 72

\bibitem[\protect\citeauthoryear{Huang et~al.,}{Huang
  et~al.}{2020a}]{Huang2020a}
Huang C.~X.,  et~al., 2020a, \mn@doi [Research Notes of the American
  Astronomical Society] {10.3847/2515-5172/ABCA2E}, 4, 204

\bibitem[\protect\citeauthoryear{Huang et~al.,}{Huang
  et~al.}{2020b}]{Huang2020b}
Huang C.~X.,  et~al., 2020b, \mn@doi [Research Notes of the American
  Astronomical Society] {10.3847/2515-5172/ABCA2D}, 4, 206

\bibitem[\protect\citeauthoryear{Husser, {Wende-Von Berg}, Dreizler, Homeier,
  Reiners, Barman  \& Hauschildt}{Husser et~al.}{2013}]{Husser2013}
Husser T.~O.,  {Wende-Von Berg} S.,  Dreizler S.,  Homeier D.,  Reiners A.,
  Barman T.,   Hauschildt P.~H.,  2013, \mn@doi [Astronomy and Astrophysics]
  {10.1051/0004-6361/201219058}, 553, A6

\bibitem[\protect\citeauthoryear{Hwang}{Hwang}{2022}]{Hwang2022}
Hwang H.-C.,  2022, eprint arXiv:2208.02257, p. arXiv:2208.02257

\bibitem[\protect\citeauthoryear{Ivanova et~al.,}{Ivanova
  et~al.}{2013}]{Ivanova2013}
Ivanova N.,  et~al., 2013, \mn@doi [Astronomy and Astrophysics Review]
  {10.1007/s00159-013-0059-2}, 21, 59

\bibitem[\protect\citeauthoryear{Jayasinghe et~al.,}{Jayasinghe
  et~al.}{2020}]{Jayasinghe2020}
Jayasinghe T.,  et~al., 2020, \mn@doi [Monthly Notices of the Royal
  Astronomical Society] {10.1093/MNRAS/STAA518}, 493, 4045

\bibitem[\protect\citeauthoryear{Jayasinghe et~al.,}{Jayasinghe
  et~al.}{2021}]{Jayasinghe2021}
Jayasinghe T.,  et~al., 2021, \mn@doi [Monthly Notices of the Royal
  Astronomical Society] {10.1093/MNRAS/STAB907}, 504, 2577

\bibitem[\protect\citeauthoryear{Khouri, Vlemmings, Tafoya,
  P{\'{e}}rez-S{\'{a}}nchez, Contreras, G{\'{o}}mez, Imai  \& Sahai}{Khouri
  et~al.}{2022}]{Khouri2022}
Khouri T.,  Vlemmings W.,  Tafoya D.,  P{\'{e}}rez-S{\'{a}}nchez A.~F.,
  Contreras C.~S.,  G{\'{o}}mez J.~F.,  Imai H.,   Sahai R.,  2022, \mn@doi
  [Nature Astronomy] {10.1038/s41550-021-01528-4}, 6, 275

\bibitem[\protect\citeauthoryear{Kirk et~al.,}{Kirk et~al.}{2016}]{Kirk2016}
Kirk B.,  et~al., 2016, \mn@doi [The Astronomical Journal]
  {10.3847/0004-6256/151/3/68}, 151, 68

\bibitem[\protect\citeauthoryear{Kiseleva, Eggleton  \& Mikkola}{Kiseleva
  et~al.}{1998}]{Kiseleva1998}
Kiseleva L.~G.,  Eggleton P.~P.,   Mikkola S.,  1998, \mn@doi [Monthly Notices
  of the Royal Astronomical Society] {10.1046/j.1365-8711.1998.01903.x}, 300,
  292

\bibitem[\protect\citeauthoryear{Knote, Caballero-Nieves, Gokhale, Johnston  \&
  Perlman}{Knote et~al.}{2022}]{Knote2022}
Knote M.~F.,  Caballero-Nieves S.~M.,  Gokhale V.,  Johnston K.~B.,   Perlman
  E.~S.,  2022, \mn@doi [Astrophysical Journal, Supplement Series]
  {10.48550/arxiv.2206.04142}, 262, 10

\bibitem[\protect\citeauthoryear{Kobulnicky, Molnar, Cook  \&
  Henderson}{Kobulnicky et~al.}{2022}]{Kobulnicky2022}
Kobulnicky H.~A.,  Molnar L.~A.,  Cook E.~M.,   Henderson L.~E.,  2022, The
  Astrophysical Journal Supplement Series, 262, 12

\bibitem[\protect\citeauthoryear{Kopal}{Kopal}{1959}]{Kopal1959}
Kopal Z.,  1959, {Close binary systems}.
Chapman and Hall, London

\bibitem[\protect\citeauthoryear{Kounkel et~al.,}{Kounkel
  et~al.}{2021}]{Kounkel2021}
Kounkel M.,  et~al., 2021, \mn@doi [Astronomical Journal]
  {10.3847/1538-3881/ac1798}, 162, 184

\bibitem[\protect\citeauthoryear{Kozai}{Kozai}{1962}]{Kozai1962}
Kozai Y.,  1962, \mn@doi [Astronomical Journal] {10.1086/108790}, 67, 591

\bibitem[\protect\citeauthoryear{Kraft}{Kraft}{1967}]{Kraft1967}
Kraft R.~P.,  1967, \mn@doi [Astrophysical Journal] {10.1086/149359}, 150, 551

\bibitem[\protect\citeauthoryear{Laos, Stassun  \& Mathieu}{Laos
  et~al.}{2020}]{Laos2020}
Laos S.,  Stassun K.~G.,   Mathieu R.~D.,  2020, \mn@doi [The Astrophysical
  Journal] {10.3847/1538-4357/abb3fe}, 902, 107

\bibitem[\protect\citeauthoryear{Lidov}{Lidov}{1962}]{Lidov1962}
Lidov M.~L.,  1962, \mn@doi [Planetary and Space Science]
  {10.1016/0032-0633(62)90129-0}, 9, 719

\bibitem[\protect\citeauthoryear{Liu et~al.,}{Liu et~al.}{2019}]{Liu2019}
Liu J.,  et~al., 2019, \mn@doi [Nature] {10.1038/S41586-019-1766-2}, 575, 618

\bibitem[\protect\citeauthoryear{Loeb \& Gaudi}{Loeb \& Gaudi}{2003}]{Loeb2003}
Loeb A.,  Gaudi B.~S.,  2003, \mn@doi [The Astrophysical Journal]
  {10.1086/375551}, 588, L117

\bibitem[\protect\citeauthoryear{Lomb}{Lomb}{1976}]{Lomb1976}
Lomb N.~R.,  1976, \mn@doi [Astrophysics and Space Science]
  {10.1007/BF00648343}, 39, 447

\bibitem[\protect\citeauthoryear{Matijevi{\v{c}}, Pr{\v{s}}a, Orosz, Welsh,
  Bloemen  \& Barclay}{Matijevi{\v{c}} et~al.}{2012}]{Matijevic2012}
Matijevi{\v{c}} G.,  Pr{\v{s}}a A.,  Orosz J.~A.,  Welsh W.~F.,  Bloemen S.,
  Barclay T.,  2012, \mn@doi [Astronomical Journal]
  {10.1088/0004-6256/143/5/123}, 143

\bibitem[\protect\citeauthoryear{Maxted}{Maxted}{2016}]{Maxted2016}
Maxted P.~F.,  2016, \mn@doi [Astronomy and Astrophysics]
  {10.1051/0004-6361/201628579}, 591, A111

\bibitem[\protect\citeauthoryear{Mazeh \& Shaham}{Mazeh \&
  Shaham}{1979}]{Mazeh1979}
Mazeh T.,  Shaham J.,  1979, Astronomy and Astrophysics, 77, 145

\bibitem[\protect\citeauthoryear{Mazeh \& Zucker}{Mazeh \&
  Zucker}{1994}]{Mazeh1994}
Mazeh T.,  Zucker S.,  1994, \mn@doi [Astrophysics and Space Science]
  {10.1007/BF00984538}, 212, 349

\bibitem[\protect\citeauthoryear{Mazeh, Faigler, Mazeh  \& Faigler}{Mazeh
  et~al.}{2010}]{Mazeh2010}
Mazeh T.,  Faigler S.,  Mazeh T.,   Faigler S.,  2010, \mn@doi [Astronomy &
  Astrophysics] {10.1051/0004-6361/201015550}, 521, L59

\bibitem[\protect\citeauthoryear{Mazeh, Nachmani, Sokol, Faigler  \&
  Zucker}{Mazeh et~al.}{2012}]{Mazeh2012}
Mazeh T.,  Nachmani G.,  Sokol G.,  Faigler S.,   Zucker S.,  2012, \mn@doi
  [Astronomy and Astrophysics] {10.1051/0004-6361/201117908}, 541, 56

\bibitem[\protect\citeauthoryear{Mazeh et~al.,}{Mazeh et~al.}{2022}]{Mazeh2022}
Mazeh T.,  et~al., 2022, arXiv, 000, arXiv:2206.11270

\bibitem[\protect\citeauthoryear{Mazzola et~al.,}{Mazzola
  et~al.}{2020}]{Mazzola2020}
Mazzola C.~N.,  et~al., 2020, \mn@doi [Monthly Notices of the Royal
  Astronomical Society] {10.1093/mnras/staa2859}, 499, 1607

\bibitem[\protect\citeauthoryear{Moe \& {Di Stefano}}{Moe \& {Di
  Stefano}}{2017}]{Moe2017}
Moe M.,  {Di Stefano} R.,  2017, \mn@doi [The Astrophysical Journal Supplement
  Series] {10.3847/1538-4365/AA6FB6}, 230, 15

\bibitem[\protect\citeauthoryear{Moe \& Kratter}{Moe \&
  Kratter}{2018}]{Moe2018}
Moe M.,  Kratter K.~M.,  2018, \mn@doi [The Astrophysical Journal]
  {10.3847/1538-4357/aaa6d2}, 854, 44

\bibitem[\protect\citeauthoryear{Moe, Kratter  \& Badenes}{Moe
  et~al.}{2019}]{Moe2019}
Moe M.,  Kratter K.~M.,   Badenes C.,  2019, \mn@doi [The Astrophysical
  Journal] {10.3847/1538-4357/ab0d88}, 875, 61

\bibitem[\protect\citeauthoryear{Morris}{Morris}{1985}]{Morris1985}
Morris S.~L.,  1985, \mn@doi [Astrophysical Journal] {10.1086/163359}, 295, 143

\bibitem[\protect\citeauthoryear{Morris \& Naftilan}{Morris \&
  Naftilan}{1993}]{Morris1993}
Morris S.~L.,  Naftilan S.~A.,  1993, \mn@doi [Astrophysical Journal]
  {10.1086/173488}, 419, 344

\bibitem[\protect\citeauthoryear{Naoz \& Fabrycky}{Naoz \&
  Fabrycky}{2014}]{Naoz2014}
Naoz S.,  Fabrycky D.~C.,  2014, \mn@doi [Astrophysical Journal]
  {10.1088/0004-637X/793/2/137}, 793

\bibitem[\protect\citeauthoryear{O'Connell}{O'Connell}{1951}]{OConnell1951}
O'Connell D. J.~K.,  1951, Publications of the Riverview College Observatory,
  2, 85

\bibitem[\protect\citeauthoryear{Offner, Moe, Kratter, Sadavoy, Jensen  \&
  Tobin}{Offner et~al.}{2022}]{Offner2022}
Offner S. S.~R.,  Moe M.,  Kratter K.~M.,  Sadavoy S.~I.,  Jensen E. L.~N.,
  Tobin J.~J.,  2022, arXiv, p. arXiv:2209.00828

\bibitem[\protect\citeauthoryear{Paczy{\'{n}}ski, Szczygie{\l}, Pilecki  \&
  Pojma{\'{n}}ski}{Paczy{\'{n}}ski et~al.}{2006}]{Paczynski2006}
Paczy{\'{n}}ski B.,  Szczygie{\l} D.~M.,  Pilecki B.,   Pojma{\'{n}}ski G.,
  2006, \mn@doi [Monthly Notices of the Royal Astronomical Society]
  {10.1111/J.1365-2966.2006.10223.X}, 368, 1311

\bibitem[\protect\citeauthoryear{Pawlak}{Pawlak}{2016}]{Pawlak2016}
Pawlak M.,  2016, \mn@doi [Monthly Notices of the Royal Astronomical Society]
  {10.1093/MNRAS/STW269}, 457, 4323

\bibitem[\protect\citeauthoryear{Pecaut \& Mamajek}{Pecaut \&
  Mamajek}{2013}]{Pecaut2013}
Pecaut M.~J.,  Mamajek E.~E.,  2013, \mn@doi [Astrophysical Journal, Supplement
  Series] {10.1088/0067-0049/208/1/9}, 208, 9

\bibitem[\protect\citeauthoryear{Pojmanski}{Pojmanski}{2002}]{Pojmanski2002}
Pojmanski G.,  2002, Acta Astronomica, 52, 397

\bibitem[\protect\citeauthoryear{Poro et~al.,}{Poro et~al.}{2022}]{Poro2022}
Poro A.,  et~al., 2022, Monthly Notices of the Royal Astronomical Society, 510,
  5315

\bibitem[\protect\citeauthoryear{Pretorius, Knigge, O'Donoghue, Henry, Gioia
  \& Mullis}{Pretorius et~al.}{2007}]{Pretorius2007}
Pretorius M.~L.,  Knigge C.,  O'Donoghue D.,  Henry J.~P.,  Gioia I.~M.,
  Mullis C.~R.,  2007, \mn@doi [Monthly Notices of the Royal Astronomical
  Society] {10.1111/J.1365-2966.2007.12461.X}, 382, 1279

\bibitem[\protect\citeauthoryear{Price-Whelan et~al.,}{Price-Whelan
  et~al.}{2020}]{Price-Whelan2020}
Price-Whelan A.~M.,  et~al., 2020, \mn@doi [The Astrophysical Journal]
  {10.3847/1538-4357/ab8acc}, 895, 2

\bibitem[\protect\citeauthoryear{Prsa \& Zwitter}{Prsa \&
  Zwitter}{2005}]{Prsa2005}
Prsa A.,  Zwitter T.,  2005, \mn@doi [The Astrophysical Journal]
  {10.1086/430591}, 628, 426

\bibitem[\protect\citeauthoryear{Pr{\v{s}}a et~al.,}{Pr{\v{s}}a
  et~al.}{2011}]{Prsa2011}
Pr{\v{s}}a A.,  et~al., 2011, \mn@doi [Astronomical Journal]
  {10.1088/0004-6256/141/3/83}, 141, 83

\bibitem[\protect\citeauthoryear{Pr{\v{s}}a et~al.,}{Pr{\v{s}}a
  et~al.}{2016}]{Prsa2016}
Pr{\v{s}}a A.,  et~al., 2016, \mn@doi [The Astrophysical Journal Supplement
  Series] {10.3847/1538-4365/227/2/29}, 227, 29

\bibitem[\protect\citeauthoryear{Pr{\v{s}}a et~al.,}{Pr{\v{s}}a
  et~al.}{2022}]{Prsa2022}
Pr{\v{s}}a A.,  et~al., 2022, \mn@doi [The Astrophysical Journal Supplement
  Series] {10.3847/1538-4365/ac324a}, 258, 16

\bibitem[\protect\citeauthoryear{Raghavan et~al.,}{Raghavan
  et~al.}{2010}]{Raghavan2010}
Raghavan D.,  et~al., 2010, \mn@doi [The Astrophysical Journal Supplement
  Series] {10.1088/0067-0049/190/1/1}, 190, 1

\bibitem[\protect\citeauthoryear{Ricker et~al.,}{Ricker
  et~al.}{2014}]{Ricker2014}
Ricker G.~R.,  et~al., 2014, \mn@doi [Proceedings of the SPIE]
  {10.1117/12.2063489}, 9143, 914320

\bibitem[\protect\citeauthoryear{Rivinius, Baade, Hadrava, Heida  \&
  Klement}{Rivinius et~al.}{2020}]{Rivinius2020}
Rivinius T.,  Baade D.,  Hadrava P.,  Heida M.,   Klement R.,  2020, \mn@doi
  [Astronomy and Astrophysics] {10.1051/0004-6361/202038020}, 637, L3

\bibitem[\protect\citeauthoryear{Rowan, Stanek, Jayasinghe, Kochanek, Thompson,
  Shappee, Holoien  \& Prieto}{Rowan et~al.}{2021}]{Rowan2021}
Rowan D.~M.,  Stanek K.~Z.,  Jayasinghe T.,  Kochanek C.~S.,  Thompson T.~A.,
  Shappee B.~J.,  Holoien T.~W.,   Prieto J.~L.,  2021, \mn@doi [Monthly
  Notices of the Royal Astronomical Society] {10.1093/mnras/stab2126}, 507, 104

\bibitem[\protect\citeauthoryear{Rowan et~al.,}{Rowan et~al.}{2022}]{Rowan2022}
Rowan D.~M.,  et~al., 2022, \mn@doi [Monthly Notices of the Royal Astronomical
  Society] {10.48550/arxiv.2205.05687}, 000, 1

\bibitem[\protect\citeauthoryear{Rucinski}{Rucinski}{1997a}]{Rucinski1997a}
Rucinski S.~M.,  1997a, \mn@doi [The Astronomical Journal] {10.1086/118263},
  113, 407

\bibitem[\protect\citeauthoryear{Rucinski}{Rucinski}{1997b}]{Rucinski1997}
Rucinski S.~M.,  1997b, \mn@doi [Astronomical Journal] {10.1086/118329}, 113,
  1112

\bibitem[\protect\citeauthoryear{Rucinski}{Rucinski}{1998}]{Rucinski1998}
Rucinski S.,  1998, \mn@doi [Astronomical Journal] {10.1086/300266}, 115, 1135

\bibitem[\protect\citeauthoryear{Rucinski}{Rucinski}{2002}]{Rucinski2002}
Rucinski S.~M.,  2002, \mn@doi [Publications of the Astronomical Society of the
  Pacific] {10.1086/342677}, 114, 1124

\bibitem[\protect\citeauthoryear{Rybicki \& Lightman}{Rybicki \&
  Lightman}{1979}]{Rybicki1979}
Rybicki G.~B.,  Lightman A.~P.,  1979, {Radiative processes in astrophysics}.
Wiley, New York

\bibitem[\protect\citeauthoryear{Scargle}{Scargle}{1982}]{Scargle1982}
Scargle J.~D.,  1982, \mn@doi [The Astrophysical Journal] {10.1086/160554},
  263, 835

\bibitem[\protect\citeauthoryear{Shahaf, Binnenfeld, Mazeh  \& Zucker}{Shahaf
  et~al.}{2020}]{Shahaf2020}
Shahaf S.,  Binnenfeld A.,  Mazeh T.,   Zucker S.,  2020, {SPARTA:
  SPectroscopic vARiabiliTy Analysis}, \url
  {https://ui.adsabs.harvard.edu/abs/2020ascl.soft07022S/abstract}

\bibitem[\protect\citeauthoryear{Shahaf, Bashi, Mazeh, Faigler, Arenou,
  El-Badry  \& Rix}{Shahaf et~al.}{2022}]{Shahaf2022}
Shahaf S.,  Bashi D.,  Mazeh T.,  Faigler S.,  Arenou F.,  El-Badry K.,   Rix
  H.-W.,  2022, eprint arXiv:2209.00828, 000, arXiv:2209.00828

\bibitem[\protect\citeauthoryear{Shenar et~al.,}{Shenar
  et~al.}{2022}]{Shenar2022}
Shenar T.,  et~al., 2022, \mn@doi [Nature Astronomy]
  {10.1038/s41550-022-01730-y}, 6, 1085

\bibitem[\protect\citeauthoryear{Soszy{\'{n}}ski et~al.,}{Soszy{\'{n}}ski
  et~al.}{2016}]{Soszynski2016}
Soszy{\'{n}}ski I.,  et~al., 2016, Acta Astronomica, 66, 405

\bibitem[\protect\citeauthoryear{Stassun et~al.,}{Stassun
  et~al.}{2019}]{Stassun2019}
Stassun K.~G.,  et~al., 2019, \mn@doi [Astronomical Journal]
  {10.3847/1538-3881/AB3467}, 158, 138

\bibitem[\protect\citeauthoryear{Tal-Or et~al.,}{Tal-Or
  et~al.}{2013}]{Tal-Or2013}
Tal-Or L.,  et~al., 2013, \mn@doi [Astronomy & Astrophysics]
  {10.1051/0004-6361/201220862}, 553, A30

\bibitem[\protect\citeauthoryear{Tal-Or, Faigler  \& Mazeh}{Tal-Or
  et~al.}{2015}]{Tal-Or2015}
Tal-Or L.,  Faigler S.,   Mazeh T.,  2015, \mn@doi [Astronomy & Astrophysics]
  {10.1051/0004-6361/201526425}, 580, A21

\bibitem[\protect\citeauthoryear{Tamuz, Mazeh  \& North}{Tamuz
  et~al.}{2006}]{Tamuz2006}
Tamuz O.,  Mazeh T.,   North P.,  2006, \mn@doi [Monthly Notices of the Royal
  Astronomical Society] {10.1111/j.1365-2966.2006.10049.x}, 367, 1521

\bibitem[\protect\citeauthoryear{Thompson et~al.,}{Thompson
  et~al.}{2019}]{Thompson2019}
Thompson T.~A.,  et~al., 2019, \mn@doi [Science] {10.1126/SCIENCE.AAU4005},
  366, 637

\bibitem[\protect\citeauthoryear{Tokovinin}{Tokovinin}{2014}]{Tokovinin2014}
Tokovinin A.,  2014, \mn@doi [Astronomical Journal]
  {10.1088/0004-6256/147/4/87}, 147

\bibitem[\protect\citeauthoryear{Tokovinin}{Tokovinin}{2022}]{Tokovinin2022}
Tokovinin A.,  2022, Astrophysical Journal, 926, 1

\bibitem[\protect\citeauthoryear{Tokovinin \& Moe}{Tokovinin \&
  Moe}{2020}]{Tokovinin2020}
Tokovinin A.,  Moe M.,  2020, \mn@doi [Monthly Notices of the Royal
  Astronomical Society] {10.1093/mnras/stz3299}, 491, 5158

\bibitem[\protect\citeauthoryear{Tokovinin, Thomas, Sterzik  \& Udry}{Tokovinin
  et~al.}{2006}]{Tokovinin2006}
Tokovinin A.,  Thomas S.,  Sterzik M.,   Udry S.,  2006, \mn@doi [Astronomy and
  Astrophysics] {10.1051/0004-6361:20054427}, 450, 681

\bibitem[\protect\citeauthoryear{Wright, Drake, Mamajek  \& Henry}{Wright
  et~al.}{2011}]{Wright2011}
Wright N.~J.,  Drake J.~J.,  Mamajek E.~E.,   Henry G.~W.,  2011, \mn@doi
  [Astrophysical Journal] {10.1088/0004-637X/743/1/48}, 743, 48

\bibitem[\protect\citeauthoryear{Zhang \& Qian}{Zhang \&
  Qian}{2020}]{Zhang2020}
Zhang X.~D.,  Qian S.~B.,  2020, \mn@doi [Monthly Notices of the Royal
  Astronomical Society] {10.1093/MNRAS/STAA2166}, 497, 3493

\bibitem[\protect\citeauthoryear{Zucker}{Zucker}{2003}]{Zucker2003}
Zucker S.,  2003, \mn@doi [Monthly Notices of the Royal Astronomical Society]
  {10.1046/J.1365-8711.2003.06633.X}, 342, 1291

\bibitem[\protect\citeauthoryear{Zucker \& Mazeh}{Zucker \&
  Mazeh}{1994}]{Zucker1994}
Zucker S.,  Mazeh T.,  1994, \mn@doi [Astrophysical Journal] {10.1086/173605},
  420, 806

\bibitem[\protect\citeauthoryear{Zucker, Mazeh  \& Alexander}{Zucker
  et~al.}{2007}]{Zucker2007}
Zucker S.,  Mazeh T.,   Alexander T.,  2007, \mn@doi [The Astrophysical
  Journal] {10.1086/521389}, 670, 1326

\makeatother
\end{thebibliography}

%\begin{thebibliography}{99}
%\end{thebibliography}

%%%%%%%%%%%%%%%%%%%%%%%%%%%%%%%%%%%%%%%%%%%%%%%%%%

%%%%%%%%%%%%%%%%% APPENDICES %%%%%%%%%%%%%%%%%%%%%

\appendix
\section{LCOGT targets}

The details of targets observed with LCOGT are given in Table~\ref{tab:lco-table}.

\begin{table*}	
\centering
\caption{The properties of 107 targets observed with LCOGT. We present the properties measured from the RV measurements: the largest difference in RV ($\Delta$RV) and its significance, and the best-fit velocity semi-amplitude ($K_1$). We also show the approximate $v_\mathrm{rot} \sin i$ of the primary star; the same value of $v_\mathrm{rot} \sin i$ was applied to the template spectrum before cross-correlation.
Lastly, we show whether the target was classified by eye as RV variable or not, and whether the companion star was also visible in the spectrum (in both cases, classifications are Y [yes], N [no], and ? [unsure]). A further five targets at the bottom of the table were not classified, either because not enough epochs were obtained, because the data were of insufficient S/N, or because a nearby star caused the target to be mis-acquired.
}
\begin{tabular}{lcccccccccccccc}
\label{tab:lco-table}
TIC ID & Score & $P_\mathrm{orb}$ [days] & $K_1$ & $\sigma_{K1}$ & $\Delta$RV &  $\Delta$RV / $\sigma_{\Delta \mathrm{RV}}$  & $\approx v_\mathrm{rot} \sin i$ & Num.\ obs. & RV variable? & Double lined? \\
\hline
10049311 & 0.63 & 0.29 & 103.4 & 5.8 & 46.6 & 135.8 & 1 & 3 & Y & Y \\
398732845 & 0.92 & 0.32 & 5.3 & 1.0 & 7.1 & 36.7 & 1 & 5 & Y & ? \\
237915523 & 0.43 & 0.33 & 115.6 & 20.9 & 93.8 & 31.2 & 141 & 3 & Y & ? \\
296449358 & 0.78 & 0.34 & 11.6 & 1.2 & 14.2 & 12.4 & 100 & 3 & Y & N \\
21442712 & 0.78 & 0.35 & 28.6 & 2.6 & 70.2 & 48.1 & 117 & 4 & Y & N \\
202442617 & 0.64 & 0.35 & 136.8 & 51.7 & 99.6 & 28.9 & 121 & 2 & Y & N \\
842442957 & 0.75 & 0.35 & -- & -- & -- & -- & 200 & 3 & Y & ? \\
143043815 & 0.95 & 0.36 & 108.5 & 1.1 & 166.2 & 121.2 & 96 & 3 & Y & Y \\
124103001 & 0.74 & 0.37 & 131.5 & 37.6 & 68.8 & 16.2 & 133 & 2 & Y & ? \\
398984915 & 0.62 & 0.38 & 22.5 & 1.8 & 35.2 & 12.5 & 80 & 3 & Y & Y \\
35654635 & 0.81 & 0.46 & 16.1 & 3.5 & 25.3 & 5.8 & 158 & 3 & Y & N \\
146349192 & 0.72 & 0.46 & 10.1 & 89.6 & 4.4 & 0.8 & 250 & 2 & Y & ? \\
149583965 & 0.91 & 0.46 & 19.7 & 5.2 & 12.5 & 17.1 & 60 & 3 & Y & N \\
232971368 & 0.90 & 0.46 & 12.7 & 1.0 & 21.3 & 47.2 & 50 & 3 & Y & Y \\
460028598 & 0.63 & 0.47 & 4.9 & 1.0 & 7.2 & 4.6 & 100 & 4 & Y & N \\
298079609 & 0.93 & 0.50 & 164.2 & 3.0 & 100.2 & 116.0 & 111 & 5 & Y & N \\
404509526 & 0.84 & 0.52 & 70.5 & 15.1 & 74.1 & 11.2 & 144 & 3 & Y & N \\
259860383 & 0.74 & 0.53 & 13.1 & 0.6 & 12.4 & 26.6 & 50 & 3 & Y & N \\
80681409 & 0.84 & 0.62 & 13.5 & 3.9 & 20.5 & 40.3 & 80 & 2 & Y & N \\
311924137 & 0.97 & 0.68 & 16.5 & 2.1 & 20.1 & 17.1 & 101 & 3 & Y & N \\
447158270 & 0.84 & 0.75 & 139.2 & 20.6 & 23.5 & 22.6 & 100 & 3 & Y & N \\
262709475 & 0.82 & 0.77 & 12.7 & 2.8 & 5.1 & 9.3 & 50 & 3 & Y & N \\
160159932 & 0.76 & 0.80 & -- & -- & -- & -- & 150 & 3 & Y & N \\
278022607 & 0.99 & 0.88 & 20.7 & 0.9 & 51.8 & 47.1 & 72 & 3 & Y & Y \\
441516862 & 0.64 & 0.93 & 64.7 & 1.4 & 86.3 & 57.1 & 100 & 3 & Y & N \\
314967370 & 0.78 & 1.02 & 7.2 & 1.1 & 10.9 & 11.0 & 63 & 5 & Y & Y \\
153935773 & 0.84 & 1.04 & -- & -- & -- & -- & -- & 1 & Y & Y \\
149473467 & 0.67 & 1.09 & -- & -- & -- & -- & 100 & 2 & Y & N \\
86016254 & 0.62 & 1.12 & 13.1 & 1.6 & 18.8 & 15.9 & 100 & 3 & Y & N \\
115471382 & 0.95 & 1.15 & 48.3 & 16.9 & 21.8 & 10.3 & 97 & 2 & Y & N \\
290598943 & 0.78 & 1.26 & -- & -- & -- & -- & 73 & 3 & Y & ? \\
89060881 & 0.92 & 1.27 & 40.8 & 0.3 & 30.2 & 203.1 & 30 & 3 & Y & N \\
77450928 & 0.61 & 1.50 & 11.2 & 0.3 & 20.6 & 48.5 & 50 & 3 & Y & N \\
198507635 & 0.78 & 1.57 & 132.5 & 1.3 & 204.7 & 176.0 & 80 & 2 & Y & Y \\
429152903 & 0.67 & 1.64 & 53.0 & 1.9 & 23.0 & 32.0 & 80 & 3 & Y & N \\
150144415 & 0.96 & 1.76 & 30.6 & 10.5 & 15.0 & 99.3 & 25 & 2 & Y & N \\
309235372 & 0.84 & 1.78 & 85.0 & 46.9 & 34.2 & 33.9 & 30 & 2 & Y & N \\
192327982 & 0.80 & 1.84 & 40.1 & 14.5 & 9.4 & 14.0 & 51 & 2 & Y & N \\
170673562 & 0.67 & 2.00 & 56.7 & 26.0 & 47.9 & 168.7 & 33 & 2 & Y & N \\
385194219 & 0.64 & 2.13 & 31.1 & 0.4 & 54.7 & 77.2 & 50 & 3 & Y & N \\
443116817 & 0.61 & 2.26 & 5.7 & 1.1 & 2.6 & 8.2 & 32 & 3 & Y & N \\
409827235 & 0.68 & 2.29 & 43.3 & 16.3 & 12.0 & 45.3 & 26 & 2 & Y & ? \\
125135715 & 0.84 & 2.34 & 42.0 & 0.3 & 71.6 & 280.0 & 32 & 3 & Y & N \\
29019282 & 0.74 & 2.77 & 53.9 & 0.7 & 93.1 & 276.6 & 50 & 3 & Y & N \\
117667585 & 0.80 & 2.80 & -- & -- & -- & -- & 24 & 2 & Y & Y \\
201740952 & 0.52 & 2.81 & 75.5 & 0.4 & 105.0 & 229.1 & 28 & 3 & Y & Y \\
435872531 & 0.40 & 2.99 & 85.0 & 0.6 & 156.7 & 140.3 & 32 & 3 & Y & Y \\
451247360 & 0.69 & 3.05 & 96.7 & 0.6 & 155.2 & 296.7 & 33 & 6 & Y & Y \\
147940462 & 0.81 & 4.92 & 93.5 & 0.2 & 160.5 & 670.6 & 29 & 3 & Y & Y \\
103802691 & 0.80 & 5.28 & 28.9 & 0.6 & 64.3 & 311.4 & 14 & 2 & Y & Y \\
137505701 & 0.73 & 0.22 & 0.6 & 1.8 & 1.4 & 2.1 & 100 & 3 & N & N \\
154067797 & 0.82 & 0.24 & 6.2 & 0.3 & 0.1 & 0.6 & 1 & 2 & N & N \\
93282548 & 0.78 & 0.27 & 0.1 & 4.6 & 0.5 & 0.7 & 100 & 2 & N & N \\
1175440578 & 0.94 & 0.27 & 0.0 & 0.1 & 0.1 & 1.2 & 1 & 2 & N & N \\
206717075 & 0.80 & 0.28 & 0.0 & 0.1 & 0.1 & 1.8 & 15 & 3 & N & N \\
\textit{continued...}\\
\hline
\end{tabular}
\end{table*}

\begin{table*}
	\centering
	\contcaption{}
\begin{tabular}{lcccccccccccccc}
TIC ID & Score & $P_\mathrm{orb}$ [days] & $K_1$ & $\sigma_{K1}$ & $\Delta$RV &  $\Delta$RV / $\sigma_{\Delta \mathrm{RV}}$  & $\approx v_\mathrm{rot} \sin i$ & Num.\ obs. & RV variable? & Double lined? \\
\hline
7211118 & 0.45 & 0.29 & -- & -- & -- & -- & 215 & 3 & N & N \\
82376749 & 0.95 & 0.29 & 0.0 & 0.2 & 0.6 & 3.7 & 5 & 5 & N & N \\
185978221 & 0.83 & 0.29 & 0.0 & 0.0 & 0.0 & 0.6 & 1 & 3 & N & N \\
127132146 & 0.79 & 0.32 & 0.1 & 1.1 & 5.5 & 2.7 & 100 & 6 & N & N \\
232222475 & 0.70 & 0.32 & 0.1 & 0.4 & 0.2 & 3.6 & 0 & 3 & N & N \\
388354171 & 0.74 & 0.32 & 0.2 & 0.4 & 0.2 & 1.6 & 1 & 3 & N & N \\
405625148 & 0.76 & 0.32 & 0.7 & 3.0 & 0.1 & 1.9 & 5 & 2 & N & N \\
68740846 & 0.61 & 0.33 & 2.7 & 4.9 & 0.4 & 0.7 & 50 & 2 & N & N \\
331606386 & 0.47 & 0.34 & 2.4 & 1.6 & 1.1 & 2.6 & 0 & 3 & N & N \\
24733359 & 0.76 & 0.38 & 0.1 & 0.5 & 1.2 & 2.3 & 80 & 3 & N & N \\
335720078 & 0.84 & 0.38 & 0.1 & 0.2 & 0.1 & 0.4 & 15 & 3 & N & N \\
142783998 & 0.74 & 0.39 & 74.7 & 62.2 & 1.3 & 5.5 & 1 & 8 & N & N \\
452955107 & 0.91 & 0.39 & 0.1 & 1.1 & 1.3 & 3.4 & 50 & 3 & N & N \\
177441091 & 0.63 & 0.43 & 6.0 & 68.0 & 17.0 & 5.2 & 100 & 3 & N & N \\
95473305 & 0.84 & 0.45 & 0.0 & 0.1 & 0.1 & 0.7 & 1 & 3 & N & N \\
364018117 & 0.80 & 0.46 & 0.6 & 3.1 & 2.7 & 2.1 & 80 & 3 & N & N \\
131239334 & 0.97 & 0.57 & 19.5 & 2.0 & 0.1 & 1.9 & 1 & 2 & N & N \\
221680497 & 0.61 & 0.69 & 4.3 & 40.4 & 13.0 & 2.0 & 134 & 2 & N & N \\
466577705 & 0.64 & 0.71 & 3.2 & 7.1 & 14.7 & 2.6 & 200 & 3 & N & N \\
250579138 & 0.88 & 0.72 & 0.1 & 0.4 & 0.4 & 1.4 & 1 & 3 & N & N \\
112931015 & 0.95 & 0.73 & 0.0 & 0.0 & 0.0 & 0.7 & 1 & 3 & N & N \\
316333039 & 0.69 & 0.83 & 0.2 & 0.1 & 0.3 & 2.2 & 10 & 3 & N & N \\
302828509 & 0.69 & 0.91 & 0.5 & 6.2 & 2.0 & 2.2 & 85 & 3 & N & N \\
68936604 & 0.76 & 1.03 & 0.0 & 0.1 & 0.0 & 0.7 & 1 & 3 & N & N \\
214529076 & 0.80 & 1.07 & 1.2 & 29.6 & 16.4 & 3.1 & 116 & 3 & N & N \\
131004153 & 0.88 & 1.11 & 0.1 & 0.3 & 0.1 & 2.1 & 1 & 3 & N & N \\
169970903 & 0.96 & 1.19 & 4.0 & 8.9 & 15.6 & 4.4 & 78 & 3 & N & N \\
391422697 & 0.62 & 1.39 & 0.0 & 0.3 & 0.4 & 0.9 & 41 & 3 & N & N \\
299885950 & 0.72 & 1.52 & 0.6 & 2.0 & 3.4 & 3.0 & 49 & 3 & N & N \\
168706700 & 0.70 & 1.59 & 13.1 & 18.9 & 4.6 & 1.2 & 150 & 3 & N & N \\
93315603 & 0.66 & 1.68 & 0.0 & 0.1 & 0.2 & 2.4 & 1 & 6 & N & N \\
3989045 & 0.57 & 1.90 & 6.4 & 7.3 & 0.0 & 0.1 & 48 & 2 & N & N \\
273313746 & 0.90 & 2.08 & 0.2 & 5.2 & 3.3 & 1.6 & 80 & 3 & N & N \\
326792846 & 0.43 & 2.14 & 2.7 & 1.6 & 4.2 & 2.6 & 100 & 4 & N & N \\
462548160 & 0.75 & 2.28 & 1.1 & 0.5 & 2.8 & 3.6 & 60 & 3 & N & N \\
193558725 & 0.75 & 0.27 & 89.9 & 39.1 & 62.1 & 8.0 & 200 & 2 & ? & ? \\
231725602 & 0.42 & 0.29 & 711.9 & 59.3 & 97.8 & 12.4 & 267 & 2 & ? & ? \\
294052459 & 0.91 & 0.39 & 6.4 & 7.1 & 42.3 & 6.9 & 150 & 3 & ? & ? \\
346315955 & 0.67 & 0.51 & 93.0 & 16.0 & 74.2 & 12.7 & 100 & 3 & ? & ? \\
427606080 & 0.49 & 0.54 & 23.4 & 2.8 & 26.1 & 9.8 & 250 & 3 & ? & ? \\
432761278 & 0.94 & 0.64 & 9.0 & 31.5 & 21.1 & 3.5 & 118 & 4 & ? & ? \\
264594607 & 0.79 & 0.82 & 356.3 & 9.3 & 143.0 & 15.8 & 150 & 3 & ? & ? \\
642625967 & 0.84 & 1.28 & 18.7 & 2.3 & 35.4 & 9.1 & 74 & 3 & ? & ? \\
254876862 & 0.74 & 1.30 & 118.7 & 80.5 & 118.6 & 18.8 & 74 & 2 & ? & ? \\
74051493 & 0.86 & 1.34 & 111.3 & 13.2 & 196.0 & 16.7 & 71 & 3 & ? & ? \\
390535698 & 0.73 & 1.76 & 14.8 & 10.9 & 13.8 & 3.1 & 46 & 3 & ? & ? \\
356209335 & 0.77 & 2.02 & 35.5 & 5.2 & 130.6 & 15.9 & 80 & 2 & ? & ? \\
8547649 & 0.95 & 0.35 & -- & -- & -- & -- & -- & 1 & -- & -- \\
22877676 & 0.82 & 0.35 & -- & -- & -- & -- & -- & 1 & -- & -- \\
374842122 & 0.94 & 0.42 & -- & -- & -- & -- & -- & 2 & -- & -- \\
37654547 & 0.89 & 1.00 & -- & -- & -- & -- & -- & 3 & -- & -- \\
245035219 & 0.77 & 1.27 & -- & -- & -- & -- & -- & 1 & -- & -- \\
\hline
\end{tabular}
\end{table*}

%%%%%%%%%%%%%%%%%%%%%%%%%%%%%%%%%%%%%%%%%%%%%%%%%%

% Don't change these lines
\bsp	% typesetting comment
\label{lastpage}
\end{document}